\documentstyle[12pt,epsfig,color,subfigure]{article}
\newcommand{\be}{\begin{equation}}
\newcommand{\ee}{\end{equation}}
\newcommand{\beq}{\begin{equation}}
\newcommand{\eeq}{\end{equation}}
\newcommand{\ba}{\begin{eqnarray}}
\newcommand{\ea}{\end{eqnarray}}
\newcommand{\bea}{\begin{eqnarray}}
\newcommand{\eea}{\end{eqnarray}}
\newcommand{\nn}{\nonumber}
\begin{document}
\baselineskip=15.5pt \pagestyle{plain} \setcounter{page}{1}
%
\begin{titlepage}

\vskip 0.8cm

\begin{center}

{\Large \bf  Deep inelastic scattering off scalar mesons in the 1/N
expansion from the D3D7-brane system}

\vskip 1.cm

{\large {{\bf David Jorrin}{\footnote{\tt
jorrin@fisica.unlp.edu.ar}}, {\bf Nicolas Kovensky}{\footnote{\tt
nico.koven@fisica.unlp.edu.ar}}, {\bf and Martin
Schvellinger}{\footnote{\tt martin@fisica.unlp.edu.ar}}}}

\vskip 1.cm

{\it Instituto de F\'{\i}sica La Plata-UNLP-CONICET \\ and \\
Departamento de F\'{\i}sica, Facultad de Ciencias Exactas,
Universidad Nacional de La Plata. \\
Calle 49 y 115, C.C. 67, (1900) La Plata, Buenos Aires, Argentina.} \\

\vspace{1.cm}

{\bf Abstract}

\vspace{1.cm}

\end{center}

Deep inelastic scattering (DIS) of charged leptons off scalar mesons
in the $1/N$ expansion is studied by using the gauge/gravity
duality. We focus on the D3D7-brane system and investigate the
corresponding structure functions by considering both the high
energy limit and the $1/N$ expansion. These limits do not commute.
From the D7-brane DBI action we derive a Lagrangian at sub-leading
order in the D7-brane fluctuations and obtain a number of
interactions some of which become relevant for two-hadron
final-state DIS. By considering first the high energy limit followed
by the large $N$ one, our results fit lattice QCD data within $1.27
\%$ for the first three moments of $F_2$ for the lightest
pseudoscalar meson.

\noindent

\end{titlepage}

\newpage

{\small \tableofcontents}

\newpage

\section{Introduction}

Dp-brane models for holographic mesons lead to very interesting
results describing masses of the low-lying mesons, meson
interactions, as well as other important properties obtained in the
large $N$ limit of their corresponding dual confining gauge field
theories at strong coupling
\cite{Kruczenski:2003be,Kruczenski:2003uq,Sakai:2004cn}. These
models include the description of quarks in the fundamental
representation of the gauge group by using flavor Dp-branes in the
probe approximation. However, we should note that there is no
holographic dual model which exactly represents all properties of
real QCD, even at large $N$. In particular, for the referred
Dp-brane modes we can comment on some of their main differences with
respect to large $N$ QCD as follows. The Sakai-Sugimoto model
\cite{Sakai:2004cn} is built out of $N$ D4-branes in type IIA
superstring theory, by adding $N_f$ D8-branes and $N_f$
$\overline{\textrm{D8}}$-branes in the probe approximation. The most
important property of this model is that it gives a geometric
realization of the chiral symmetry breaking. Recall that
antiperiodic boundary conditions for the fermions are imposed on an
$S^1$ where one of the spatial directions of the D4-brane is
wrapped, thus supersymmetry is completely broken. At low energy this
model describes several properties as QCD does for mesons in the
large $N$ limit. At high energy, on the other hand, the size of the
circumference grows, which implies that the Kaluza-Klein modes
become relevant. This is a signal that the dual gauge theory becomes
a five-dimensional one. Moreover, the existence of an $S^4$ in the
ten-dimensional superstring theory background leads to a global
$SO(5)$ symmetry which is absent in QCD. Neither the model based on
$N$ D4-branes and flavor D6-branes in type IIA superstring theory
represents real QCD at its full extent \cite{Kruczenski:2003uq}. In
addition, the D3D7-brane model in type IIB superstring theory that
we investigate in the present work is the holographic dual
description of the $SU(N)$ ${\cal {N}} = 2$ supersymmetric
Yang-Mills theory in four dimensions, in the multicolor limit and at
strong coupling, with quarks in the fundamental representation of
the gauge group \cite{Kruczenski:2003be}. Another property which
distinguishes between the D3D7-brane model and QCD is that the
D3D7-brane system does not lead to any geometric realization of
chiral symmetric breaking.

Thus, none of the mentioned Dp-brane models are exact holographic
dual models of the large $N$ limit of QCD. However, as we have shown
in \cite{Koile:2011aa,Koile:2013hba,Koile:2014vca,Koile:2015qsa}, it
is possible to investigate the internal structure of the
corresponding scalar and polarized vector mesons of the models
\cite{Kruczenski:2003be,Kruczenski:2003uq,Sakai:2004cn} by using the
gauge/string duality at large $N$ and strong coupling. This is very
interesting because in references
\cite{Koile:2011aa,Koile:2013hba,Koile:2014vca,Koile:2015qsa} it has
been found that the behavior of the corresponding structure
functions is {\it model independent} in the sense that relations of
the Callan-Gross type, as well as generalizations of it to other
structure functions for polarized vector mesons, hold independently
of which Dp-brane model one considers. This means that there is a
sort of universal behavior for the meson structure functions which
should be shared by the large $N$ limit of QCD. This universal
property is due to the fact that the dynamics of mesons in the
string theory dual model is accounted for by the Dirac-Born-Infeld
(DBI) action of the corresponding flavor Dp-brane. We have chosen
the D3D7-brane model for several reasons. Firstly, as we have
already mentioned their meson structure functions display universal
behavior for relations of the Callan-Gross type. Secondly, the
relative simplicity of the geometry of this background allows one to
perform a detailed analysis and obtain explicit expressions of the
structure functions of scalar mesons in the $1/N$ expansion, which
in the end permits to obtain results to compare with lattice QCD and
phenomenology. Thirdly, the D3D7-model is the one which compares
better with lattice QCD results for the pion and the rho meson (for
$N \rightarrow \infty$)
\cite{Best:1997qp,Brommel:2006zz,Chang:2014lva,Koile:2015qsa}. In
addition, the fall-off of the structure functions obtained from the
D3D7-brane model at large $N$ for $x \rightarrow 1$ leads to a
factor $(1-x)^2$, in agreement with phenomenological results
\cite{Wijesooriya:2005ir,Holt:2010vj,Reimer:2011,Chang:2014gga,Detmold:2003tm,Aicher:2011ai,Aicher:2010cb,Koile:2015qsa}.

Also, it is interesting to notice that by using the gauge/gravity
duality in the case of the $SU(N)$ ${\cal {N}}=4$ SYM theory
glueballs have been studied in the large $N$ limit, while lattice
gauge theory simulations have permitted to investigate their
properties at finite $N$
\cite{Meyer:2004jc,Meyer:2004hv,Meyer:2004gx}. Moreover, meson
spectrum and decay constants have been obtained in the quenched
approximation with the Wilson fermion action for $N = 2, 3, 4, 5, 6,
7$ and $17$ and then extrapolated to $N \rightarrow \infty$
\cite{Bali:2013kia, Bali:2013fya}.

Recently, the structure of holographic mesons in the
\cite{Kruczenski:2003be,Kruczenski:2003uq,Sakai:2004cn} models has
been investigated in the large $N$ limit and at strong coupling
\cite{Koile:2011aa,Koile:2013hba,Koile:2014vca,Koile:2015qsa}, which
corresponds to considering single-hadron final states. The process
under investigation is the deep inelastic scattering (DIS) of a
charged lepton from a hadron. Its differential cross section is
obtained in terms of the forward Compton scattering (FCS) by using
the optical theorem in quantum field theory (QFT). In the strong
coupling limit of the QFT, the appropriate framework to calculate
the structure functions is the gauge/string duality
\cite{Polchinski:2002jw}. In particular, within the Bjorken
parameter range $1/\sqrt{\lambda} \ll x < 1$ for scalar mesons it
turns out that the structure function $F_1$ vanishes since it is
proportional to the corresponding Casimir operator of the Lorentz
group, which has been confirmed by direct calculation using
supergravity \cite{Koile:2011aa,Koile:2013hba}. On the other hand,
for $\exp{(-\sqrt{\lambda})} \ll x \ll 1/\sqrt{\lambda}$, $F_1$ does
not vanish and it is obtained in terms of superstring theory
\cite{Koile:2014vca}. Recall that $\lambda$ is the 't Hooft coupling
and $x$ the Bjorken parameter. The reason for this behavior of $F_1$
comes from the fact that at strong coupling the virtual photon
probes the entire hadron, thus within the supergravity framework no
partons are found in this limit, $1 \ll \lambda \ll N$. The
non-vanishing structure function $F_2$, on the other hand, has also
been calculated in \cite{Koile:2011aa,Koile:2013hba,Koile:2014vca}
in the corresponding parametric regimes of $x$, and it has been
shown how its first moments agree with the corresponding results
from lattice QCD simulations
\cite{Best:1997qp,Brommel:2006zz,Chang:2014lva} with an accuracy of
$10.8 \%$ \cite{Koile:2015qsa}. Similar results have been obtained
for the Sakai-Sugimoto model and for the D4D6-brane model
\cite{Koile:2015qsa}. The result of \cite{Koile:2015qsa} strictly
corresponds to the tree-level Feynman diagram for FCS, i.e. by
considering a single-hadron final state DIS, which in terms of the
following discussion corresponds to considering first the $N
\rightarrow \infty$ limit and then the high energy limit.

Beyond the large $N$ limit, within the gauge/string duality
framework one must consider the $1/N$ expansion for which there are
two possible approaches which work for different regimes of the
Bjorken parameter. For $\exp{(-\sqrt{\lambda})} \ll x <
1/\sqrt{\lambda}$ it is required a genus expansion in superstring
theory, while for $1/\sqrt{\lambda} \ll x \leq 1$ it is enough to
include Feynman loop diagrams in the supergravity calculation. In
both situations, a genus-one world-sheet in superstring theory and
the corresponding one-loop diagrams in supergravity, lead to the
holographic dual description of one-loop FCS in the dual QFT. This
corresponds to a two-hadron final state in DIS within two different
kinematical regimes of $x$. Specifically, we can look at the
longitudinal structure function of a scalar meson in the $1/N$
expansion, and simultaneously we can also perform an expansion in
inverse powers of the momentum transfer of the virtual photon $q$,
which leads to
\begin{eqnarray}
F_L &=& F_2 - 2 \, x \, F_1 \nonumber \\
&=& f^{(0)}_2 \, \left(\frac{\Lambda^2}{q^2}\right)^{\Delta_{in}-1}
+ \frac{1}{N} \, \left( f^{(1)}_2 - 2 \, x \, f^{(1)}_1 \right) \,
\left(\frac{\Lambda^2}{q^2}\right) + \frac{1}{N^2} \, \left(
f^{(2)}_2 - 2 \, x \, f^{(2)}_1 \right) \,
\left(\frac{\Lambda^2}{q^2}\right) + \cdot \cdot \cdot \label{FL}
\nonumber \\
&&
\end{eqnarray}
where $\Lambda$ is an IR confining scale of the QFT. Notice that
$\Delta_{in}$ is the conformal dimension of the incident scalar
state in supergravity, while $f^{(n)}_i$'s stand for the structure
functions at the corresponding order in $1/N^n$, with $i=1, 2$ and
$n=0, 1, \dots$, where $n$ indicates the number of loops of the FCS
Feynman diagram (i.e. the number of hadrons in the final state DIS).

Recall that for glueballs there is a $1/N^{2n}$ expansion instead of
the $1/N^n$ one shown in the previous equation, which simply
reflects the fact that glueballs in the calculations
\cite{Jorrin:2016rbx} are made of ${\cal {N}}=4$ SYM theory fields
in the gauge supermultiplet (thus all of them belong to the adjoint
representation of $SU(N)$). On the other hand, the mesons considered
here correspond to fields of the hypermultiplet of ${\cal {N}}=2$
SYM theory, thus being in the fundamental representation. For the
glueballs of ${\cal {N}}=4$ SYM theory it turns out that the large
$N$ limit and the high energy limit, i.e. $q^2 \gg \Lambda^2$ do not
commute \cite{Gao:2014nwa,Jorrin:2016rbx}. This leads to an
important consequence on the longitudinal structure function for
glueballs $F_L^{glueball}$, which shows a rich structure for the
currents which contain spin-1, spin-$1/2$ and spin-0 fields from the
${\cal {N}}=4$ SYM theory.

We would naively expect that the results for scalar mesons should
not change substantially in comparison with those for glueballs.
Still it is really worth to carry out these explicit calculations
because there are both lattice QCD
\cite{Best:1997qp,Brommel:2006zz,Chang:2014lva} as well as
phenomenological results
\cite{Wijesooriya:2005ir,Holt:2010vj,Reimer:2011,Chang:2014gga} to
compare with for scalar mesons, in particular for the pion.

In fact, we will show how by considering first the high energy limit
$q^2 \gg \Lambda^2$ and then the $N \rightarrow \infty$ limit, we
obtain expressions for the structure functions for scalar mesons
which lead to results for the moments of $F_2$ which compare very
well with lattice QCD simulations\footnote{Notice that for the
comparison with lattice QCD data we consider $F_L \sim F_2$ since
$F_1$ is sub-leading in the large energy limit.}. In particular, for
the case of the pion the agreement with lattice QCD results for the
first three moments of $F_2$
\cite{Best:1997qp,Brommel:2006zz,Chang:2014lva} is within $1.27 \%$
accuracy. This shows the importance of taking these limits in the
correct order to obtain physically sensible results. The reason for
the difference between $10.8 \%$ accuracy obtained in
\cite{Koile:2015qsa} and the $1.27 \%$ accuracy obtained in the
present work comes from the fact that in \cite{Koile:2015qsa} we
considered the large $N$ limit first, which implies that in the
previous equation only the first term contributes to the structure
functions. On the other hand, when we consider first the high energy
limit, the second term of equation (\ref{FL}) is the relevant one,
while the first term leads to a smaller contribution. In order to
have an idea of the level of accuracy of the present results notice
for instance that within the gauge/gravity duality two-point
functions usually lead to $10 \%$ differences with respect to
observables for mesons \cite{Erlich:2005qh,DaRold:2005mxj}, while
four-point functions lead to about $30 \%$ differences
\cite{Hambye:2005up,Hambye:2006av}, which is reasonable taking into
account that these calculations have been done in the large $N$
limit, in comparison with real QCD, i.e. $N=3$.

Also, it is worth mentioning that for the ${\cal {N}}=4$ SYM plasma
the DC electrical conductivity, spectral functions and photoemission
rates are also calculated from the correlation functions of two
electromagnetic currents. In fact, these properties have been
calculated in \cite{CaronHuot:2006te} in the strong coupling limit
for $1 \ll \lambda \ll N$, while in
\cite{Hassanain:2011ce,Hassanain:2011fn,Hassanain:2010fv,Hassanain:2012uj}
the ${\cal {O}}(\alpha'^3)$ corrections from type IIB superstring
theory have been calculated. Although these calculations strictly
hold in the large $N$ limit and the strong coupling expansion, i.e.
$1/\lambda$, it turns out that by setting $N=3$ and $\lambda \approx
15$ there is a good agreement with lattice QCD simulations
\cite{Aarts:2007wj}. In addition, for the ${\cal {N}}=4$ SYM plasma
at strong coupling the structure functions $F_1$ and $F_2$ have been
obtained in \cite{Hatta:2007he,Hatta:2007cs}, while ${\cal
{O}}(\alpha'^3)$ corrections from type IIB string theory have been
calculated in \cite{Hassanain:2009xw}.

~

This paper is organized as follows. In Section 2 we carry out a
detailed derivation of the interaction Lagrangian at different
orders in terms of the D7-brane fluctuations. This is done by
starting from the Dirac-Born-Infeld action of the D7-brane in the
probe approximation. We also describe the solutions of the
corresponding equations of motion. In Section 3 we calculate the
leading one-loop Feynman-Witten diagram in type IIB supergravity,
which corresponds to the Bjorken parameter range $1/\sqrt{\lambda}
\ll x < 1$. Then, from this one-loop supergravity diagram we obtain
the structure functions for scalar mesons. In Section 4 we perform a
comparison with lattice QCD simulations and with phenomenological
results, and carry out the discussion and conclusions. Some details
of the calculations are described in the appendices.

\section{The interaction Lagrangian}

\subsection{Derivation of the interaction Lagrangian from the
D7-brane DBI-action}

In this section we begin with the derivation of the interaction
Lagrangian corresponding to scalar mesons from the Dirac-Born-Infeld
action of a single D7-brane\footnote{We consider a single-flavor
calculation in the dual gauge field theory. The multi-flavor
generalization can be easily done following \cite{Koile:2013hba},
where a single-hadron final state has been considered.} in the
AdS$_5 \times S^5$ background obtained from the backreaction of $N$
D3-branes in type IIB superstring theory
\begin{equation}
ds^2=\frac{r^2 }{R^2}ds^2(E^{(1,3)})+\frac{R^2}{r^2}d\vec{Z} \cdot
d\vec{Z} \, . \label{metricGAB}
\end{equation}
Let us call the metric (\ref{metricGAB}) $G_{AB}$, with $A, B = 0,
1, \cdot \cdot \cdot, 9$. The Dirac-Born-Infield action of the
D7-brane is given by
\begin{equation}
S=-\mu_7 \int d^8\xi \ \ \sqrt{-\det(P[G]_{a b}+2\pi \alpha'
F_{ab})}  + \frac{(2\pi \alpha')^2}{2} \mu_7 \int P[C^{(4)}]\wedge F
\wedge F \, ,
\end{equation}
where the relevant part of the Ramond-Ramond potential $C^{(4)}$ is
given in \cite{Kruczenski:2003be}, while $P$ stands for the pullback
of the metric
\begin{equation}
P[G]_{a b}= G_{A B} \frac{d x^A}{d\xi^a} \frac{d x^B}{d\xi^b} \, ,
\end{equation}
being $a, b = 0, 1, \cdot \cdot \cdot, 7$ the indices which
parameterize the D7-brane coordinates. The coordinates perpendicular
to the D7-brane are $Z^5$ and $Z^6$, and following
\cite{Kruczenski:2003be} one can parameterize the transversal
fluctuations in terms of two scalar fields $\chi$ and $\phi$ by
\begin{eqnarray}
Z^5=2 \pi \alpha' \chi  \, , \ \ \ \ Z^6=L+2 \pi \alpha' \phi \, ,
\end{eqnarray}
which represent the holographic scalar mesons. On the other hand,
$Z^i$ with $i=1, \dots, 4$ are parameterized in terms of spherical
coordinates with radius $\rho$ and angles $\psi$, $\theta$ and
$\omega$. The radial coordinate $r$ of the AdS$_5$ can be written in
terms of the new coordinates as
\begin{equation}
r^2=\rho^2+(L+ 2\pi \alpha ' \phi)^2+(2 \pi \alpha' \chi)^2 \, ,
\end{equation}
and the metric induced by the D7-brane fluctuations is then
\begin{equation}
ds^2=\frac{r^2}{R^2}ds^2(E^{(1,3)})+\frac{R^2}{r^2}[(2 \pi \alpha'
)^2(d\chi^2+d\phi^2)+d\rho^2+\rho^2d\Omega_3] \, . \label{metric8}
\end{equation}
In order to solve the equations of motion (EOM) we consider the
static gauge with $x^i=\xi^i$ for $i=0, \dots, 3$, while
$\rho=\xi^4$, $\psi=\xi^5$, $\theta=\xi^6$ and $\omega=\xi^7$.
Therefore, the holographic scalar mesons are functions of these
coordinates: $\phi(\xi_i)$ and $\chi(\xi_i)$. Moreover, in order to
obtain the interaction vertices one has to carry out a Taylor series
expansion in $\phi$ and $\chi$ around the classical solution
$\phi=0$ and $\chi=0$. We identify two kinds of fluctuations of the
pullback
\begin{eqnarray}
P[G]_{a b}&=&\left(G_{M N}|_{\chi,\phi=0} + \frac{\partial G_{M
N}}{\partial\chi}|_{\chi,\phi=0} \ \chi + \frac{\partial G_{M
N}}{\partial\phi}|_{\chi,\phi=0} \ \phi + {\cal{O}}(\phi^2,
\phi\chi, \chi^2)\right) \nonumber \\
&& \times\left(\delta^{M}_{a} \delta^N_{b}+ \delta^M_8
\delta^N_8\partial_a \phi \partial_b \phi +\delta^M_9 \delta^N_9
\partial_a \chi
\partial_b \chi\right). \label{PGab}
\end{eqnarray}
Recall that the zeroth order term, $P[G]^{(0)}_{ab}$, is given by
$g_{ab} \equiv G_{M N}|_{\chi,\phi=0} \delta^{M}_{a} \delta^N_{b}$,
and it is obtained from the induced metric on the D7-brane
\begin{equation}
ds^2 = \frac{r_0^2}{R^2}\eta_{\mu\nu}dx^\mu dx^\nu +
\frac{R^2}{r_0^2}\left(d\rho^2 + \rho^2 d\Omega_3^2\right) \, ,
\label{metric-on-D7-brane}
\end{equation}
where $\mu, \nu = 0, 1, 2, 3$, and $r_0^2 = \rho^2 + L^2$. Thus, due
to fluctuations perpendicular to the D7-brane the pullback changes
as $P[G]^{(0)}_{ab} \rightarrow P[G]_{ab} = P[G]^{(0)}_{ab} + h_{ab}
+ X_{ab}$. We can write $h_{ab}=\sum_i h^{(i)}_{ab}$, with $i=1,
\dots, 4$ indicating at which order the scalar fluctuations appear
in the metric. Fluctuations $h_{ab}$ come from the $\delta^{M}_{a}
\delta^N_{b}$ terms in equation (\ref{PGab}). In addition, we must
consider the contributions due to the product of the metric
expansion times the derivatives of the scalar fluctuations. These
generate the kinetic terms of the effective Lagrangian. They are
also induced by the perturbations in the transverse directions to
the D7-brane and are denoted by $X_{ab}=\sum_j X^{(j)}_{ab}$ with
$j=2, 3, 4$ being $j$ the order at which the fluctuation appears.

In order to calculate $h_{ab}$ and $X_{ab}$, let us focus on the
fluctuations of metric tensor. We only need to consider the
following metric warp factors:
\begin{eqnarray}
\frac{r^2}{R^2}&=&\frac{r_0^2}{R^2} +
\frac{1}{R^2}\left[2(2\pi\alpha')L\phi + (2\pi\alpha')^2(\phi^2 + \chi^2)\right] \\
\frac{R^2}{r^2}&=&\frac{R^2}{r_0^2} +
\frac{R^2}{r_0^4}\left\{-2(2\pi\alpha')L\phi + (2\pi\alpha')^2
\left[\left(4\frac{L^2}{r_0^2}-1\right)\phi^2 - \chi^2\right]
\right. \nonumber \\
&& \left. + \frac{(2\pi\alpha')^3}{r_0^2}4L\left[
\left(1-2\frac{L^2}{r_0^2}\right)\phi^3 + \phi \chi^2\right] \right)
+ {\cal{O}}(\phi^4,\phi^3 \chi,...) \, .
\end{eqnarray}
The expansion is written up to fourth order terms indicated by
${\cal{O}}$. By plugging these expressions in the induced metric
(\ref{metric8}) we obtain the $h_{ab}$ and the $X_{ab}$
contributions. The latter are given by
\begin{eqnarray}
X_{ab} &=& (2\pi\alpha')^2 \left[\frac{R^2}{r_0^2} +
\frac{R^2}{r_0^4}\left\{-2(2\pi\alpha')L\phi + (2\pi\alpha')^2
\left[\left(4\frac{L^2}{r_0^2}-1\right)\phi^2 -
\chi^2\right]\right\}\right] \times \nonumber \\
&& \left(\partial_a \phi \partial_b \phi + \partial_a \chi
\partial_b \chi\right) \equiv X_{ab}^{(2)} + X_{ab}^{(3)} +X_{ab}^{(4)}
\ .
\end{eqnarray}

Now, let us consider a generic background metric $M_{ab}$ with
perturbations of the form $m_{ab}$. One can write the following
expression
\begin{eqnarray}
\sqrt{\det\left(M_{ab}+ m_{ab}\right)} &=& \sqrt{M}\left[ 1+
\frac{1}{2}m + \left(\frac{1}{8}m^2 -\frac{1}{4}m \cdot m\right)
\right.
\nonumber \\
&& \left. + \left( \frac{1}{48}m^3 - \frac{1}{8} m (m\cdot m)
+ \frac{1}{6} m\cdot m \cdot m \right)\right. \nonumber \\
&&\left. + \left( \frac{1}{384} m^4 + \frac{1}{32} (m\cdot m)^2 -
\frac{1}{32} m^2 (m\cdot m) + \frac{1}{12}m (m\cdot m \cdot m)
\right. \right. \nonumber \\
&& \left. \left. - \frac{1}{8} m \cdot m \cdot m \cdot m\right)
\right] \, ,
\end{eqnarray}
where all indices are raised and lowered with the unperturbed metric
$M$. We use the following notation:
\begin{eqnarray}
m \equiv m^{a}_a = M^{ab}m_{ab} \ , \ \ \ \ \ m^2  =
(M^{ab}m_{ab})^2 \ , \ \ \ \ \ m\cdot m \equiv m^a_b m^b_a =
M^{bc}M^{ad} m_{ab}m_{cd}.
\end{eqnarray}
In the present case we set $M_{ab}=g_{ab}$, i.e. the unperturbed
metric induced on the D7-brane, and consider the following matrix
perturbation $m_{ab} = h_{ab} + X_{ab} + \tilde{F}_{ab}$, where
$\tilde{F}_{ab} = 2 \pi \alpha' F_{ab}$. Recall that $F_{ab}$ are
the contributions from the fluctuations along
the D7-brane directions associated with vector mesons.

Now, we can derive the Lagrangian terms order by order in the
perturbations as follows.

~

\subsubsection*{First order effective Lagrangian}

As expected, there are no linear terms in the fluctuations of the
metric, thus this Lagrangian vanishes as shown below\footnote{Note
that $X^{(1)}_{ab}=0$.}
\begin{eqnarray}
L_{1} =  -\mu_7 \sqrt{-g} \left[\frac{1}{2}m^{(1)}\right] =
-\frac{\mu_7}{2}\sqrt{-g} \, g^{ab} \left(h^{(1)}_{ab} +
\tilde{F}_{ab}\right) = -\frac{\mu_7}{2}\sqrt{-g} \, h^{(1)} = 0 \,
,
\end{eqnarray}
where obviously $g^{ab} \tilde{F}_{ab}=0$, while the trace of
$h^{(1)}$ also vanishes. \\

\subsubsection*{Second order effective Lagrangian}

This leads to the kinetic terms for the scalar and vector
fluctuations which correspond to the kinetic terms of the
holographic scalar and vector mesons as in \cite{Kruczenski:2003be},
\begin{eqnarray}
L_{2} &=& -\mu_7 \sqrt{-g} \left[\frac{1}{2}m^{(2)} - \frac{1}{4}
m^{(1)}\cdot m^{(1)}
+ \frac{1}{8}(m^{(1)})^2\right] \nonumber \\
&=& -\mu_7 \sqrt{-g} \left[ \frac{1}{2}(h^{(2)} + X^{(2)})  -
\frac{1}{4}(\tilde{F} \cdot \tilde{F} + h^{(1)}\cdot h^{(1)}
+h^{(1)}\cdot \tilde{F}) + \frac{1}{8}( (h^{(1)})^2 + \tilde{F}^2 + h^{(1)} \tilde{F}) \right] \nonumber \\
&=& -\mu_7 \sqrt{-g} \left[ \frac{1}{2}X^{(2)} + \frac{1}{4} \tilde{F} \cdot \tilde{F}\right] \nonumber \\
&=& -\mu_7 (2 \pi \alpha')^2 \sqrt{-g}\left[\frac{1}{2}
\frac{R^2}{\rho^2 + L^2} g^{ab}\left(\partial_a \phi \partial_b \phi
+\partial_a \chi
\partial_b \chi \right) -\frac{1}{4} F_{ab} F^{ab}\right] \, .
\label{cuadratic-action}
\end{eqnarray}
Note that there are several vanishing terms due to the antisymmetric
character of $F_{ab}$, and in addition the sum of all terms coming
exclusively from $h_{ab}$ at any order in the scalar fluctuations
vanishes. \\

\subsubsection*{Third order effective Lagrangian}

The non-vanishing terms are
\begin{equation}
L_3 = -\mu_7 \sqrt{-g} \left[\frac{1}{2}(X^{(3)}-h^{(1)}\cdot
X^{(2)}) + \frac{1}{2}h^{(1)} \cdot \tilde{F} \cdot \tilde{F}
\right] \, , \label{L3}
\end{equation}
which after looking for the explicit dependence of the scalar
fluctuations becomes
\begin{equation}
L_3 = -\mu_7 (2 \pi \alpha')^3 \sqrt{-g} \left[\frac{R^4L}{(\rho^2 +
L^2)^3}\phi (\partial_\mu\phi
\partial_\nu \phi+ \partial_\mu\chi
\partial_\nu \chi)\eta^{\mu\nu} + \frac{L}{\rho^2+L^2}\phi
(F_{aI}F^{aI} - F_{a\mu}F^{a\mu}) \right],
\end{equation}
as reported in \cite{Kruczenski:2003be}. \\

\subsubsection*{Fourth order effective Lagrangian}

The fourth order Lagrangian is
\begin{eqnarray}
L_4 &=& -\mu_7 \sqrt{-g} \left[ \frac{1}{2}X^{(4)}  +
\frac{1}{8}(X^{(2)})^2 - \frac{1}{4}X^{(2)}\cdot X^{(2)} +
\frac{1}{32}(\tilde{F} \cdot \tilde{F})^2
- \frac{1}{8}\tilde{F}\cdot \tilde{F}\cdot \tilde{F}\cdot \tilde{F}\right. \nonumber \\
&&- \frac{1}{2} h^{(2)}\cdot X^{(2)} + \frac{1}{2} h^{(1)}\cdot
h^{(1)}\cdot X^{(2)}
+ \frac{1}{2} \tilde{F}\cdot \tilde{F}\cdot X^{(2)} - \frac{1}{8}X^{(2)} (\tilde{F}\cdot \tilde{F}) \\
&&- \frac{1}{2}h^{(1)}\cdot X^{(3)} + h^{(1)} \cdot X^{(2)} \cdot
\tilde{F}
+ \frac{1}{2}h^{(2)}\cdot \tilde{F} \cdot \tilde{F} - \frac{1}{2} h^{(1)}\cdot h^{(1)}\cdot \tilde{F} \cdot \tilde{F} \nonumber \\
&&\left. - \frac{1}{4} h^{(1)}\cdot \tilde{F} \cdot h^{(1)} \cdot
\tilde{F} \right] \, . \nonumber
\end{eqnarray}
While these vertices could appear in different one-loop diagrams the
related amplitudes turn out to be sub-leading in the high energy
limit. Thus, it is not necessary to write down the explicit form of
this Lagrangian in terms of the meson fields.

~

All these are terms of the Lagrangian which are obtained by
considering scalar and vector fluctuations on the D7-brane
contributing to the one-loop Feynman-Witten diagrams on the gravity
side of the calculations. \\

\subsubsection*{Effective interaction Lagrangian for gravitons
propagating on the D7-brane}

In addition to the effective interaction Lagrangian due to
transversal and longitudinal fluctuations of the D7-brane, there are
also tensor fluctuations of the induced metric on the D7-brane,
$g_{ab}$. They are obtained by considering small perturbations
$H_{ab}$ corresponding to gravitons propagating within the D7-brane
worldvolume. The effect of this type of tensor fluctuations is
reflected both on the squared root of the determinant of the metric,
as well as, on the metric used to raise indices on the D7-brane
directions. These perturbations couple to fluctuations on the
D7-brane. We can consider the fluctuations of the metric of the form
$G_{M N} + H_{M N}(x^{\mu})$ leading to extra contributions to the
pullback $\delta P[G]^H$ due to the ten-dimensional bulk metric
fluctuations $H_{MN}$,
\begin{eqnarray}
\delta P[G]^H_{a b}&=& \sum_{i=1}^3 \ H_{ab}^{(i)} = H_{a b}+ H_{a
8} (2\pi \alpha')
\partial_{b}\phi + H_{8 b} (2\pi \alpha')   \partial_{a}\phi +H_{a
9} (2\pi \alpha')   \partial_{b}\chi \nonumber \\ &&+ H_{9 b} (2\pi
\alpha')   \partial_{a}\chi +H_{8 9} (2\pi \alpha')^2
\partial_{a}\phi \partial_{b}\chi \label{Pgrav} \, .
\end{eqnarray}
Note that these contributions only include a single graviton. Thus,
cubic vertices having a single graviton come from the following
Lagrangian
\begin{eqnarray}
L_{graviton} &=& - \mu_7 \sqrt{-g}\left[\frac{1}{2}H^{(3)} -
\frac{1}{2}H^{(2)}\cdot h^{(1)} -\frac{1}{2} H^{(1)}\cdot h^{(2)} +
\frac{1}{2}H^{(1)}\cdot h^{(1)}\cdot h^{(1)}
+ \frac{1}{4} H^{(1)}X^{(2)}  \right. \nonumber \\
&& \left. - \frac{1}{2} H^{(1)}\cdot X^{(2)} + \frac{1}{2}
H^{(1)}\cdot \tilde{F} \cdot \tilde{F} - \frac{1}{8}
H^{(1)}(\tilde{F} \cdot \tilde{F})\right].
\end{eqnarray}
By using the DIS {\it Ansatz} for the graviton $H_{m i} \sim A_m
v_i$, where $A_m$ is a five-dimensional gauge field on AdS$_5$ while
$v_i$ is a Killing vector of $S^3$, all terms containing
$H^{(2)}_{ab}$ and $H^{(3)}_{ab}$ vanish. Similarly, the trace of
$H^{(1)}$ is zero. Also terms like $H^{(1)} \cdot h^{(2)}$ and
$H^{(1)} \cdot h^{(1)} \cdot h^{(1)}$ vanish because $h^{(i)}_{ab}$
and $g^{ab}$ are diagonal. Therefore, the above Lagrangian becomes
\begin{eqnarray}
L_{graviton} &=& -\mu_7 \sqrt{-g} \left[- \frac{1}{2} H^{(1)}\cdot
X^{(2)} + \frac{1}{2} H^{(1)}\cdot \tilde{F} \cdot \tilde{F} \right]
. \label{Lgrav}
\end{eqnarray}
In principle, we have quartic vertices\footnote{These vertices come
from $H^{(1)} \cdot X^{(3)}$ or $H^{(1)} \cdot h^{(1)} \cdot
X^{(2)}$.} which contain a gravi-photon and three mesons. They have
the same contribution in powers of $N$ that the diagram with two
cubic vertices (one of them contains the graviton)\footnote{See
Subsection 3.2.}. We do not write them explicitly because these
diagrams produce sub-leading contributions in powers of
$\Lambda^2/q^2$ and will be more suppressed as in
\cite{Jorrin:2016rbx}.

\subsection{Solutions of the equations of motion}

The equations of motion of the mesons are obtained from the second
order fluctuations calculated in last subsection. For scalar mesons,
the quadratic Lagrangian is given by $X^{(2)}_{ab}$ and the EOM is
\begin{eqnarray}
\partial_a \left(\frac{\rho^3 \sqrt{-g}}{\rho^2+L^2} g^{ab} \partial_b \phi
\right)=0  \ .
\end{eqnarray}
The solutions have been calculated in terms of Hypergeometric
functions in \cite{Kruczenski:2003be}. $L \sim \Lambda R^2$ plays
the role of a cut-off in the radial coordinate. In the limit $L \to
0$ the conformal symmetry is recovered, the coordinate $\rho \to r$
and the induced metric becomes $AdS_5 \times S^3$.

We are interested in the high energy limit, $q^2 \gg \Lambda^2$,
therefore we shall consider finite but small values of $L$ such that
the background is approximately $AdS_5\times S^3$ and the solutions
are somehow similar to those for glueballs in $AdS_5 \times S^5$.
This allows one to have cubic vertices which are linearly
proportional to $L$, and use the solutions of meson fields obtained
in \cite{Koile:2011aa}. By solving the EOM in the limit $\rho \gg L$
and by imposing a hard cut-off, we obtain the solutions for the
scalar fields $\Phi=\phi, \, \chi$ with some four-momentum $k^\mu$
as
\begin{equation}
\Phi^{(l)}(x^{\mu}, z, \Omega) = c e^{i k\cdot x}  z J_{\Delta-2}(k
z) Y^l(\Omega) \ , \label{Sol-norm}
\end{equation}
where we have introduced the variable $z=R^2/\rho$ and a
normalization constant $c \sim \sqrt{\Lambda k}$, while
$Y^l(\Omega)$ is a scalar spherical harmonic on $S^3$. Notice that
for closed strings there is a factor $z^2$ instead of $z$
multiplying the Bessel function $J$. The EOM for the gauge fields
($F_{ab}=
\partial_a B_b-\partial_b B_a$) on the D7-brane which follows from
the second term in (\ref{cuadratic-action}) plus the Wess-Zumino
term, is
\begin{eqnarray}
\partial_{a}\left( \sqrt{-g} F^{ab}\right)-\frac{4 \rho(\rho^2+L^2)}{R^4}
\epsilon^{bjk}\partial_{j}B_{k} =0 \ ,
\end{eqnarray}
where $\epsilon_{ijk}$ is the Levi-Civita pseudo-tensor density, the
indices $a, \ b , \ c, \dots $ run over all directions of the
D7-brane world-volume, and $i, \ j, \ k, \dots$ belong to $S^3$. The
second term is the contribution from the Wess-Zumino action and it
is nonzero only if $b$ is one of the $S^3$ indices. We can expand
$B_{M}$ in scalar and vector spherical harmonics on $S^3$, and
obtain three modes
\begin{eqnarray}
\hbox{type \  I} \ \  &:&\ \ \ B_{\mu}=0, \ \ B_{\rho}=0, \ \ \ \
B_{i}=\phi_I^{\pm}(\rho) \ e^{i k \cdot x} \ Y_i^{l,\pm1}(\Omega), \\
&& \nonumber \\
\hbox{type \ II} \ &:& \ \ \ B_{\mu}=\zeta_{\mu} \ \phi_{II}(\rho)
\ e^{i k \cdot x} \ Y^l(\Omega), \ \ \ k \cdot \zeta =0,  \ \ \ B_{\rho}=0, \ \ \ B_{i}=0, \\
&& \nonumber \\
\hbox{type \ III}&:& \ \ \ B_{\mu}=0, \ \ \ B_{\rho}=
\phi_{III}(\rho) \ e^{ik \cdot x} \ Y^l(\Omega), \ \ \
B_{i}=\tilde{\phi}_{III}(\rho) \ e^{ik \cdot x} \ \nabla_i
Y^l(\Omega).
\end{eqnarray}
$Y_i^{l,\pm1}(\Omega)$ and $\nabla_i Y^l(\Omega)$ are the different
vector spherical harmonics on $S^3$. The solutions are associated
with different representations of the isometry group of $SO(4)
\approx SU(2) \times SU(2)$. The properties of the modes and their
on-shell solutions are shown in table 1. For type I and III modes,
which interact with the scalar meson, we obtain the following $\rho$
dependence (in the limit $q^2 \gg \Lambda^2$)
\begin{eqnarray}
\phi_{I}^{\pm1} (\rho)= c_I \frac{J_{\Delta-2}( \frac{M
R^2}{\rho})}{\rho^2}, \ \ \phi_{III}(\rho)= c_{III}
\frac{J_{\Delta-2}(\frac{M R^2}{\rho})}{\rho^2}, \ \
\tilde{\phi}_{III}(\rho)=\frac{ \frac{1}{\rho} \partial_\rho(\rho^3
\phi_{III}(\rho))}{l(l+2)}, \nonumber \\
&&
\end{eqnarray}
where the constants are $c_{I} \sim R^4 \sqrt{M \Lambda}$ and
$c_{III} \sim R^2 \sqrt{\frac{l(l+2)\Lambda}{M}}$. In this work we
use the propagator of the type I mode. Since the solution in the AdS
is analogous to the glueball solution, we can use a similar
propagator as in references \cite{Jorrin:2016rbx,Gao:2014nwa}
\begin{equation}
G_\Delta (x,z;x',z') = - \int \frac{d^4k}{(2\pi)^4} \int \frac{d\omega \, \omega}{\omega^2 + k^2 - i \varepsilon}
z^2 J_{\Delta-2}(\omega z)(z')^2 J_{\Delta-2}(\omega z'),
\end{equation}
together with the corresponding vector spherical harmonics.

{\renewcommand{\arraystretch}{1.4}
\begin{table}
\begin{center}
\centering
\begin{tabular}{|c|c|c|c|c|}
\hline
Field & Type of field in $5D$ & Built from & $\Delta(l)$ & $SU(2)\times SU(2)$ irrep \\
\hline
$\phi,\chi$ & scalars & $\phi,\chi$ & $l+3$, $l\geq 0$ & $\left(\frac{l}{2},\frac{l}{2}\right)$ \\
\hline
$B_\mu$ & vector & $B_\mu^{II}$ & $l+3$, $l\geq 0$ & $\left(\frac{l}{2},\frac{l}{2}\right)$ \\
\hline
$\phi_I^{-}$ & scalar & $B_i^{I}$ & $l+1$, $l\geq 1$ & $\left(\frac{l+1}{2},\frac{l-1}{2}\right)$ \\
\hline
$\phi_I^+$ & scalar & $B_i^{I}$ & $l+5$, $l\geq 1$ & $\left(\frac{l-1}{2},\frac{l+1}{2}\right)$ \\
\hline
$\phi_{III}$ & scalar & $B_{i,z}^{III}$ & $l+3$, $l\geq 1$ & $\left(\frac{l}{2},\frac{l}{2}\right)$ \\
\hline
\end{tabular}
\caption{\small Some features of D7-brane fluctuations around the
AdS$_5\times S^3$ background which are relevant to this work. The
integer $l$ indicates the $SO(4)\sim SU(2) \times SU(2)$ irreducible
representation (irrep) and defines the corresponding Kaluza-Klein
mass. The relation between the scaling dimension of the associated
operator $\Delta$ and $l$ is written.}
\end{center}
\end{table}

Using the gauge/string duality the current operator inserted at the
boundary of the AdS excites a non-normalizable mode which propagates
within the bulk. The perturbations (gravi-photons) take the form
$\delta G_{mi}= A_m v_i$. The field $A_m$ is derived from a Maxwell
Lagrangian in the AdS space with the boundary condition $A_{\mu}(y,
\infty) = n_{\mu} \ e^{iq \cdot y}$. In the Lorentz-like gauge the
solution is given by
\begin{eqnarray}
A_{\mu}= n_{\mu} \ e^{i q \cdot y} \ q \ z \ K_1(q \ z) \ , \ \ \ \
A_z= i \ n \cdot q \ e^{i q \cdot y} \ z \ K_0(q \ z) ,
\end{eqnarray}
where $K_0$ and $K_1$ are modified Bessel functions of the second
kind.

\section{The leading diagram in the $1/N$ expansion}

\subsection{DIS and comments on the FCS tree-level calculation}

The DIS cross section is related to the matrix element of a product
of two electromagnetic currents $J_\mu(y) \, J_\nu(0)$ inside the
hadron. Through the so-called optical theorem, one has to
calculate the FCS process associated with the DIS one. Specifically,
there are two steps to follow. The first one is given by the
operator product expansion of $J_\mu(y) \, J_\nu(0)$, which is
obtained within an un-physical region of the Bjorken parameter, $x
\gg 1$. For the second step one needs to consider the dispersion
relations in order to connect the un-physical calculation with the
physical DIS process for $0 \leq x < 1$.

Recall that the matrix element of two electromagnetic currents
$J_\mu(y) \, J_\nu(0)$ inside a hadron can be expressed by using the
$T_{\mu\nu}$ tensor. For the incoming and outgoing hadrons with
polarizations $h$ and $h'$, we can write
\begin{equation}
T_{\mu\nu}(q^2, x)=i \int d^4y \, e^{iq\cdot y} \, \langle P,
h'|{\hat T}( J_{\mu}(y) \, J_{\nu}(0))|P,h \rangle \, .
\end{equation}
By using the optical theorem the $T_{\mu\nu}$ tensor is related to
the hadronic tensor $W_{\mu\nu}$ as follows,
\begin{equation}
W_{\mu\nu}(q^2, x)=i \int d^4y \, e^{iq\cdot y} \langle P,
h'|[J_{\mu}(y),J_{\nu}(0)]|P,h \rangle \, .
\end{equation}
In the present work we investigate scalar mesons, which reduces this
tensor to only two terms
\begin{equation}
W_{\mu\nu}= F_1(x,q^2) \left(\eta_{\mu\nu} - \frac{q_\mu
q_\nu}{q^2}\right) + \frac{2 x}{q^2} \, F_2(x,q^2) \left(P_\mu +
\frac{q_\mu}{2x} \right)\left(P_\nu + \frac{q_\nu}{2x} \right) \, ,
\label{tensorW}
\end{equation}
where it has been used the Bjorken parameter defined as
\begin{equation}
x=\frac{-q^2}{2 P \cdot q} \, ,
\end{equation}
while $F_1(x,q^2)$ and $F_2(x,q^2)$ are the so-called structure
functions. At weak coupling these functions are obtained within the
parton model, and they are related to the parton distribution
functions (PDFs). The PDFs represent the probability of finding a
parton with a fraction $x$ of the target hadron momentum, $P$.
Particularly, from the optical theorem one learns that $2 \pi$ times
the imaginary part of the structure functions associated with FCS
gives exactly the DIS structure functions. Based on this Polchinski
and Strassler \cite{Polchinski:2002jw} proposed a way to calculate
structure functions at strong coupling by using the gauge/string
theory duality.

The tree-level type IIB supergravity calculation, which in terms of
the $1/N$ expansion implies taking the large $N$ limit from the
beginning, has been done in \cite{Koile:2011aa} and
\cite{Koile:2013hba}. In the range $1/\sqrt{\lambda} \ll x < 1$ the
results using the D3D7-brane model for scalar mesons have been
obtained in \cite{Koile:2011aa} for one flavor (and in
\cite{Koile:2013hba} for the multi-flavor case) obtaining
\begin{equation}\label{kru23}
F_{1}=0, \,\,\,\,\,\,\,\,\:\,\,\,\,\,\,\,\,\:\:\:\: F_{2}=
{\tilde{A}}_{0} \, {\mathcal{Q}}^2 \,
\bigg(\frac{\Lambda^2}{q^2}\bigg)^{l+2} \, x^{l+4} \, (1-x)^{l+1} \,
,
\end{equation}
where ${\tilde{A}}_{0} = 2^{2 l - 4} \,  \pi^{-7} \, (l+2)!^2 \,
|c_i|^2 \, |c_X|^2$ is a dimensionless normalization constant, while
$c_i$ and $c_X$ are the normalization constants of the incident and
intermediate (in FCS) scalar mesons. We consider that the integer
$l>0$, which means that the scalar fields are charged under a $U(1)$
group. Notice that ${\mathcal{Q}}$ labels the charge under the
$U(1)$ symmetry group induced by transformations on the three-sphere
in the direction of the Killing vector $v^j$.

On the other hand, although in the present work we only consider the
calculation within the validity range of supergravity, we can also
write the structure functions for scalar mesons for small $x$ values
within the range $\exp(-\sqrt{\lambda}) \ll x \ll 1/\sqrt{\lambda}$.
In this case the calculation has been done by considering type IIB
string theory scattering amplitudes of two open strings
(representing the scalar mesons) and two closed strings (which
represent the two virtual photons in the FCS). This has been done in
\cite{Koile:2014vca}, obtaining
\bea
F_1 &=& \frac{\pi^2}{16 x^2} \ \rho_3 \ |c_i|^2 \
\left(\frac{\Lambda^2}{q^2}\right)^{l+1}
\frac{1}{\sqrt{4 \pi \lambda}} I_{1,2 l+5} \ , \\
F_2 &=& \frac{\pi^2}{8 x} \ \rho_3 \ |c_i|^2 \
\left(\frac{\Lambda^2}{q^2}\right)^{l+1} \frac{1}{\sqrt{4 \pi
\lambda}} (I_{0,2 l+5}+I_{1,2 l+5}) \, .
\eea
Notice that $\rho_3$ is defined through the normalization condition
for the spherical harmonics on $S^3$
\bea
\int d\Omega_3 \ \sqrt{\tilde{g}} \ v_i \ v^i \ Y(\Omega_{3}) \
Y^{*}(\Omega_{3}) = \rho_3 \ R^2 \, . \label{angular0}
\eea
Also the definition of $I_{j,n}$ is given in terms of the integral
of the square of the modified Bessel functions of the second kind
times integer powers of its argument $\omega = \frac{q R^2}{r}$,
\bea
I_{j,n} = \int_{0}^{\infty} d\omega \, \omega^n \, K_j^2(\omega) \,
.
\eea

Next we focus on the type IIB supergravity calculation of one-loop
diagrams which are the holographic dual representation of the
one-loop FCS corresponding to a DIS process with two outgoing hadron
states. The following calculations hold in the kinematical range
$1/\sqrt{\lambda} \ll x < 1$, and at strong coupling.

\subsection{The leading diagram for the one-loop FCS calculation}

We can redefine the fields in such a way that their kinetic terms
become canonically normalized in terms of $N$, i.e. they do not
depend on $N$. Through this field redefinition it is possible
perform the $1/N$ counting of each Feynman-Witten diagram. By noting
that $ \mu_7 = [(2 \pi)^7 g_s \alpha'^4]^{-1} = 2 N [R^4 (2\pi)^6
\alpha'^2]^{-1}$, the scalar and vector meson fields are redefined
as
\begin{equation}
\phi \rightarrow \frac{\phi}{\sqrt{N}} , \ \ \ \chi \rightarrow
\frac{\chi}{\sqrt{N}} , \ \ \ F_{ab} \rightarrow
\frac{F_{ab}}{\sqrt{N}} \, .
\end{equation}
Notice that with this field redefinition the cubic and quartic
vertices have the factors $1/\sqrt{N}$ and $1/N$, respectively.

In addition, the normalization for the graviton modes (closed
strings) implies a different power of $N$ in comparison with the
meson fields, since in order to obtain canonically normalized
quadratic terms one has to re-scale
\begin{equation}
H_{ab}\rightarrow \frac{H_{ab}}{N} ,
\end{equation}
since the Newton's constant in type IIB supergravity is $1/k_{10}^2
=N^2/(4 \pi^5 R^8)$.

The arguments used to select the leading diagram contributing to the
$1/N$ expansion are similar to those exposed in
\cite{Jorrin:2016rbx} and \cite{Gao:2014nwa}. The idea is to
understand what changes when considering large but finite values of
the number of color degrees of freedom in the holographic
calculation of the structure functions $F_i(x,q^2)$ of scalar mesons
obtained from the D3D7-brane model, with respect to the methods and
results of the tree-level calculation performed in
\cite{Koile:2011aa,Koile:2013hba}. Moreover, we want to compare both
the results of the present $1/N$ expansion and those of
\cite{Koile:2011aa,Koile:2013hba,Koile:2015qsa} in the large $N$
limit with lattice QCD simulations for the first three moments of
the $F_2$ structure function of the pion
\cite{Best:1997qp,Brommel:2006zz,Chang:2014lva}. Since we focus on
the $1/\sqrt{\lambda} \ll x < 1$ range the supergravity description
is accurate enough. Therefore, in order to obtain the leading $1/N$
correction\footnote{In fact, at the end of the calculation, after
considering the high energy limit first, it will become clear that
the result will not be a correction but the leading contribution.
Obviously, if one considers the large $N$ limit first, which in the
high energy limit is not the physical situation, it can be seen as a
$1/N$ correction, see equation (1).} we have to consider all
one-loop diagrams that can be drawn for the holographic dual FCS
process. This involves two non-normalizable gauge bosons $A_\mu$
coming from the boundary, plus two normalizable modes on the
D7-brane, in this case the scalar mesons that describe the D7-brane
transverse fluctuations. By using the optical theorem we just need
to calculate the imaginary parts of these diagrams, which means that
we have to introduce a vertical cut in the one-loop supergravity
Feynman-Witten diagrams and, therefore, we have to consider
two-particle on-shell intermediate states. All these fields and
their interactions have been described in the previous section for
small but non-vanishing $L$. This permits to unveil important
physical aspects of the process.

The crucial point is that we are working in the large-$q^2$ limit.
This allows one to classify diagrams in a $\Lambda^2/q^2$ series
expansion. The $q^\mu$ four-moment is carried by a gauge field,
which is the holographic dual representation of a virtual photon on
the boundary gauge theory. As in the case of the AdS$_5 \times S^5$
fields, which we carefully studied in \cite{Jorrin:2016rbx}, the
vertices coupling to the $A_\mu$ field (which is a five-dimensional
field after integrating over the $S^3$) with scalar bulk fields are
always of the same form
\begin{equation}
S_{\Phi \Phi A} \propto \int d^{p+1}x \ \sqrt{-g} \ h^{ab} \
\partial_a \Phi \ \partial_b \Phi \ ,  \ \ \ \ h^{a b} \sim (A^{a} v^{b} +
A^{b} v^{a}) \, ,
\end{equation}
where $\Phi$ represents some generic scalar field. This implies that
in the on-shell evaluation of this vertex for a given field
$\Phi_{\Delta}$ coming from the IR region we will find a suppression
factor $(\Lambda^2/q^2)^{\Delta-1}$ in the structure functions. The
physical reason for this suppression is understood as follows.
Bessel-$J$ functions of bulk fields solutions (incident holographic
hadrons) mainly live in the IR region near $z\sim\Lambda^{-1}$,
while the Bessel-$K$ function decreases exponentially from the
boundary towards the interior. This fall-off is characterized by
$q$, hence the $\Lambda^2/q^2$ factor and its $\Delta$ power are
related with the probability for the $\Phi_\Delta$ hadron to tunnel
from the IR to the UV region where it can interact with the gauge
field \cite{Polchinski:2002jw}.

We should mention that after studying other types of vertices, such
as quartic interactions, and checking that they do not change this
analysis, we can conclude that the $\Lambda/q$ expansion will be
dominated by processes where the non-normalizable gauge field
interacts with the scalars with the smallest possible value of
$\Delta$. In the $N\rightarrow \infty$ limit there is only one
allowed interaction vertex, and this index (conformal dimension) is
fixed by the incoming hadron with its associated scaling dimension
$\Delta_{in}$, but when the number of
color degrees of freedom becomes finite one must take into account
the one-loop processes where the initial hadron splits into two
other particles, and only one of them interacts with $A_\mu$. The
leading contribution comes from the case where this splitting
happens in IR region. Furthermore, among the fields that one obtains
within the D3D7-brane model we can see that the type I gauge fields
will play a key role since they can have the lowest $\Delta_{min} =
2$ index. Note that this is the same $\Delta_{min}$ that we have
obtained in \cite{Jorrin:2016rbx} in the AdS$_5\times S^5$ context
from the scalar fields usually called $s$-scalars, a particular
combination of graviton and 4-form perturbations \cite{Kim:1985ez}.
Thus, the $1/q^2$ dependence of the final result will be the same as
in our paper \cite{Jorrin:2016rbx}, however the dependence on the
Bjorken parameter $x$ will be significantly different.

The conclusion from this analysis is that the fields involved in the
calculation and the diagram rendering the leading contribution to
the on-shell scattering amplitude and the structure functions of the
scalar mesons are given schematically by the diagram of figure 1.
\begin{figure}
\centering
\includegraphics[scale=1.1,trim = 0 0 0 0 mm]{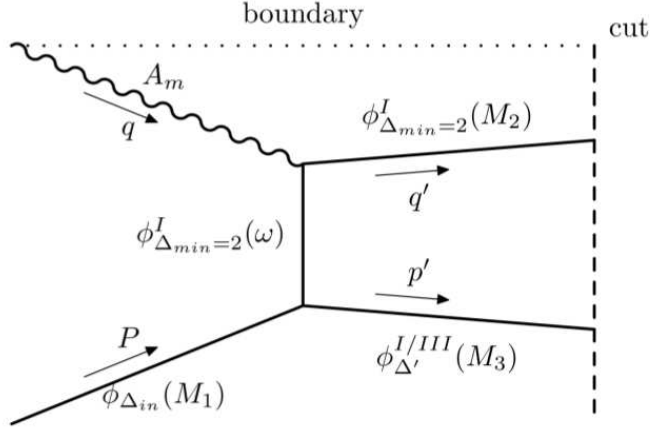}
\caption{\small Feynman-Witten diagram corresponding to the left
side of the cut (vertical long-dashed line) of the leading one-loop
FCS related through the optical theorem to the DIS diagram with a
two-hadron final state. The field associated with each line is
explicitly written with the corresponding four-dimensional momenta,
$\Delta$ indices and AdS masses. The solutions are described in
Section 2.} \label{DIS}
\end{figure}

In the following subsection we will analyze explicitly the two
interaction vertices and the propagator that appears in this
diagram. The interaction vertices are: the IR vertex where the
initial hadron splits into two intermediate hadrons, and the UV
vertex where one of the resulting fields interacts with the gauge
field near the boundary. We will also carry out the final steps of
the calculation and, within some approximations, obtain the explicit
form of the longitudinal structure function $F_L = F_2 - 2x F_1$.

\subsection{The UV interaction vertex}

This vertex comes from the second term in the Lagrangian of equation
(\ref{Lgrav}), where the metric fluctuation couples to two vector
modes. In principle, both $F_{ab}$ can be built out from one of the
vector modes of type I$^\pm$, II or III. The one associated with the
vertical propagator in figure 1 must be the type I$^-$ mode, which
has the lowest possible index $\Delta_{min} =2$. The relevant
interaction is of the form $A \phi_I \phi_I$. The situation where
the second vector involved in the UV interaction vertex is a type II
or type III mode is excluded since in that case the interaction
Lagrangian vanishes because of the angular integral. The effective
action associated with this vertex is
\begin{equation}
S_{A\phi_I\phi_I}= -\frac{\mu_7}{N} (2\pi \alpha')^2 \int d^4x \
d\rho \ d\Omega_3 \ \sqrt{-g} \ \frac{1}{2} H^{(1)}\cdot F^I \cdot
F^{*I} \ ,
\end{equation}
where
\begin{equation}
F_{\mu\nu}^I = 0 \ , \ \ \ \ F_{\mu z}^I=0 \ , \ \ \ \ \ F_{\mu i}^I =
\partial_\mu B^I_i \ , \ \ \ \  F_{z i}^I = \partial_z B^I_i \ ,
\end{equation}
thus, we have
\begin{eqnarray}
H^{(1)}\cdot F^I \cdot F^{*I}  &=&   g^{bc}g^{de}g^{af} h_{ab} F_{cd}^I F_{ef}^{*I} \nonumber \\
&=&    A^{\mu} v^i g^{de} F_{\mu d}^I F_{ei}^{*I}+ A^{\mu} v^i
g^{de} F_{i d}^I F_{e\mu}^{*I} \nonumber \\
&=&  A^{\mu} v^i \partial_{\mu}B^j_I \left( \partial_j
B^{*I}_i-\partial_i B_j^{* I} \right)+ A^{\mu} v^i
\partial_{\mu}B_I^{*j}\left( \partial_j B_i^I -\partial_i B_j^I \right) \nonumber \\
&=& -\left(A^{\mu} \partial_{\mu} B^j v^i  \partial_i B^*_j +A^{\mu}
\partial_I B^{*j} v^i  \partial_i B_j \right)  \ .
\end{eqnarray}
In order to evaluate the on-shell vertex one must insert the actual
form of the solutions described in the previous section and
integrate over the eight-dimensional space. The integration over the
first four coordinates $x^\mu$ is trivial since it always renders
the momentum conservation condition associated with the
corresponding momenta. The integrals over the spherical harmonics
can be simplified by considering the charge
eigenstates\footnote{Notice that the charge ${\cal{Q}}$ does not
need to be the one carried by the initial hadron, ${\cal{Q}}_i$
because of the hadron splitting process.}
\begin{equation}
v^i \ \partial_i Y^j = i {\cal{Q}} \ Y^j \, .
\end{equation}
Finally, by changing variables $z=\frac{R^2}{\rho}$, the effective
action becomes
\begin{eqnarray}
S_{A\phi_I\phi_I} &=& i \ {\cal{Q}} \ \frac{\mu_7}{N} \ 2 \ (\pi
\alpha')^2 \int
d^4x \ dz \ d\Omega_3 \ \sqrt{-g} \ A^{m}(z) \times \nonumber \\
&& \left( B^{I i}(z,\Omega) \
\partial_{m} B_i^{ * I}(z,\Omega) - B_i^{*I}(z,\Omega) \
\partial_{m} B^{I i} (z,\Omega)\right) \, . \label{int-UV}
\end{eqnarray}
Type I modes labeled with $(\pm)$ are orthogonal and therefore they
do not couple to each other. The only outgoing particle is a type I
scalar mode with label $(-)$ and with the same quantum numbers with
$l, m, m'$ as the incoming scalar. Hence, even if in the full
$8$-dimensional theory the type I modes come from gauge fields and
the existence of their solution rely on the presence of the
Wess-Zumino term in the action together with the DBI term in
\cite{Kruczenski:2003be} the angular integral only leads to charge
conservation, which also indicates that there is no mixing with
other particles in this vertex. Then, the on-shell action that we
obtain is exactly the same found for glueballs
\cite{Polchinski:2002jw}, scalar mesons
\cite{Koile:2011aa,Koile:2013hba} and $s$ scalars
\cite{Jorrin:2016rbx}.

After integration of equation (\ref{int-UV}) on $S^3$, by using the
orthogonality relations of the vector spherical harmonics, we obtain
\begin{eqnarray}
S_{A\phi_I\phi_I}= i \ {\cal{Q}} \ \frac{\mu_7}{N} \ 2 \ (\pi
\alpha')^2 \int d^4x \ dz \ \sqrt{-g} \ A^{m}(z) \ \left(
\phi_I(z,\Omega) \
\partial_{m} \phi_I^{*}(z) - \phi^{*}_I(z) \ \partial_{m} \phi_{I}
(z)\right) \ ,
\end{eqnarray}
where $\phi^I= \sqrt{\Lambda \omega} \, e^{i (P-p_\omega) \cdot x}
\, z^2 \, J_{\Delta_{\omega}-2}(\omega z) $ and $\phi^{I*}=
\sqrt{\Lambda M_3} \, e^{- i q'\cdot x} \, z^2 \,
J_{\Delta_{\omega}-2}(M_3 z)$.

\subsection{The IR interaction vertex}

The relevant vertex couples the incident scalar meson to a scalar
mode of type I$^-$ having the smallest conformal dimension
$\Delta_{min}= 2$ corresponding to $l=1$. From the Lagrangian at
cubic order the only term which couples the scalar meson $\phi$ to
type I$^{\pm}$ eight-dimensional vector modes is the second term of
equation (\ref{L3}). Thus, for small $L$ we have\footnote{Note that
in the conformal case, i.e. $L=0$, this vertex does not exist. Here,
we analyze the situation for the non-conformal background and keep a
non-vanishing but small $L$ in order to approximate the
Hypergeometric functions by Bessel functions. The $L=0$ case should
be analyzed in a different way.}
\begin{equation}
L_{\phi F F} = \frac{\mu_7}{N^{3/2}} (2\pi\alpha')^3\sqrt{-g} \
\frac{L}{\rho^2} \ \phi \
\left(F_{IJ}F^{IJ}-F_{\mu\nu}F^{\mu\nu}\right) \nn \ .
\end{equation}
Since one of the vector modes must be of type I$^-$, its field
strength is such that $F^{I}_{\mu\nu}=0$. Hence, we are left with
the term proportional to $F_{IJ}F^{IJ}$ only. Note that this implies
that the on-shell mode produced in this process (whose mass is
denoted by $M_3$) cannot be a type II mode. This means that we only
have to consider scalar modes from the five-dimensional point of
view. The remaining two-form field strength contraction can be
decomposed in terms of angular coordinates on $S^3$, and the radial
coordinate $\rho$ as
\begin{equation}
F_{IJ}F^{IJ} = F_{ij}F^{ij} + 2 F_{i\rho}F^{i\rho} = 2\left[\nabla_i
B_j \nabla^i B^j - \nabla_i B_j \nabla^j B^i + \nabla_\rho B_i
\nabla^\rho B^i - \nabla_\rho B_i \nabla^i B^\rho \right] \ ,
\end{equation}
where in the last step we have used the fact that for a type I mode
$B_\rho=0$. Plugging it in the action and taking the complex
conjugate field for the outgoing field we obtain
\begin{eqnarray}
S_{\phi \phi_I \phi_I}&=&-\frac{\mu_7}{N^{3/2}} \ (2 \pi\alpha')^3
\times \nonumber
\\
&& \int d^8\xi \, \sqrt{-g} \, \frac{2 L \phi}{ \rho^2} \left(
\nabla_i B_j \nabla^i B^{j *} - \nabla_i B_j \nabla^j B^{i *}+
\nabla_{\rho} B_i \nabla^{\rho} B^{i *} -\nabla_{\rho} B_i
\nabla^{i} B^{\rho *}\right), \nonumber \\
&&  \label{Int-IR}
\end{eqnarray}
where $B$ corresponds to the type I$^-$ scalar mode with mass
$\omega$ which comes from the propagator and interacts with the
virtual photon in the UV region. On the other hand, $B^{*}$ is the
outgoing mode with mass $M_3$.
We will analyze in detail the case where this mode is of type
I$^{\pm}$. The possibility for the on-shell mode outgoing from this
IR vertex to be of type III is considered in Appendix B.

If the
outgoing normalized mode corresponds to a type I scalar one has
$B^*_{\rho}=0$ and, therefore, the last term in equation
(\ref{Int-IR}) vanishes. Plugging the solutions of the modes in the
action and taking into account that $\Delta = \Delta_{min} = 2$ for
the scalar that corresponds to the vertical propagating line in the
diagram of figure 1 we find the following interaction action,
\begin{eqnarray}
S^{I^{\pm}}_{\phi \phi_I\phi_I}&=& -\frac{\mu_7}{N^{3/2}} (2
\pi\alpha')^3 2 L \ C \int d^4x \ e^{i (p_1+p_{\omega}-p_3)x} \left(
\int_0^{\frac{1}{\Lambda}} dz z^2
J_{\Delta_{i}-2} (M_1 z)  J_0 (\omega z) J_{\Delta_{3}-2} (M_3 z) I_1\right. \nonumber \\
&& \left. +  \int_0^{\frac{1}{\Lambda}} dz \ J_{\Delta_{\Phi}-2}
(M_1 z) \ \partial_z(z^2 J_0 (\omega z)) \ \partial_z(z^2
J_{\Delta_{3}-2} (M_3 z) ) \ I_2\right) \ , \label{Int-IR2}
\end{eqnarray}
where $\Delta_{in}$ and $\Delta_3$ are associated with the spherical
harmonic representation index of the incident and outgoing on-shell
modes, respectively, and $C=\sqrt{\Lambda^3 M_1 M_3 \omega}$ is the
product of the corresponding normalization constants discussed in
Section 2. In addition, $I_1$ and $I_2$ are integrals of the
spherical harmonics on $S^3$ defined as follows
\begin{eqnarray}
I_1=\int d\Omega_3 \left(  \nabla_i \vec{Y}^{l''} \cdot \nabla^i
\vec{Y}^1 \ Y^{l_{in}} - \nabla_i Y^{l''}_j \cdot \nabla^j Y^{1, i}
\ Y^{l_{in}} \right) \ , \ \ \ \ \ I_2=\int d\Omega_3 \vec{Y}^{l''}
\cdot \vec{Y}^1 \ Y^{l_{in}} \ ,
\end{eqnarray}
where $l''$ and $l_{in}$ are related to the conformal dimensions as
shown in table 1. By using properties of vector spherical harmonics
we obtain the following identity
\begin{eqnarray}
\pm (l+1) \epsilon_{ilm}Y^{l,\pm}_{i}=\epsilon_{ilm}
\epsilon_{ijk}\nabla_j Y^{l,\pm}_{k} =\nabla_l
Y^{l,\pm}_{m}-\nabla_m Y^{l,\pm}_{l} ,
\end{eqnarray}
which allows us to express one of these integrals in terms of the
other as
\begin{eqnarray}
I_1=\int d\Omega_3 \ Y^{l_{in}} \ \nabla^i Y^{l'',\pm,j} \left(
\nabla_i Y^{1,-}_j - \nabla_j Y^{1,-}_i \right) = \mp 2 (l''+1) \
I_2.
\end{eqnarray}
The result of the integral $I_1$ is presented in Appendix A and it
restricts the conformal dimension of the outgoing mode. In order to
calculate the structure functions we have to sum over indices $m$
and $n$ of the spherical harmonics of the intermediate field by
using the optical theorem. Note that there are many vanishing terms
due to the $U(1)$ charge conservation associated with these indices.

Hitherto we have worked from first principles, finding the leading
diagram and studying the needed on-shell vertices and propagators.
Once we have dealt with the angular integrals, we are left with
definite $z$-integrals (within the integration region given by $0
\leq z \leq z_0$, where $z=0$ is the AdS-boundary and
$z_0=\Lambda^{-1}$ corresponds to the IR cut-off) of products of
three Bessel functions of the first kind times some positive integer
power of $z$. Since these integrals are not known analytically,
there are two ways to proceed. The first one would be a numerical
approach, simplified by the fact that since the splitting occurs
mainly in the IR region, the Bessel functions can replaced by their
asymptotic expression
\begin{equation}
J_m(a z) \approx \sqrt{\frac{2}{\pi a z}} \cos\left(az - m
\frac{\pi}{2}- \frac{\pi}{4}\right) \, . \label{Jcos}
\end{equation}
However, the intricate $x$-dependence of the scattering amplitude
difficults the extraction of the $x$-dependence of the
structure functions. In this work we will proceed as in
\cite{Jorrin:2016rbx} and attempt to obtain these functions
$F_i(x,q^2)$ semi-analytically within the range of validity of some
reasonable approximations. Most of the details of the following
calculations can be found in our previous work
\cite{Jorrin:2016rbx}, and the new ingredients that appear due to
the different structure of the IR vertex are analyzed in this
section and are collected in Appendix C.

\subsection{Calculation of the structure functions}

The $1/N$ corrections to the structure functions can be obtained
from the hadronic tensor as in \cite{Jorrin:2016rbx} where glueballs
have been considered. One focuses on the DIS process in the boundary
theory, and isolates the contribution from two-particle intermediate
states to the hadronic tensor $W^{\mu\nu}$ in terms of the
corresponding electromagnetic current $J^\mu$ one-point functions,
which is related to the FCS tensor $T^{\mu\nu}$ by the optical
theorem. In this context we can schematically write
\begin{eqnarray}
\textrm{Im}\left(T_2^{\mu\nu}\right) &=& \pi \sum_{X_1,X_2}
\langle P,Q|\tilde{J}^\mu (q)|X_1,X_2\rangle\langle X_1,X_2|J^\nu (0)|P,Q\rangle  \\
&=& \pi \sum_{M_2, M_3}\int \frac{d^3p'}{2E_{p'}(2\pi)^3}
\frac{d^3q'}{2E_{q'}(2\pi)^3}\langle P,Q|\tilde{J}^\mu (q)|X_1,X_2\rangle\langle X_1,X_2|J^\nu (0)|P,Q\rangle \nonumber \\
&=& 4\pi^3 \sum_{M_2,M_3} \int \frac{d^4q'}{(2\pi)^4} \delta
\left(M_2^2-q'^2\right) \delta \left(M_3^2-(P+q-q')^2\right)|\langle
P,Q|J^\nu (0)|X_1,X_2\rangle|^2,\nonumber
\end{eqnarray}
where the subindex in $T^{\mu\nu}_2$ indicates that we are
considering only processes with two-particle intermediate states
$X_1$ and $X_2$ associated with the momenta $p'$ and $q'$ (see
figure 1), and
\begin{equation}
n_\mu \langle P,Q|\tilde{J}^\mu (q)|X_1,X_2 \rangle = (2\pi)^4
\delta^{(4)}\left(P+q-p'-q'\right) \langle P,Q|n \cdot J
(0)|X_1,X_2\rangle \, ,
\end{equation}
is identified in the AdS/CFT duality with the amplitude of our
diagram of figure 1. We refer the reader to our previous paper
\cite{Jorrin:2016rbx} for details of the rest of the calculation
since there are several common steps. As in references
\cite{Jorrin:2016rbx,Gao:2014nwa} the dominant diagram is the
$t$-channel one. Therefore, the tensor structure of the amplitude is
governed by\footnote{Note that $y'$ plays the role of the Bjorken
parameter for the scattering of the scalar $\phi_I$ mode and the
gauge field $A^\mu$.}
\begin{equation}
v_s^\mu \equiv \frac{1}{q}\left(P^\mu + \frac{q^\mu}{2x}\right) \ \
\ \ \textrm{and} \ \ \ v_t^\mu \equiv \frac{1}{q}\left(q'^\mu +
\frac{q^\mu}{2y'}\right) \ \ \ \ \textrm{with} \ \ \
y'=\frac{-q^2}{2q'\cdot q} \, , \label{defvt}
\end{equation}
which means that the structure functions are obtained from
\begin{eqnarray}
F_1(x,q^2) &=& \pi \sum_{M_2,M_3} \int \frac{d^3p'}{2
E_{p'}(2\pi)^3} \frac{d^3q'}{2 E_{q'}(2\pi)^3} (2\pi)^4
\delta^{(4)}\left(P+q-p'-q'\right) |C_t|^2  \nonumber \\
&& \,\,\,\,\,\,\,\,\,\,\,\,\,\,\,\,\,\,\,\,\,\,\,\,\,\,\,\,
\times 2q^2 \left[v_t^2 + 4x^2 (v_s \cdot v_t)^2\right] ,\label{F1}\\
F_2(x,q^2) &=& \pi \sum_{M_2,M_3} \int \frac{d^3p'}{2
E_{p'}(2\pi)^3} \frac{d^3q'}{2 E_{q'}(2\pi)^3} (2\pi)^4
\delta^{(4)}\left(P+q-p'-q'\right) |C_t|^2 \nonumber \\
&& \,\,\,\,\,\,\,\,\,\,\,\,\,\,\,\,\,\,\,\,\,\,\,\,\,\,\,\, \times 4
x q^2 \left[v_t^2 + 12 x^2 (v_s \cdot v_t)^2\right] \, , \label{F2}
\end{eqnarray}
where $C_t$ is given by
\begin{eqnarray}
C_t(M_2, M_3, p', q') &=& \int dz \, dz' \left[ V_{IR}(z) \times
V_{UV}(z') \times G(z,z')\right] \nonumber\\
&=&  \int d\omega  \frac{\omega}{\omega^2+(P-p')^2}  S^{(z)}_{\phi
\phi_I \phi_I}(M_1, M_3, \omega) \ S^{(z')}_{A \phi_I
\phi_I}(M_2,q,\omega)\ ,
\end{eqnarray}
where the momentum conservation Dirac delta functions have been
written in equations (\ref{F1}) and (\ref{F2}). This means that we
can identify a first term in the $F_i$ functions that fulfills
exactly the Callan-Gross relation $F_2 = 2 x F_1$, and a second term
which contributes to the longitudinal structure function
\begin{equation}
\left(\begin{array}{c l}
   F_1\\
    F_2\\
    F_L
\end{array} \right)  = \frac{1}{N} \sum_{M_2 M_3} \frac{q |\vec{p'}|}{8}\sqrt{\frac{x}{1-x}} \int d\theta \sin
\theta \left(\begin{array}{c l}
    v_t^2 + 4 x^2 (v_s \cdot v_t)^2\\
    2x [v_t^2 + 12 x^2 (v_s \cdot v_t)^2]\\  16 x^3 (v_s \cdot v_t)^2
\end{array} \right) |C_t|^2 \, .
\end{equation}
One should keep in mind that the $N^{-1}$ pre-factor carries all the
dependence on the number of colors once the fields have been
re-scaled in order to obtain canonically normalized kinetic term in
the Lagrangian. The term corresponding to $F_1$ turns out to be
sub-leading in the large-$q^2$ expansion, thus we focus on the
calculation of $F_L(x,q^2)$. The constant $C_t$ contains the
integrals in $z$ of each vertex as well as the contribution from the
propagator of type I scalars $G(z,z')$, which has the same form as
in the case of glueballs.

The main difference with respect to the glueball case of
\cite{Jorrin:2016rbx} comes from the integrals of the IR interaction
vertex, containing integrals of three Bessel functions of the first
kind multiplied by $z^\kappa$, with positive integer values
$\kappa$. Recall that the $\kappa=1$ case appears for the glueballs.
In the present case, although the $z$-integrals are difficult to be
solved analytically, due to the presence of the cut-off
$z_0=\Lambda^{-1}$ we can approximately relate them to the one with
$\kappa=1$ and then analyze them by using techniques inspired in the
case studied by Auluck \cite{Auluck:2012}. These approximations are
described in detail in Appendix C. The resulting formulas that will
be used in the rest of this section in order to obtain the structure
functions are given by the equations
\begin{eqnarray}
&& I^{(\kappa)} (a,b,c,\Lambda) \equiv \int_0^{\Lambda^{-1}} dz \,
z^\kappa \, J_m (a z) \ J_n (b z) \ J_l (c z)  \nn \\
&\Rightarrow&\Lambda^3 I^{(4)} (a,b,c,\Lambda) \approx \Lambda^2
I^{(3)} (a,b,c,\Lambda) \approx \Lambda I^{(2)} (a,b,c,\Lambda)
\approx I_1(a,b,c, \Lambda), \label{aproxIn}
\end{eqnarray}
which is written up to certain ${\cal{O}}(1)$ numerical constants
that are not relevant in studying the leading $x$-dependence of the
structure functions. $I^{(1)}(M_1, M_3, \omega, \Lambda)$ is the
integral which appears in the glueball case. Equation (\ref{aproxIn}) implies that since
we are working in the small $\Lambda$ regime the larger contribution
comes from the $\kappa=4$ case, thus in what follows we will focus
on this case. However, as we will see the contribution of the other
integrals will become important in the $x \rightarrow 1$ limit. In
addition, we can perform an approximation similar to the one we have
used in \cite{Jorrin:2016rbx}
\begin{eqnarray}
I^{(4)} (a,b,c,\Lambda) &=& \int_0^{\Lambda^{-1}} dz \, z^4 \ J_m (a
z)
\ J_n (b z) \ J_l (c z) \nonumber \\
&\approx&
\left(\frac{1}{\Lambda}\right)^3\frac{1}{\sqrt{ab}}\left[(-1)^{\alpha}\delta
(c-(a+b))  + (-1)^\beta \delta(c-(a-b))\right] \, , \label{aproxI4}
\end{eqnarray}
for some integer powers $\alpha$ and $\beta$ that carry all the
dependence on the indices of the Bessel functions\footnote{See
Appendix C and also \cite{Auluck:2012}.}. The similarities between
the different integrals come from the fact that, regardless of the
integration limit, as functions of $c$ their largest contribution
comes from the region near $c=|a\pm b|$. In this context we have
$\omega = M_1 \pm M_3$. This kind of behavior where bulk
interactions in AdS act as some sort of energy conservation
restriction has been noted before \cite{Gao:2014nwa,Jorrin:2016rbx},
and in a sense it is an intuitive interpretation for the Dirac delta
functions approximation (\ref{aproxI4}).

Considering this approximation for the IR vertex and, since both the
UV vertex and the propagator $G(z,z')$ can be treated in the same
way as for the glueball calculation, one can square the amplitude,
perform the $\omega$ integration and carry out the angular integration in
$\theta$. The leading amplitude is given by $\omega = M_1 - M_3$.
Then, the sum over $M_3$ indicates that the important contributions
are given when the mass $M_3$ takes values near $\alpha M_1$, with
$\alpha = |\vec{p'}|/|\vec{p}|$. All of these results indicate that
the splitting occurs at small angles and that the ratio between the
momentum carried by the on-shell resulting particle of mass $M_3$
and the momentum $p$ of the incoming hadron is similar to that of
the AdS masses\footnote{As in \cite{Jorrin:2016rbx} we call $m^2 =
R^{-2}\Delta(\Delta-4)$ the Kaluza-Klein mass and $M_i$ ($i=1,2,3$)
as the AdS masses.}. Finally, we are left with the following $M_2$
sum
\begin{equation}
F_L^{I}=\frac{B^2}{\lambda N} \frac{M_1^6}{\Lambda^3}
\sum_{M_2}\frac{M_2}{q^{14}} (M_2^2 + q^2)^2 x^6 (q^2(1-x)-x
M_2^2)^3 x^6 \left( 1 + \frac{M_2^2}{q^2} \right)^6 \, ,
\end{equation}
where $B$ is a numerical constant.

The difference between this sum and the glueball one is given
by some constant factors $\Lambda$, but also $M_3$ and $\omega = M_1
- M_3$ which change the result as a function of $x$. Now, the
leading contribution comes from the case where $M_2$ takes values of
order $q$, which means that we can treat this sum as an integral
with measure $dM_2/\Lambda$ \cite{Polchinski:2002jw}. This integral
gives the final result for the longitudinal structure function
\begin{equation}
F_L^{(I)}(x,q^2) = \frac{1}{\lambda
N}\frac{B^2}{120}\left(\frac{M_1}{\Lambda}\right)^6
\frac{\Lambda^2}{q^2} \ x^3 (1-x)^4 (1+2x(2+5x)) \, .
\label{FLfinal}
\end{equation}
This is the most important result of this paper. However, since
equation (\ref{FLfinal}) behaves as $(1-x)^4$ when $x \rightarrow 1$
one has to keep in mind that there are other sub-leading
$z$-integrals. These contributions render terms proportional to $x^3
(1-x)^2 (1+x(2+3x))$ and $x^3 (1-x)^2$. This last term is exactly
the one that appears in the glueball case. Notice that the $x$
dependence is independent of the conformal dimension of the initial
state. This is very different in comparison with the large $N$
limit, as it has been noted for the glueball case
\cite{Jorrin:2016rbx}. Among the contributions coming from these
terms, the one coming from the $z^2$ and $z^3$ integrals are the
leading ones in this limit: when $x \rightarrow 1$ they behave as
$(1-x)^2$. All other terms are sub-leading. The appearance of this
asymptotic $(1-x)^2$ behavior is an important observation in terms
of the comparison with phenomenology.

Notice that the upper index in equation (\ref{FLfinal}) indicates that
this is the leading contribution we have from the type I mode. Since we
are using the optical theorem, we must add to this the other
leading-order contribution that we have from the possibility that
one of the intermediate states is associated with a type III mode.
Details of the calculation are shown in Appendix B, being the final
result of the same form as (\ref{FLfinal}).

Finally, it is worth noticing that as expected in the leading
structure function (\ref{FLfinal}) (and also in the rest of the
contributions) the $q$-dependence that one obtains is the same: the
amplitude fall off is $\Lambda^2/q^2$. In the case of the glueballs
this is predicted by OPE arguments on the quantum field theory side
\cite{Polchinski:2002jw}. The gravitational interpretation of this
is clear, when
\begin{equation}
q^2 > \Lambda^2 N^{2/(\tau_{\cal {Q}}-\tau_c)} \, , \label{qc}
\end{equation}
the $1/N^2$ suppression for large (but finite) $N$ of the one-loop
level process and the $(\Lambda^2/q^2)^{\Delta_{in}-1}$ suppression
factor of the tree-level amplitude becomes comparable, and for
larger $q$ the former calculation associated with two-particle final
states DIS is the leading one\footnote{Note that $\tau_{\cal {Q}}$
is the minimum twist of a single-trace operators of charge ${\cal
{Q}}$, and $\tau_c$ is the minimum twist of all electrically charged
single-trace operators. In both cases the operator's anomalous
dimensions are order 1.}. In that case, one obtains a suppression
factor $\Lambda^2/q^2$. Thus, initial hadron splitting and particle
creation are allowed. The scalar mode with lower Kaluza-Klein mass
(i.e. with $\Delta_{min}=2$) is the one propagating along the
vertical line and interacting with the non-normalizable gauge field
representing the holographic virtual photon. We expect the same
interplay between these two terms in the ${\cal {N}}=2$ SYM theory
dual to the D3D7-brane model. An obvious difference is that instead
of the factor $N^{2/(\tau_{\cal {Q}}-\tau_c)}$ in equation
(\ref{qc}) we should have $N^{1/(\tau_{\cal {Q}}-\tau_c)}$ due to
the fact that the fields in the hypermultiplet of the ${\cal {N}}=2$
SYM theory transform in the fundamental representation of $SU(N)$.

\section{Discussion and conclusions}

In this work we have investigated the longitudinal structure
function $F_L(x, q^2)$ for scalar mesons derived from the D3D7-brane
model, at strong coupling and in the $1/N$ expansion. Equation
(\ref{FL}) shows that the large $N$ limit and the high energy limit
($\Lambda^2 \ll q^2$) do not commute. This is because $\Delta \geq
3$ for scalar mesons (see table 1), therefore in the $\Lambda^2 \ll
q^2$ limit the first term is suppressed by (at least) an additional
factor $\Lambda^2/q^2$ in comparison with the rest of terms. This
implies that in this limit the second term in that equation becomes
the leading contribution. Similarly to what happens with DIS of
charged leptons off glueballs, we find that for scalar mesons
two-hadron final states dominate DIS processes. In terms of the FCS
this implies that certain one-loop Feynman-Witten diagrams in the
supergravity calculation are the most relevant ones. Using some
reasonable approximations explained in the preceding sections, we
have obtained $F_L(x, q^2)$ in the high energy limit:
\begin{eqnarray}
F_L &=& \frac{1}{N} \, \left( f^{(1)}_2 - 2 \, x \, f^{(1)}_1
\right) \, \left(\frac{\Lambda^2}{q^2}\right) \ . \nonumber
\end{eqnarray}
Specifically we have obtained the expression (\ref{FLfinal}), where
we have calculated the explicit dependence on the Bjorken parameter
when the FCS intermediate state corresponds to a type I mode, and a
similar expression when the exchanged particle is a type III mode.
These two expressions behave as $(1-x)^4$ as $x$ approaches 1. There
are additional contributions from the sub-leading $z$-integrals
which behave as $(1-x)^2$, however they are only relevant for $x$
very close to 1, therefore their contribution to the moments of the
structure functions is very small. Also, we have obtained the
explicit dependence on the virtual photon momentum transfer $q^2$.

We observe that the one-loop structure of the DIS amplitude leads to
a non-vanishing $F_1$ structure function even for scalar hadrons,
where this contribution is sub-leading in $1/N$. The leading term
contribution to the DIS amplitude is given by $F_2$, or in this case
the longitudinal structure function $F_L = F_2 - 2 x F_1$. We have
obtained the full $x$-dependence for the Kaluza-Klein tower of
scalar (and pseudoscalar) mesons. In all cases the key element comes
from the analysis of the $z$-integral of the Bessel functions
involved in the splitting process of the incoming hadron, followed
by the sum over the intermediate masses $M_2$ and $M_3$ (see figure
1). A remarkable effect is that in equation (\ref{FLfinal}) there is
a factor $1/\lambda$ in addition to the $1/N$ factor. This is
expected since the cubic interaction vertex involving three mesons
has a coupling strength proportional to
\begin{equation}
g_{cubic} \propto \frac{1}{\sqrt{N}} \frac{\alpha'}{L} \, ,
\end{equation}
where $L=\Lambda R^2$ and since $R^2=\sqrt{\lambda} \alpha'$ then
\begin{equation}
g_{cubic} \propto \frac{1}{\sqrt{N}} \frac{1}{\sqrt{\lambda}} \, .
\end{equation}

There are several interesting aspects that we should emphasize.
Firstly, the $l$ dependence of the structure function appears only
in the coefficients, but not in the powers of $x$ or $(1-x)$. This
is an important difference with respect to the structure functions
in the $N\rightarrow \infty$ limit
\cite{Koile:2013hba,Koile:2015qsa}, where $F_2 \propto
x^{l+4}(1-x)^{l+1}$. This behavior has also been found for glueballs
\cite{Jorrin:2016rbx}. Secondly, for all mesons the structure
function behaves as $F_L \sim (1-x)^2$ in the $x \approx 1$ region.
This has already been pointed out in \cite{Koile:2015qsa} for the
pion, and the fact that it holds for the one-loop correction and
extends to the rho meson constitutes an important test for the
validity of our results. In the context of the valence structure
functions it has been found a fall-off $(1 - x)^{2 \pm 0.1}$
\cite{Aicher:2010cb}.

The idea of this work is to show that the contribution of certain
one-loop Feynman-Witten diagrams of FCS lead to a better agreement
with lattice QCD simulations and phenomenological results for scalar
mesons, in comparison with the tree-level calculations. Since the
Bjorken parameter dependence of the results for $F_L$ is independent
of $\Delta_{in}$ it should hold for different scalar and
pseudoscalar mesons. Thus, we have compared our results for the
lightest pseudoscalar mesons from the D3D7-brane model with the
pion, for which there are more available data. In fact the structure
functions of the pion, and the associated parton distribution
functions have been extensively studied allowing us to compare with
data coming from experiments and also from different
phenomenological models
\cite{Wijesooriya:2005ir,Holt:2010vj,Reimer:2011,Chang:2014gga,Detmold:2003tm,Aicher:2011ai,Aicher:2010cb},
as well as from lattice QCD simulations
\cite{Best:1997qp,Brommel:2006zz,Chang:2014lva}.

The experiments carried out in order to analyze the internal
structure of the pion are generally based on the Drell-Yan process
within the parametric region $0.2 \leq x \leq 1$. This is
approximately the range of values of $1/\sqrt{\lambda} \ll x < 1$
where the supergravity description is accurate, since the
center-of-mass energy is not high enough in order to produce excited
string states in the intermediate channels \cite{Polchinski:2002jw}.
This is true at tree level, and in this work we have assumed that
the absence of excited strings also holds at one-loop level. For
smaller values of $x$ supergravity is not a good description and one
has to take into account the full string theoretical description in
the holographic dual model, in that case string loop effects become
important and, eventually, it could lead to black hole formation.
Nevertheless, it is worth noticing that even when $x$ is small some
approximations can be done in order to describe the curved-space
string theory scattering amplitude at high energy in terms of the
flat-space ten-dimensional string theory scattering amplitude
\cite{Polchinski:2002jw}. We have done this for scalar and polarized
vector mesons in the $N \rightarrow \infty$ limit in
\cite{Koile:2014vca}, i.e. for single-hadron outgoing states. As we
have pointed out in \cite{Koile:2015qsa}, in the multi-color limit
the results in the $1/\sqrt{\lambda} \ll x < 1$ region, which one
obtains from the holographic dual description of DIS of charged
leptons off mesons in terms of supergravity, are well described in
terms of the valence distribution functions. On the other hand, for
smaller values of $x$ the structure functions obtained from the
inclusion of string theory effects seem to be associated with the
contribution emerging from the soft gluons and the sea of quarks
\cite{Koile:2014vca,Koile:2015qsa}.

Another important data to compare with are the first moments of the
structure functions obtained from QCD lattice simulations. These
moments are defined as
\begin{equation}
M_n [F_i] = \int_0^1 dx \, x^{n-1} F_i(x,q^2) \ ,
\end{equation}
for a generic structure function $F_i$. In \cite{Koile:2015qsa} we
have already analyzed our $N \rightarrow \infty$ results in terms of
the moments of the pion and rho meson in comparison with the QCD
lattice simulations results presented in \cite{Best:1997qp}. In our
holographic description of scalar mesons the lightest pseudoscalar one
corresponds to the
case where the initial and final states are described by the scalar
fluctuation $\phi$ whose solution has the smallest Kaluza-Klein mass
and couples to the $U(1)$ gauge field $A_m$ given by a graviton. In
the D3D7-brane model, the smallest $\Delta=l+3$ corresponds to the
case $l=1$, where $l$ indicates the irreducible representation of
$SO(4)\sim SU(2)\times SU(2)$ of the associated scalar spherical
harmonic.

\begin{figure}
\centering
\includegraphics[scale=1.2,trim = 0 0 0 0 mm]{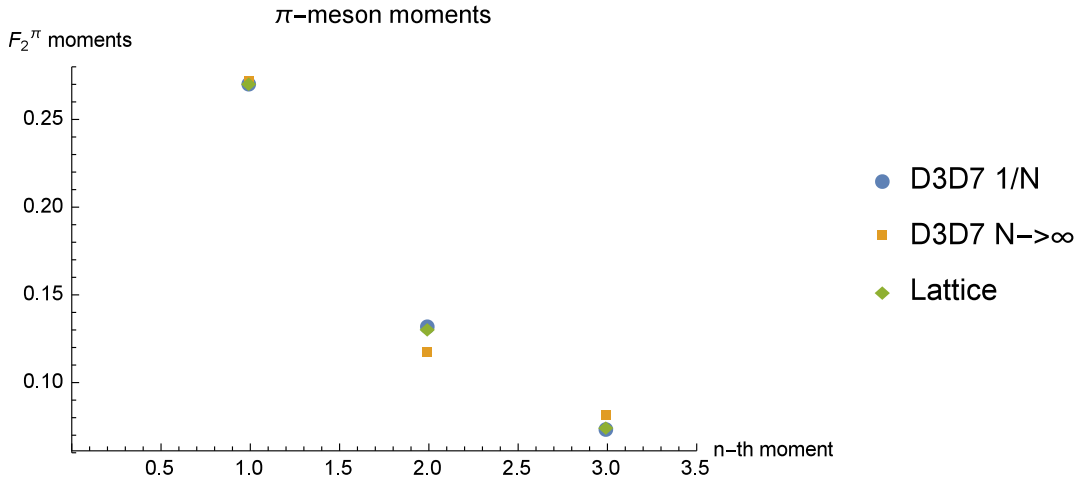}
\caption{\small The first three moments of $F_2$ are shown for the
pion. The free parameters of the D3D7-brane model are chosen in
order to fit the results of \cite{Brommel:2006zz,Chang:2014lva},
obtained with lattice QCD simulations, and labeled by "Lattice QCD".
Also, for comparison we have included the best fitting corresponding
to the moments of the structure function in the large $N$ limit,
with errors up to $10.8 \%$ with respect to lattice QCD results.
These are labeled by "D3D7 $N \rightarrow \infty$". "D3D7 $1/N$"
labels the best fitting corresponding to the moments of the
structure function in the high energy limit, with errors up to $1.27
\%$ with respect to lattice QCD simulations \cite{Brommel:2006zz,
Chang:2014lva}.} \label{First-Moments-Pion}
\end{figure}

The first three moments of $F_1(x, q^2)$ and $F_2(x, q^2)$ have been
calculated from lattice QCD in \cite{Best:1997qp} for the pion and
the rho meson. We consider the moments of $F_2(x, q^2)$ for the pion
and compare them with the moments of our $F_L(x, q^2)$ since in our
case $F_1(x, q^2)$ is sub-leading. Thus, we integrate our result
between $x=0.1$ and $x=1$, i.e. within the range of validity of the
supergravity calculations (we refer this parametric region as the
large-$x$ region). However, since we are analyzing the first moments it is
important to take into account the region for smaller values of $x$
in order to be able to integrate the structure function for lower
$x$ values as well. We will assume that the small-$x$ behavior is
similar to the one we found in \cite{Koile:2014vca} and used in
\cite{Koile:2015qsa}, i.e., $F_L^{small}(x, q^2) \propto x^{-1}$.
The reason is given as follows. The main difference between our
result for large-$x$ in the one-loop calculation from the leading
diagram of figure 1 and the previous one obtained in the planar
limit is the fact that in the tree-level FCS calculation both the $q$-
and $x$-dependence are determined by $\Delta_{in}$ of the target
hadron, while at one-loop it is determined by $\Delta_{min}$.
Recall that $\Delta_{min}$ is given by the lowest conformal
dimension available among the supergravity excitations. Nevertheless,
at low-$x$ the string theoretical calculation is independent of
$\Delta_{in}$. In this way, we may conjecture that in this aspect
this will not be very different in comparison with the one-loop
level situation. Thus, we consider this $1/x$ behavior and add it to
the moment calculation by integrating it from $x=0.0001$ and $x=0.1$
as before \cite{Koile:2015qsa}. We rewrite the rest of the structure
function in two dimensionless constants: one in front of the
small-$x$ $F_L$ part and the other one multiplying the large-$x$
$F_L$ part. Then, we carry out the best fitting for these two
constants in comparison with the lattice QCD calculations of three
lowest moments for the pion.

The results of that fitting are presented in figure 2 compared with
the known results and the previous fitting performed with the
$N\rightarrow \infty$ structure function $F_2$. The first constants
are approximately 0.0017 and 14.47. They are similar to the ones
found in our previous work \cite{Koile:2015qsa} in the large $N$
limit, for which the constants associated with the small-$x$ $F_L$
and with the large-$x$ $F_L$ are 0.0143 and 28.89, respectively.
Another interesting point is the ratio of the third and second
moments of $F_2$, which in large $N$ limit gives $M_3[F_2]/M_2[F_2]
= 0.69$ \cite{Koile:2015qsa}, while in the high energy limit gives
$M_3[F_2]/M_2[F_2] = 0.55$. The last result is closer to the
expected ratio near 0.5 \footnote{We thank Andreas Schafer for this
comment.}.

Table 2 shows a comparison of our new results for the first three
moments of the structure function $F_2$ of the lightest pseudoscalar
meson with respect to the average results of the lattice QCD
computations in \cite{Brommel:2006zz, Chang:2014lva} and in
comparison with the results presented in \cite{Koile:2015qsa} at
large $N$. Uncertainties in the lattice computations are omitted.
%
\begin{table}
\def\arraystretch{1.5}
\begin{center}
\begin{tabular}{|c|c|c|c|}
\hline
Model / Moment & $M_1(F_2)$ & $M_2(F_2)$ & $M_3(F_2)$ \\
\hline
Lattice QCD & 0.27 & 0.13 & 0.074 \\
\hline
D3D7 ($1/N$) & 0.2699 & 0.1326 & 0.0731 \\
\hline
Percentage error & 0.04 & -1.27 & 1.27 \\
\hline
D3D7 ($N\rightarrow \infty$) & 0.2708 & 0.1161 & 0.0803 \\
\hline
Percentage error & -0.3 & 10.8 & -8.5 \\
\hline
\end{tabular}
\caption{\small Comparison of our new results for the first moments
of the structure function $F_2$ of the lightest pseudoscalar meson
for a suitable choice of the normalization constants with respect to
the average results of the lattice QCD simulations in
\cite{Brommel:2006zz, Chang:2014lva} and in comparison with the
results presented in \cite{Koile:2015qsa}. Uncertainties in the
lattice computations are omitted.}\label{momentstablepi}
\end{center}
\end{table}

Also, the shape of the $F_2$ structure function as a function of the
Bjorken parameter for fixed virtual photon momentum transfer is
shown in figure 3. The darker line represents the present $1/N$
calculations, while the other curve corresponds to the previous ones
reported in \cite{Koile:2015qsa} in the large $N$ limit. For low-$x$
we consider our previous result from \cite{Koile:2014vca}. The
difference between the two low-$x$ curves is due to the slightly
different constants needed for the best fitting in each situation.
\begin{figure}
\centering
\includegraphics[scale=1.,trim = 0 0 0 0 mm]{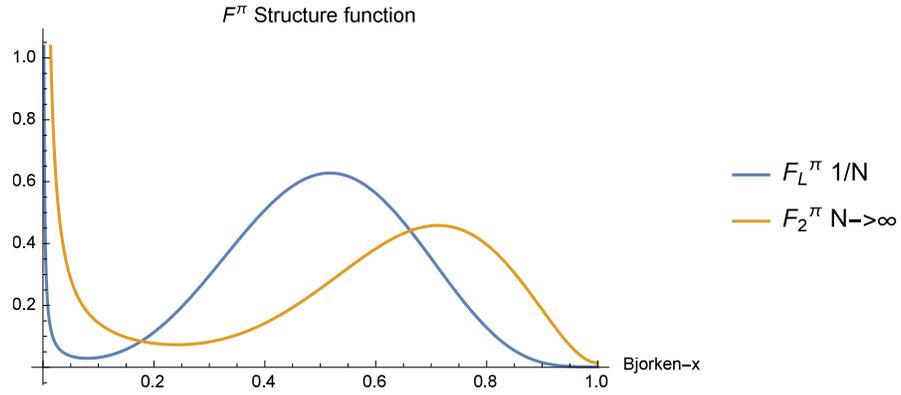}
\caption{\small $F_2$ as a function of the Bjorken parameter $x$. We
consider the values of the constants which give the moments
indicated in table 2.} \label{First-Moments-Pion}
\end{figure}
Also, notice that for the $1/N$ expansion for small-$x$ $F_L$ is
smaller in comparison with the large $N$ limit, while there is an
opposite trend for the larger $x$ region. In addition, as expected
from phenomenological results the peak in $F_L$ in the $1/N$
calculation moves toward smaller values of $x$.

Higher order moments can also be calculated from our results. We
display these in figure 4 in comparison with \cite{Nam:2012vm}. The
difference between the corresponding first moments of figure 4 and
table 2 is due to the fact higher order moments are generally
calculated from the valence structure functions and they do not
include the small-$x$ contributions.
\begin{figure}
\centering
\includegraphics[scale=1.2,trim = 0 0 0 0 mm]{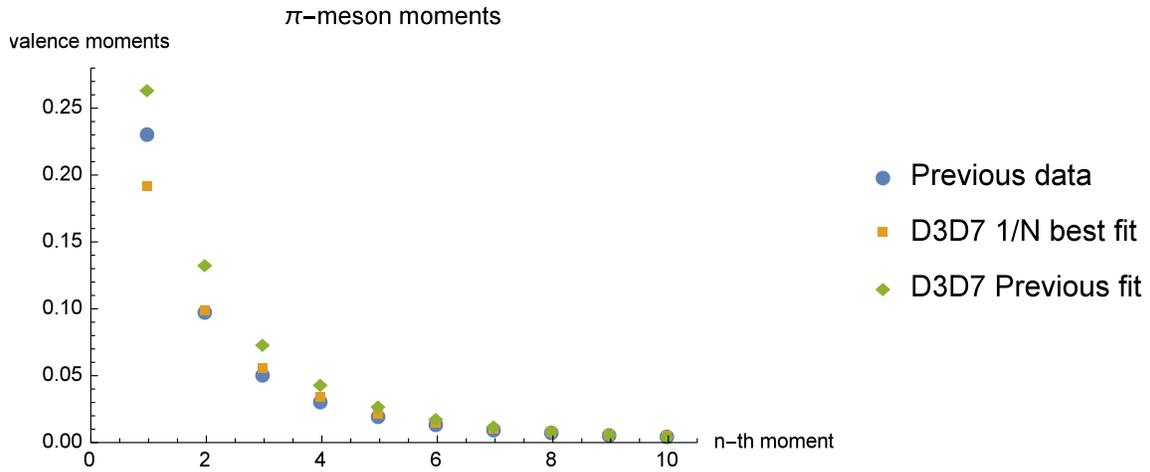}
\caption{\small Higher moments of $F_2$ are shown for the pion.
Previous data correspond to reference  \cite{Nam:2012vm}.}
\end{figure}

The analysis of this work is restricted to the $1/N$ corrections to
the holographic dual description of DIS of a charged lepton off
scalar mesons. The Bjorken parameter dependence of
the structure functions for higher orders in this expansion is
difficult to calculate explicitly. However, we
can comment on the $q^2$-dependence of these terms. This dependence
for the leading diagram is dictated by the UV interaction vertex,
which has the same form as for the glueball case. In that situation
the propagating mode is an $s$-scalar, also with the lowest
conformal dimension \cite{Jorrin:2016rbx}. Hence, it is reasonable
to expect that at higher order in $1/N$ the splitting process will
be more complicated, but still it will be restricted to the IR
region. Note that for these ladder type higher-order diagrams there
are in principle two possibilities: a type I$^-$ D7-brane field and
the $s$-scalar bulk field. In any case the leading $1/q^2$
contribution should not change.

Let us very briefly comment on the $L=0$ case which is very
different for several reasons. Formally a null separation between
the D7-brane and the stack of $N$ D3-branes implies that conformal
symmetry is restored, therefore the quarks become massless. This is
because for $k$ D7-branes the beta function for the 't Hooft
coupling is proportional to $k/N$, which vanishes in the probe limit
\cite{Kruczenski:2003be}. From the computational point of view, the
crutial IR interaction vertex that couples the scalar mesons with
any type of gauge modes is absent. In fact, except for the
non-Abelian case (number of flavors larger than one) all three-point
vertices vanish. This means that our leading diagram of figure 1
does not exist in this case. Thus, the results will be conceptually
very different in this limit. For example, as in the previous
paragraph one should go to higher orders in the $1/N$ expansion in
order to find a diagram with a propagating mode carrying the lowest
dimension. If the number of loops is increased one should also have
higher powers of $1/\lambda$ multiplying the $1/N$ ones.

Possible extensions of this work to other gauge field theories can
be done by considering the DpDp+4-brane models, which are duals to
gauge theories in p+1 dimensions such as the ones discussed in
\cite{Myers:2006qr}.

~

~

~

\centerline{\large{\bf Acknowledgments}}

~

We thank S. K. H. Auluck for correspondence on reference
\cite{Auluck:2012}, and Ezequiel Koile, Gustavo Michalski and Carlos
N\'u\~nez for comments and discussions. The work of D.J., N.K. and
M.S. is supported by the CONICET. This work has been supported in
part by
the CONICET-PIP 0595/13 grant and UNLP grant 11/X648.

\newpage

\setcounter{section}{0}

\appendix

\section{Spherical harmonics on $S^3$ }
\label{appendix-SH}

In this appendix we list some properties and formulas relevant for
the type IIB supergravity Feynman diagram calculation at one-loop
level that we use in order to obtain the structure functions of
scalar mesons in the $1/N$ expansion. Several of the basic results
involving scalar and vector spherical harmonics have been derived in
\cite{Cutkosky:1983jd} and \cite{Aharony:2006rf}.

\subsection{Basic properties of spherical harmonics}

Spherical harmonics belong to representations of the isometry group
of the three-sphere, i.e. $SO(4) \approx SU(2) \times SU(2)$. The
scalar spherical harmonics transform in the
$(\frac{l}{2},\frac{l}{2})$ representation, where $l$ is a
non-negative integer, while $-\frac{l}{2} \leq m, n \leq
\frac{l}{2}$. They satisfy an orthogonality condition
\begin{eqnarray}
\int_{S^3} Y_l^{m,n} \, Y_{l'}^{m',n'} &=& \delta_{l l'} \,
\delta_{m n} \, \delta_{m' n'} \, ,
\end{eqnarray}
and their complex conjugate are calculated from
\begin{equation}
(Y_{l}^{m,n})^*=(-1)^{m+n} \, Y_{l}^{-m,-n} \, .
\end{equation}
Spherical harmonics are eigenfunctions of the Laplace operator on
the sphere
\begin{equation}
\nabla^{2}Y_{l}^{m,n}= -l(l+2) \, Y_{l}^{m,n} \, .
\end{equation}
Under parity transformation their eigenvalues are $(-1)^l$.

A vector field on $S^3$ can be spanned by a combination of gradients
of the scalar spherical harmonics $\nabla_i Y$ plus a set of vector
spherical harmonics $Y_i^{\pm}$, which transform in the $(\frac{l\mp
1}{2}, \frac{l\pm 1}{2} )$ representation of the $SO(4) \approx
SU(2) \times SU(2)$ group, with $l \geq 1$. In order to make the
notation simpler, the indices $m$ and $n$ can be omitted. Whenever it is necessary to write them
explicitly, we use the following notation $\vec{Y}_{l, \epsilon}^{m,
n}$, where $\epsilon=\pm1$ indicate the representation.
They satisfy the eigenvalue equations
\begin{eqnarray}
\nabla_i\nabla^i Y^{l,\pm}_{j}-R_j^k Y^{l,\pm}_{k} &=&-(l+1)^2 \, Y^{l,\pm}_{j}  \, , \\
\epsilon_{ijk} \nabla_j Y^{l,\pm}_{k}&=&\pm (l+1) \, Y^{l,\pm}_{i}  \, , \\
\nabla^i Y_i^{l, \pm}&=&0  \, , \label{prop-SH}
\end{eqnarray}
where $R_{ij}= 2\delta_{i j}$ is the Ricci tensor of an $S^3$ of unit
radius. Also, they satisfy the following relation
\begin{eqnarray}
\vec{Y}_{l,\epsilon}^{*,m,n}=(-1)^{m+n+1} \,
\vec{Y}_{l,\epsilon}^{-m,-n} \, .
\end{eqnarray}
The vector spherical harmonics satisfy orthogonality relations,
\begin{eqnarray}
\int_{S^3} \vec{Y}_{l,\epsilon}^{m,n} \cdot
\vec{Y}_{l',\epsilon'}^{m',n'} &=& \delta_{l,l'} \,
\delta_{m,m'} \, \delta_{n,n'} \, \delta_{\epsilon, \epsilon'}  \, , \nonumber \\
\int_{S^3} \vec{Y}_{l,\epsilon}^{m,m'} \cdot
\vec{\nabla}Y_{l'}^{n,n'} &=& 0 \, .
\end{eqnarray}
The $\vec{Y}_{l,\epsilon}^{m,n}$ harmonics does not mix with other vector
spherical harmonics since they belong to different representations
of $SO(4)$.

\subsection{Integrals of spherical harmonics}

The interaction vertices we consider have coefficients involving
integrals over three spherical harmonics. These integrals lead to
selection rules for the outgoing modes and introduce a dependence
in $l$. The relevant integrals are,
\begin{eqnarray}
\int_{S^3}   Y_{l}^{m,n} \, \vec{Y}_{l',\epsilon}^{m',n'} \cdot
\vec{Y}_{l'',\epsilon'}^{m'',n''} &=& \left(\begin{array}{ccc}
\frac{l'+\epsilon}{2} & \frac{l''+\epsilon'}{2} & \frac{l}{2} \\
m' & m'' & m \\ \end{array}  \right) \left(\begin{array}{ccc}
\frac{l'-\epsilon}{2} & \frac{l''-\epsilon'}{2} & \frac{l}{2} \\
n' & n'' & n \\ \end{array}  \right)  R_{1,\epsilon,\epsilon'}(l',l,l'') \\
&& \nonumber \\
\int_{S^3}   Y_{l}^{m,n} \, \vec{Y}_{l',\epsilon}^{m',n'} \cdot
  \nabla Y_{l''}^{m'',n''} &=& \left(\begin{array}{ccc}
\frac{l''}{2} & \frac{l'+\epsilon'}{2} & \frac{l}{2} \\
m'' & m' & m \\ \end{array}  \right) \left(\begin{array}{ccc}
\frac{l''}{2} & \frac{l'-\epsilon'}{2} & \frac{l}{2} \\
n'' & n' & n \\ \end{array}  \right)  R_{2}(l',l,l''),
\end{eqnarray}
where the matrices are the 3$j$-symbols, while the functions $R_1$
and $R_2$ are defined as
\begin{eqnarray}
R_{1,\epsilon,\epsilon'}(x,y,z)&=& \frac{(-1)^{\sigma+(\epsilon+\epsilon')/2}}{\pi}
\left(\frac{(y+1)}{32 (x+1)(z+1)}\right)^{1/2} \left( ( \epsilon(x+1)+ \epsilon' (z+1)+ y+2) \right.\nonumber \\
&&  (\epsilon(x+1) +\epsilon'(z+1) +y ) (\epsilon(x+1)+\epsilon'(z+1)-y )\nonumber\\
&&\left .(\epsilon(x+1)+\epsilon'(z+1)-y-2 ) \right)^{1/2}  \, , \\
&& \nonumber \\
R_{2}(x,y,z)&=&\frac{(-1)^{\sigma'}}{\pi}\left[
\frac{(x+1)(z+1)(\sigma'-x)(\sigma'-y)(\sigma'-z)(\sigma'+1)}{
(y+1)} \right]^{\frac{1}{2}}.
\end{eqnarray}
The right-hand sides of these equations are defined to be
non-vanishing only if the inequality $|x-z| \leq y \leq x + z$ is
fulfilled, and if $\sigma = \frac{x+y+z}{2}$ in $R_1$  and
$\sigma'=\frac{x+y+z+1}{2}$ in $R_2$ are integers.

The leading diagram in the $1/N$ expansion has an incoming scalar
meson, a vector type I ($\epsilon=-1$) mode with $l=1$, and a third
field which could be a type I or type III mode. The intermediate
meson with $l=1$ only admits $m'=0$ and the $n'$ index can take
three possible values ($\pm 1,0$). Firstly, we consider the case
with a type I ($\epsilon=-1$) scalar as the third field in the
vertex. The angular integral is
\begin{eqnarray}
\int_{S^3} Y_{l}^{m,n} \vec{Y}_{1,-1}^{m',n'} \cdot
\vec{Y}_{l'',-1}^{m'',n''}= \left(\begin{array}{ccc}
\frac{l''-1}{2} & 0 & \frac{l}{2} \\
m'' & 0 & m \\ \end{array}  \right) \left(\begin{array}{ccc}
\frac{l''+1}{2} & 1 & \frac{l}{2} \\
n'' & n' & n \\ \end{array}  \right)  R_{3,-1,-1}(l'',l,1) \, .
\end{eqnarray}
The first $j$-symbol imposes a selection rule on $l''$
\begin{eqnarray}
\left(\begin{array}{ccc}
\frac{l'-1}{2} & 0 & \frac{l}{2} \\
m & 0 & -m \\ \end{array} \right) =  \begin{array}{c c}
(-1)^{-m}  \frac{i^{-l}}{\sqrt{l+1}} \ \ & \textrm{if} \ l''=l+1  \\
0 \ \ & \textrm{if}  \   l'' \not= l+1 \, .
\end{array} \
\end{eqnarray}
From the conservation conditions $m+m'=0$ and $n+n'+n''=0$, we can
simplify the integral and calculate the sum of the square terms by
using the optical theorem
\begin{eqnarray}
\sum_{n'=-1}^1 \left(\left(\begin{array}{ccc}
\frac{l}{2} & 0 & \frac{l}{2} \\
-m & 0 & m \\ \end{array}  \right) \left(\begin{array}{ccc}
\frac{l+2}{2} & 1 & \frac{l}{2} \\
-n-n' & n' & n \\ \end{array}  \right)
R_{3,-1,-1}(l,l+1,1)\right)^2= \frac{1}{2 \pi^2} \, .
\end{eqnarray}
The result is independent of the conformal dimension related to the
incoming field ($\Delta \sim l$). If the third mode is a type I
scalar with $\epsilon=1$ the $j$-symbols change and the selection
rule is $l''=l-1$, but the result is the same. However, for a type
III scalar we obtain a selection rule $l=l''$ and the result depends
on the conformal dimension of the incoming field
\begin{eqnarray}
\sum_{n'=-1}^1 \left( \left(\begin{array}{ccc}
\frac{l}{2} & 0 & \frac{l}{2} \\
m & 0 & -m \\ \end{array}  \right) \left(\begin{array}{ccc}
\frac{l}{2} & 1 & \frac{l}{2} \\
-n-n' & n' & n \\ \end{array}  \right)  R_{2}(l,l,1)\right)^2=
\frac{l(l+2)}{2 \pi^2} \, .
\end{eqnarray}

\section{Contribution of the type III mode}

Let us consider an outgoing type III mode. In eight dimensions the
solution is defined by two functions $\phi_{III}(\rho)$ and
$\tilde{\phi}_{III}(\rho)$, and the scalar spherical harmonics
$Y^{(l)}(\Omega)$ of $S^3$. It takes the form
\cite{Kruczenski:2003be}
\begin{equation}
B_\mu = 0 \ \ \ , \ B_\rho = e^{i k\cdot x} \, \phi_{III}(\rho) \,
Y^{(l)}(\Omega) \ \ \ , \ B_i = e^{i k\cdot x} \,
\tilde{\phi}_{III}(\rho) \, \nabla_i Y^{(l)}(\Omega) \ ,
\end{equation}
where the associated conformal dimension is $\Delta = l+3$ and the
relation between the radial functions is
\begin{equation}
l(l+2) \, \tilde{\phi}_{III} = \frac{1}{\rho} \,
\partial_{\rho}\left( \rho^3 \phi_{III} \right) \, . \label{phi3-1}
\end{equation}
The equation of motion is given by
\begin{equation}
\partial_{\rho}\left( \frac{1}{\rho} \partial_{\rho} (\rho^3 \phi_{III} (\rho)) \right)- l(l+2)
\, \phi_{III}(\rho)- \frac{M^2 R^2}{\rho^2} \, \phi_{III}(\rho)=0 \,
, \label{eqn91}
\end{equation}
where we have taken $L$ very small. Since $L$ appears only at
quadratic order in equation (\ref{eqn91}), we can use the $L=0$
solutions for the on-shell evaluation instead of the Hypergeometric
ones corresponding to non-zero values of $L$. Then, the normalizable
solutions are given in terms of Bessel functions is
$\phi_{III}=c_{III}J_{\Delta-2}(MR^2/\rho)$ where
$c_{III}=R^2\sqrt{l(l+2) \Lambda/M}$. Note that these modes have a
different normalization constant in comparison with the mode I.
Plugging this expression in the on-shell interaction action given in
equation (\ref{Int-IR}), the first two terms of the effective action
vanish. Thus
\begin{equation}
\int d\Omega_3 \, Y^{l''} \, \nabla^i Y^{1 j} \left( \nabla_i
\nabla_j Y^{*l}-\nabla_j \nabla_i Y^{*l}\right)=0 \, .
\end{equation}
Therefore, we obtain the following expression for the on-shell
evaluation of the interaction action
\begin{equation}
S_{int} =-\frac{\mu_7}{N^{3/2}} (2 \pi\alpha')^3 2 L \int d\rho
d\Omega_3 \sqrt{-g} \frac{\phi(\rho)}{\rho^2}
\nabla^{\rho}(\phi_{I}(\rho)) \left( \partial_{\rho}
\tilde{\phi}_{III} (\rho)- \phi_{III} (\rho) \right) Y^{in}Y^{I
i}\nabla_i Y^{III},
\end{equation}
where we have omitted the momentum conservation factor as before.
This can be simplified by using the relation between
$\tilde{\phi}_{III}$ and $\phi_{III}$ of equation (\ref{phi3-1}) and
the equation of motion, since they imply
\begin{equation}
\nabla_{\rho} (\tilde{\phi}_{III}(\rho))=\frac{\partial_{\rho}
\left(\frac{1}{\rho} \partial_{\rho}(\rho^3 \phi_{III}(\rho))
\right)}{l(l+2)} =\phi_{III}\left(1+\frac{M_3^2 R^4}{l(l+2) \rho^2}
\right) .
\end{equation}
The resulting action in terms of $z = R^2/\rho$ for the outgoing
type III mode is
\begin{eqnarray}
S_{int}&=&-\frac{\mu_7}{N^{3/2}} (2\pi\alpha')^3 2 L
\sqrt{\frac{M_3 M_1 \omega}{l(l+2)}} M_3 \int dz z^2
\partial_z(z^2 J_{0}(\omega z)) J_{\Delta_{in}-2}(M_1 z) J_{\Delta_3-2}(M_3 z)) \
I_3, \nonumber \\
&&
\end{eqnarray}
where $I_3=\int d\Omega_3 Y^{(in)} \vec{Y}^I \cdot \nabla Y^{III} $
is the angular integral on the sphere which it is performed in
Appendix A. As we can see, the $z$-integrals have powers of order 3
and 4 as for the type I case. Now, one can insert this on-shell
$S_{int}$ in the holographic expression for the electromagnetic
current one-point function with the corresponding two-particle final
state. This leads to a non-vanishing contribution to the
longitudinal structure function of the same form as in the type I
case. The $M^2_3$ factor is canceled with the normalization constant
of the modes III. In addition, the factor $\sqrt{l(l+2)}$ in the
denominator, which introduces a $\Delta$ dependence, is canceled by
the angular integral on $S^3$.

\section{Integrals of products of Bessel functions}

In this appendix we discuss the approximations of the $z$-integrals
of three Bessel functions at the IR interaction vertex that we use
to obtain the structure functions in Section 3.5.

In the case of glueballs \cite{Jorrin:2016rbx}, the IR interaction
vertex describes a process where an incoming hadron, whose
holographic dual representation is given by a normalizable
Kaluza-Klein mode of the dilaton, splits into two other hadrons.
Similarly, in the present case for scalar mesons, in the leading
contribution one of the two resulting fields has the minimal
conformal dimension $\Delta_{min}$. This is $\Delta_{min}=2$, and it
is the same both for glueballs \cite{Jorrin:2016rbx} and for scalar
mesons. Considering the change of variable $z=R^2/r$, the on-shell
interaction action involves a $z$-integral of the form
\begin{equation}
I^{(1)} = \int_0^{z_{max}} dz \, z \ J_{\Delta_{in}-2}(M_1 z)
J_{\Delta'-2}(M_3 z) \ J_{0}(\omega z) \ , \ \ \ \ z_{max} =
\Lambda^{-1} \ . \label{glueballint}
\end{equation}
This kind of integrals are not explicitly known for arbitrary
integration limits. The known analytic results are obtained when the
upper limit is $z \rightarrow \infty$, and they can be written in
terms of Hypergeometric functions and Appell series \cite{Libro}.
However, an interesting numerical analysis has been developed by
Auluck. In \cite{Auluck:2012}, it has been proposed that as a
function of one of the AdS masses, $\omega$, and in the limit of
large $z_{max}$ the integral of equation (\ref{glueballint}) behaves
approximately as a sum of Dirac delta functions, up to some
normalization constant. It is easy to see how these functions arise.
By using the asymptotic expression\footnote{Since the hadron
splitting occurs in the IR region this approximation makes sense
because the main contribution to the $z$-integral comes from values
of $z$ far away from zero.} (\ref{Jcos}) it allows one to rewrite a
general product of three Bessel functions $J_{m}(az) \ J_{n}(bz) \
J_{l}(cz)$ as
\begin{eqnarray}
&&\left(\frac{2}{\pi z}\right)^{3/2} \frac{1}{\sqrt{abc}}
\cos\left(a z - m\frac{\pi}{2}-\frac{\pi}{4}\right) \cos\left(b z -
n\frac{\pi}{2}-\frac{\pi}{4}\right)
\cos\left(c z -l\frac{\pi}{2} -\frac{\pi}{4}\right) \label{cosenocubo}\\
&=& \left(\frac{2}{\pi z}\right)^{3/2} \frac{1}{\sqrt{abc}}
\sum_{\alpha=\pm 1, \beta=\pm 1} \cos \left[ (c-\alpha a - \beta b)z
+ (-m+\alpha n+\beta l)\frac{\pi}{2} +
(-1+\alpha+\beta)\frac{\pi}{4} \right].  \nonumber
\end{eqnarray}
Now, the integration of each term multiplied by $z$ leads to the
appearance of the square of the corresponding frequency, plus some
signs and Fresnel sine and cosine functions. These frequencies are
given by $|c\pm a \pm b|$, and each term describes the correct
behavior near the region where one of these factors vanishes. In our
case it means that the integral has two divergencies, namely: at
$\omega = (M_1 \pm M_3)$. Another way to see this is to proceed as
in \cite{Hung:2010pe} by using the analytic continuation of the
series expansion of the Bessel functions
\begin{eqnarray}
J_m(az) & \approx & \frac{1}{\sqrt{2\pi a z}} \left[e^{i \left(a z -
m\frac{\pi}{2}-\frac{\pi}{4}\right)} \sum_{j=0}^{\infty}
\frac{i^j(m,j)}{(2az)^j}+e^{-i \left(a z -
m\frac{\pi}{2}-\frac{\pi}{4}\right)} \sum_{j=0}^{\infty}
\frac{i^{-j}(m,j)}{(2az)^j} \right] \ ,
\end{eqnarray}
where
\begin{eqnarray}
&& (m,j) \equiv
\frac{\Gamma\left(\frac{1}{2}+m+j\right)}{n!\Gamma\left(\frac{1}{2}+m-j\right)}
\ . \nonumber
\end{eqnarray}
By combining this expression for each Bessel function, multiplying
by $z$ and integrating term by term, one obtains the same poles as
before plus finite terms. Thus, scaling arguments for the behavior
of the integral under the change $(a,b,c) \rightarrow (k a, k b, k
c)$ for some constant $k$, together with a numerical analysis
similar to the one of reference \cite{Auluck:2012} around each
singularity, lead to an approximation in terms of two Dirac delta functions. It takes the form
\begin{equation}
I^{(1)} \approx \frac{1}{\sqrt{M_1 M_3}}
\left[(-1)^{\gamma_-}\delta\left(\omega-(M_1 - M_3)\right) +
(-1)^{\gamma_+}\delta\left(\omega-(M_1 + M_3)\right)\right] \ ,
\end{equation}
where $\gamma_\pm$ can be $0$ or $1$ according to the phases
determined by the $(m, n, l)$ indices in the asymptotic
approximation (\ref{cosenocubo}). They are not important for the
present calculation, since we only need the square of the first
term. There is a simple physical interpretation for this
Dirac delta function behavior: it is associated with some sort of
mass-conservation condition in the IR process
\cite{Gao:2014nwa,Jorrin:2016rbx}.

In the present case, the situation seems to be more complicated since we
have a linear combination of different integrals of the form
\begin{equation}
I^{(\kappa)} = \int_0^{z_{max}} dz \, z^\kappa J_{\Delta_{in}-2}(M_1
z) \ J_{\Delta'-2}(M_3 z) \ J_{0}(\omega z) \ , \label{nuevasint}
\end{equation}
with $\kappa = 2, 3, 4$. Naively, it seems that we might have a
problem since for $z_{max} \rightarrow \infty$ the integrand grows
(and oscillates) with $z$ for $\kappa \geq 3/2$. However, we are not
integrating up to $z=\infty$ and the fact that there is an upper
limit given by the cut-off is important.  In addition, there would
be no problem even if there was no cut-off: one has to keep in mind
that the Bessel function solutions are only an approximation. The
background is not exactly $AdS_5\times S^3$ and the exact form of
the solutions is given in \cite{Kruczenski:2003be} in terms of
Hypergeometric functions. The product of three of them times
$z^{\kappa}$ falls off for any $\kappa$ for large $z$ for all the
values of $\kappa$ we are dealing with. We do not see this
explicitly because this behavior occurs at distances larger than $z
\sim R^2/L$, where the approximation breaks down. On the other hand,
a similar analysis singles out the same singularities on the
$\omega$-plane. Now, all of this encourages us to consider an analogous approximation to what was
used in the glueball calculation for $\kappa=1$, and we only need to
study the behavior near $\omega = M_1 \pm M_3$.

We find that these integrals behave very similarly when they are
divided by an appropriate power of the upper limit of integration.
The observed numerical behavior is depicted by some examples where
each integral is studied as a function of $\Lambda^{-1}$ for
different values of $\omega$ (figures 5, 6 and 7). From these figures we can see that $\Lambda I^{(2)}$,
$\Lambda^2 I^{(3)}$ and $\Lambda^3 I^{(4)}$ behave in the same way,
up to some ${\cal{O}}(1)$ numerical constants, and very similarly to $I^{(1)}$.

\begin{figure}
\centering
\begin{subfigure}[$\kappa=1$,$\omega<M_1-M_3$]{
\includegraphics[scale=0.4]{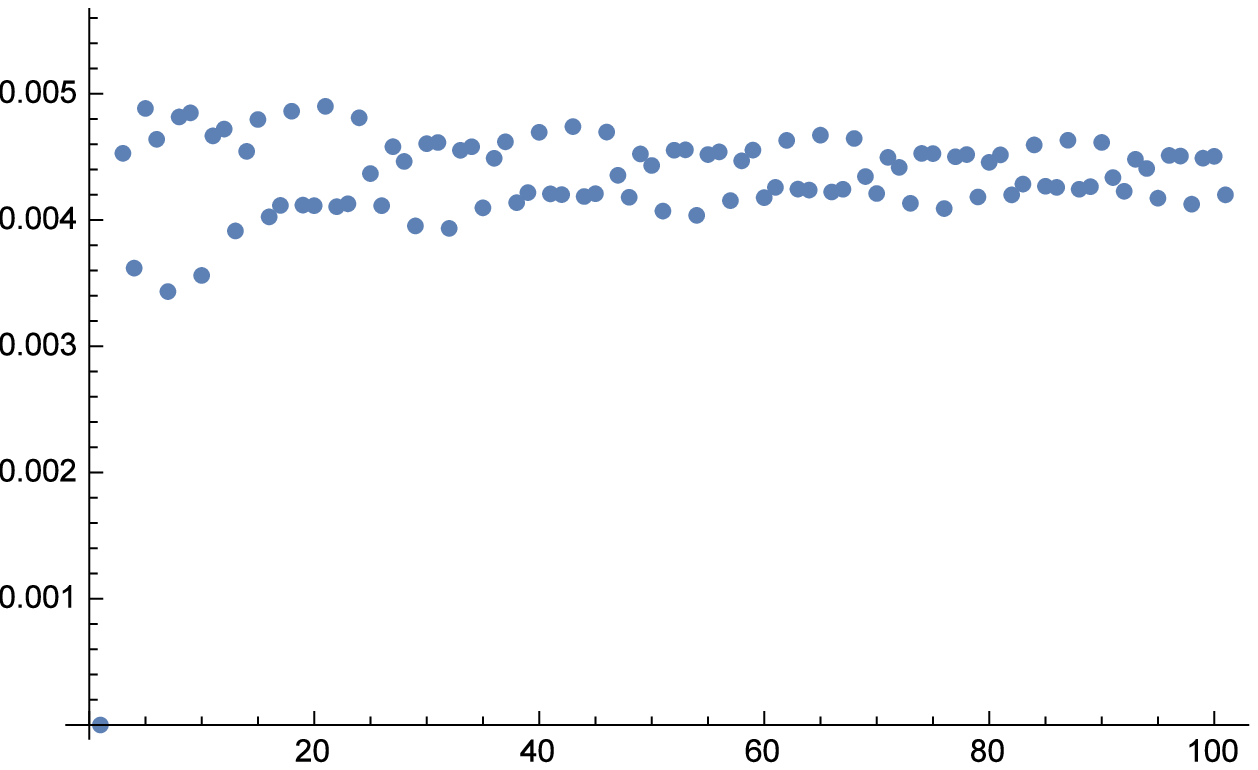}}
\end{subfigure}
\begin{subfigure}[$\kappa=1$,$\omega=M_1-M_3$]{
\includegraphics[scale=0.4]{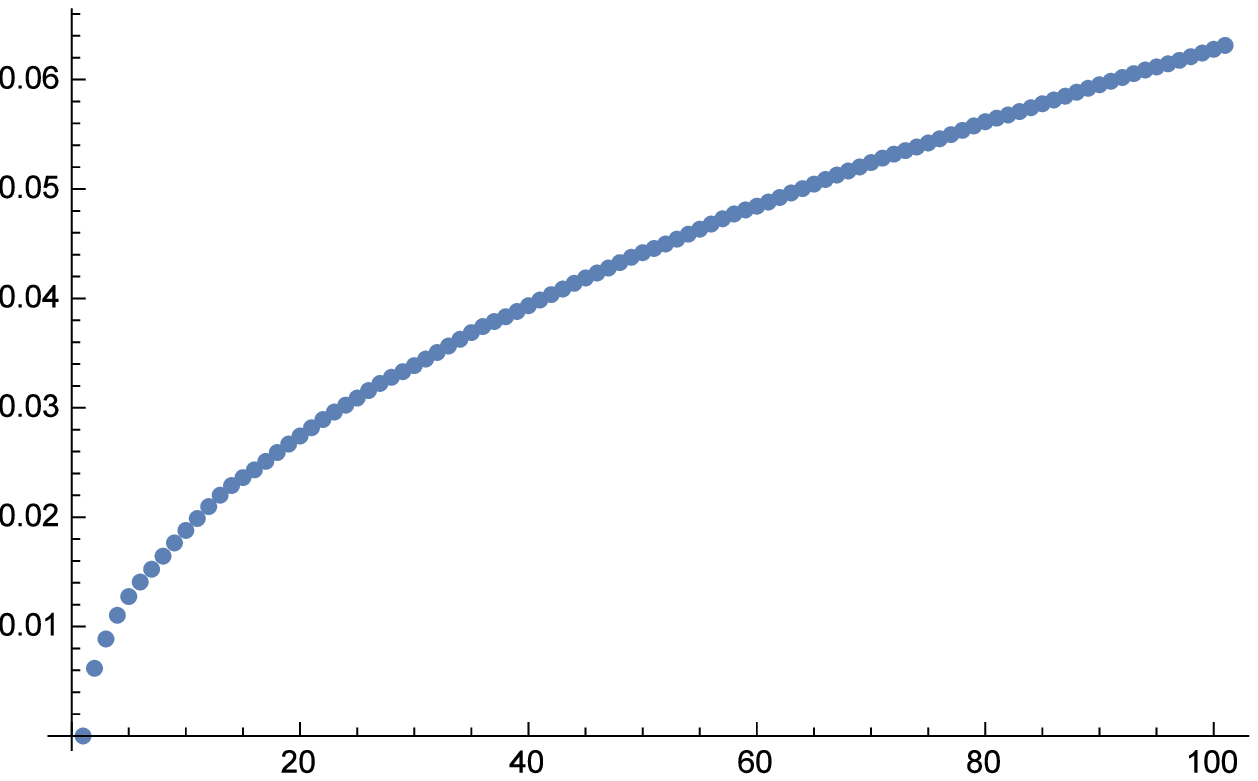}}
\end{subfigure}
\begin{subfigure}[$\kappa=1$,$\omega>M_1-M_3$]{
\includegraphics[scale=0.4]{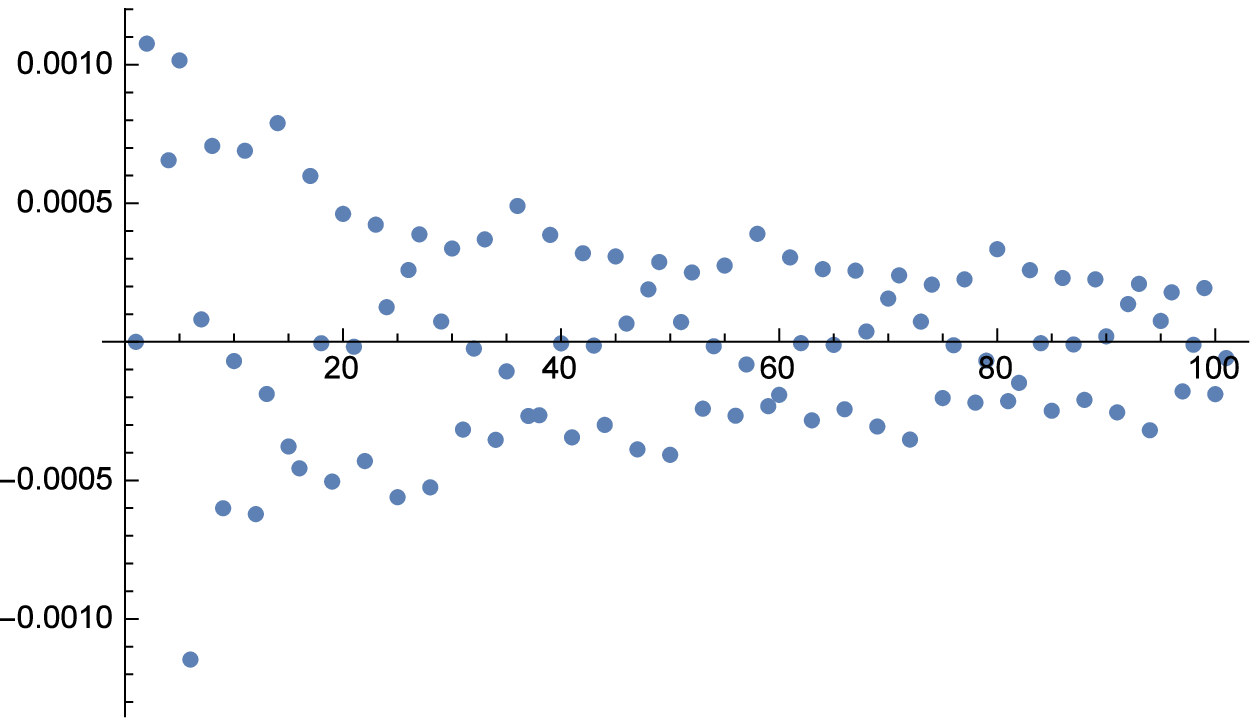}}
\end{subfigure}
\begin{subfigure}[$\kappa=2$,$\omega<M_1-M_3$]{
\includegraphics[scale=0.4]{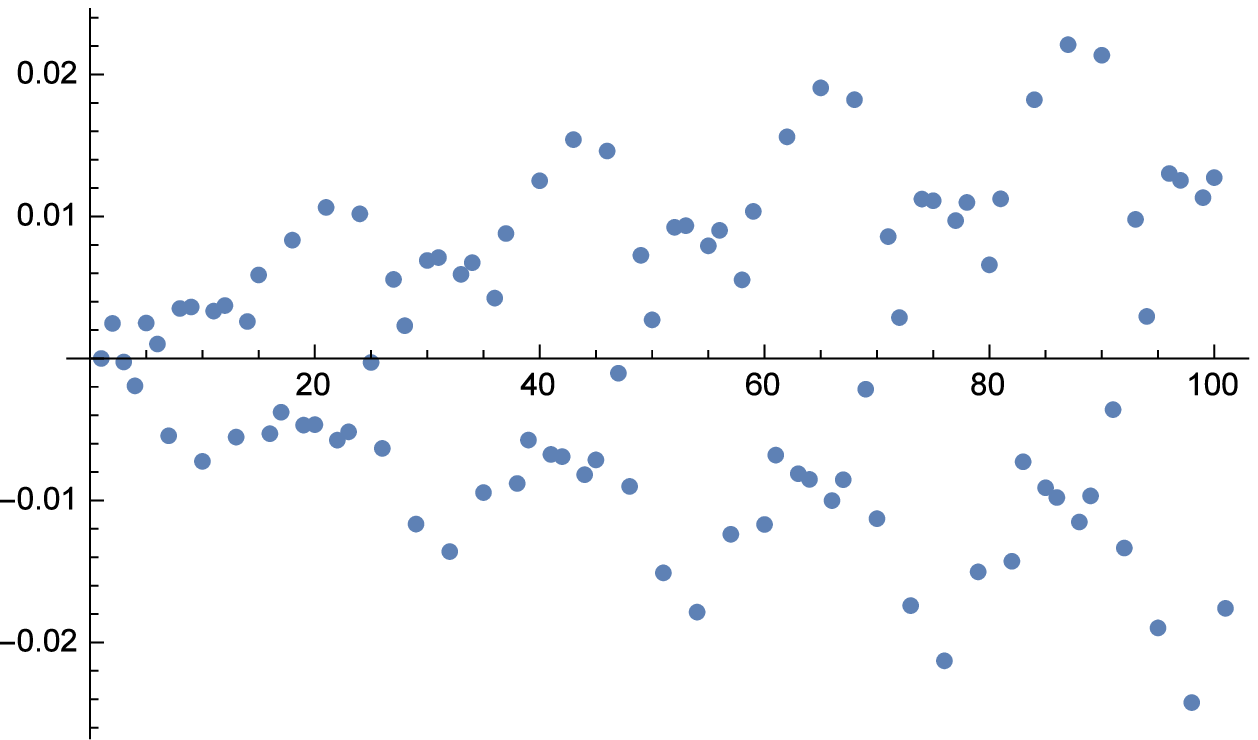}}
\end{subfigure}
\begin{subfigure}[$\kappa=2$,$\omega=M_1-M_3$]{
\includegraphics[scale=0.4]{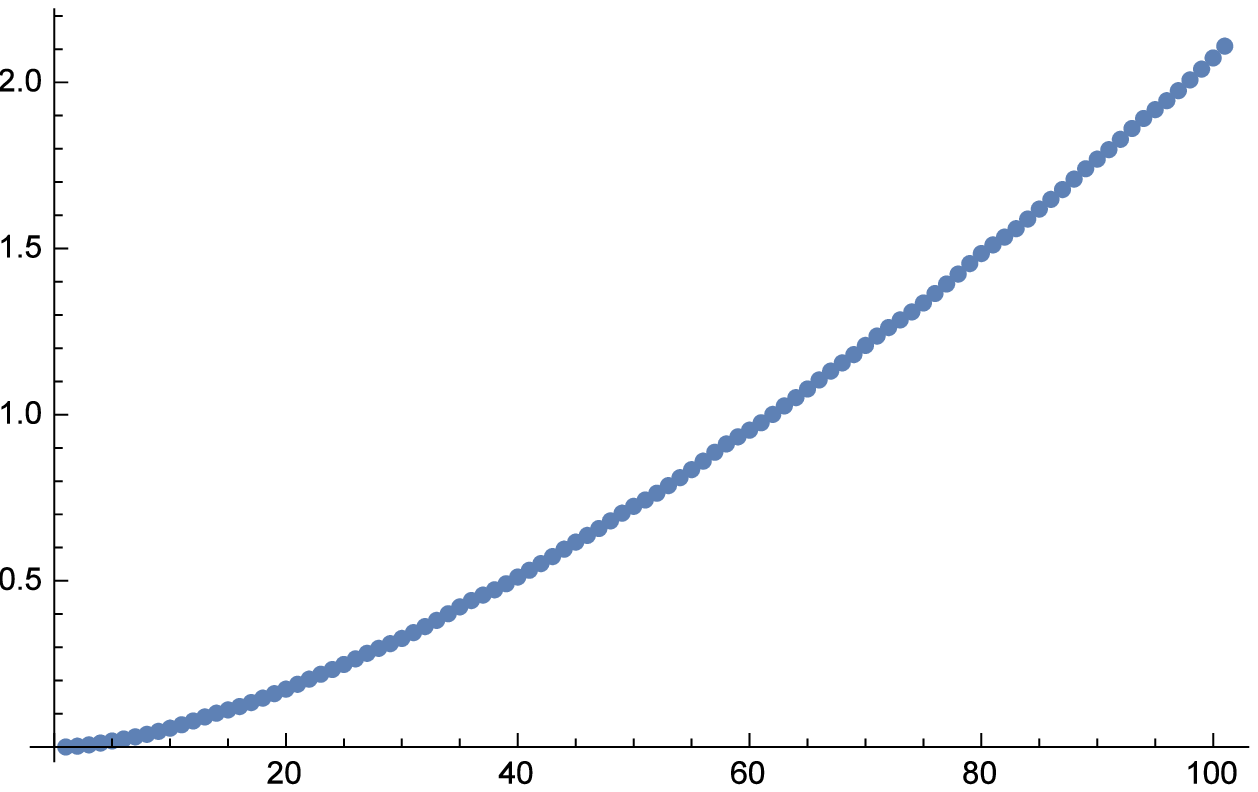}}
\end{subfigure}
\begin{subfigure}[$\kappa=2$,$\omega>M_1-M_3$]{
\includegraphics[scale=0.4]{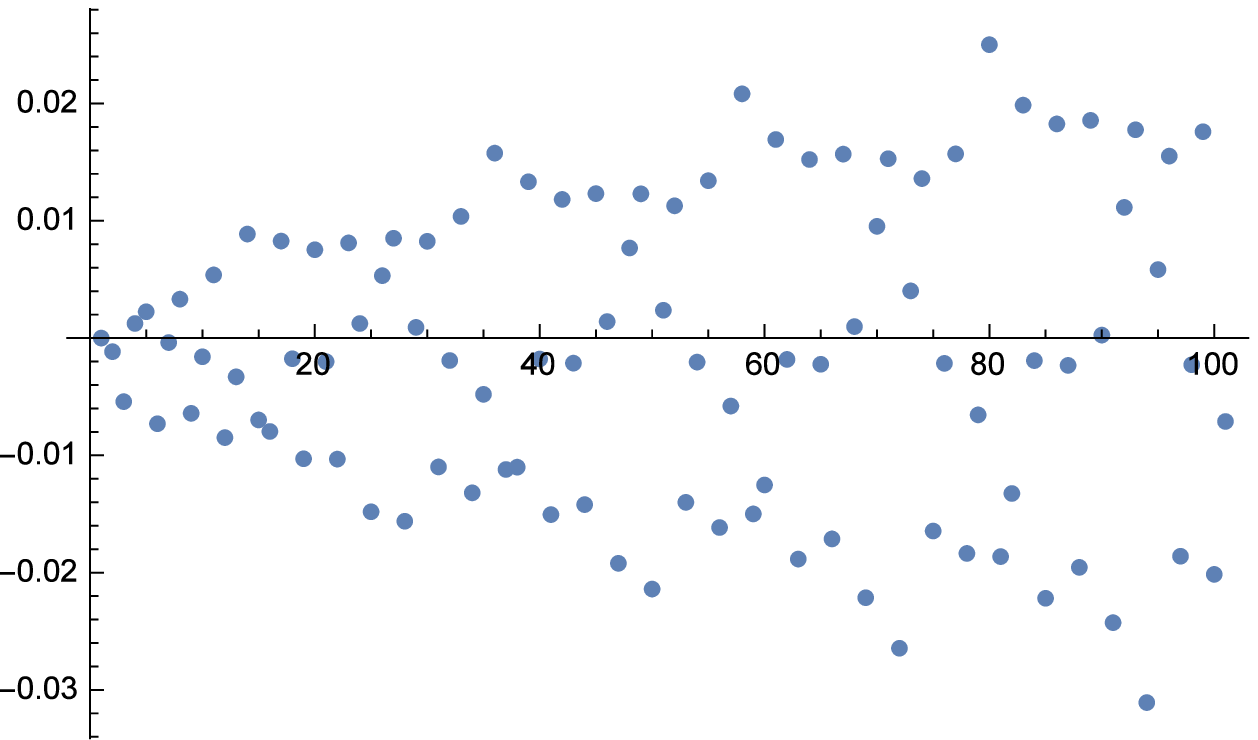}}
\end{subfigure}
\begin{subfigure}[$\kappa=3$,$\omega<M_1-M_3$]{
\includegraphics[scale=0.4]{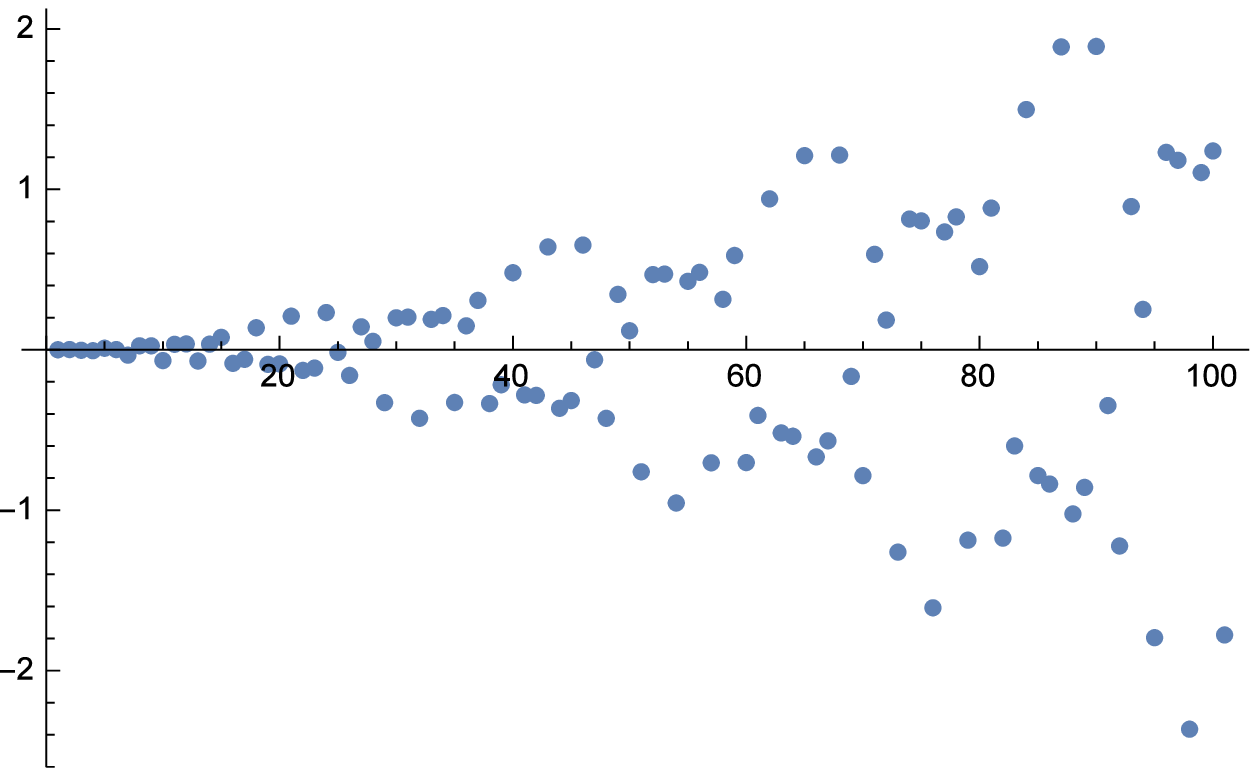}}
\end{subfigure}
\begin{subfigure}[$\kappa=3$,$\omega=M_1-M_3$]{
\includegraphics[scale=0.4]{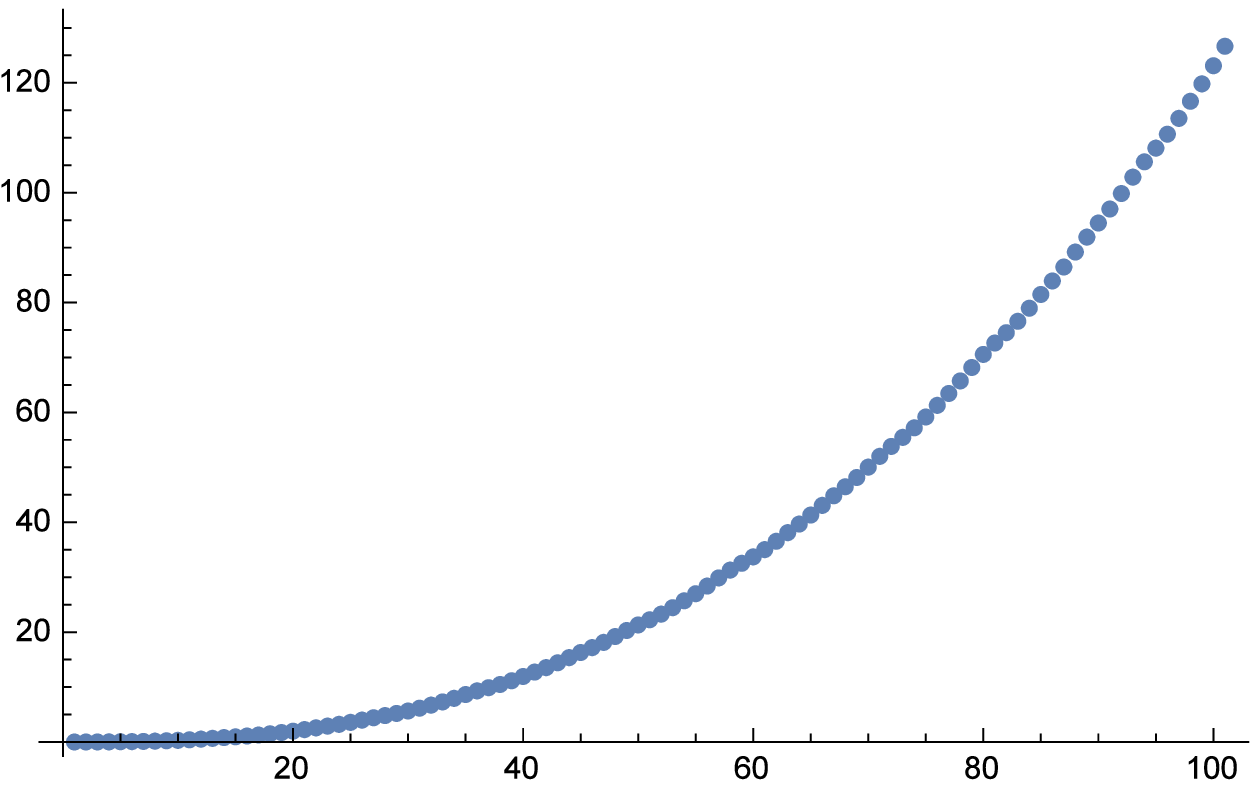}}
\end{subfigure}
\begin{subfigure}[$\kappa=3$,$\omega>M_1-M_3$]{
\includegraphics[scale=0.4]{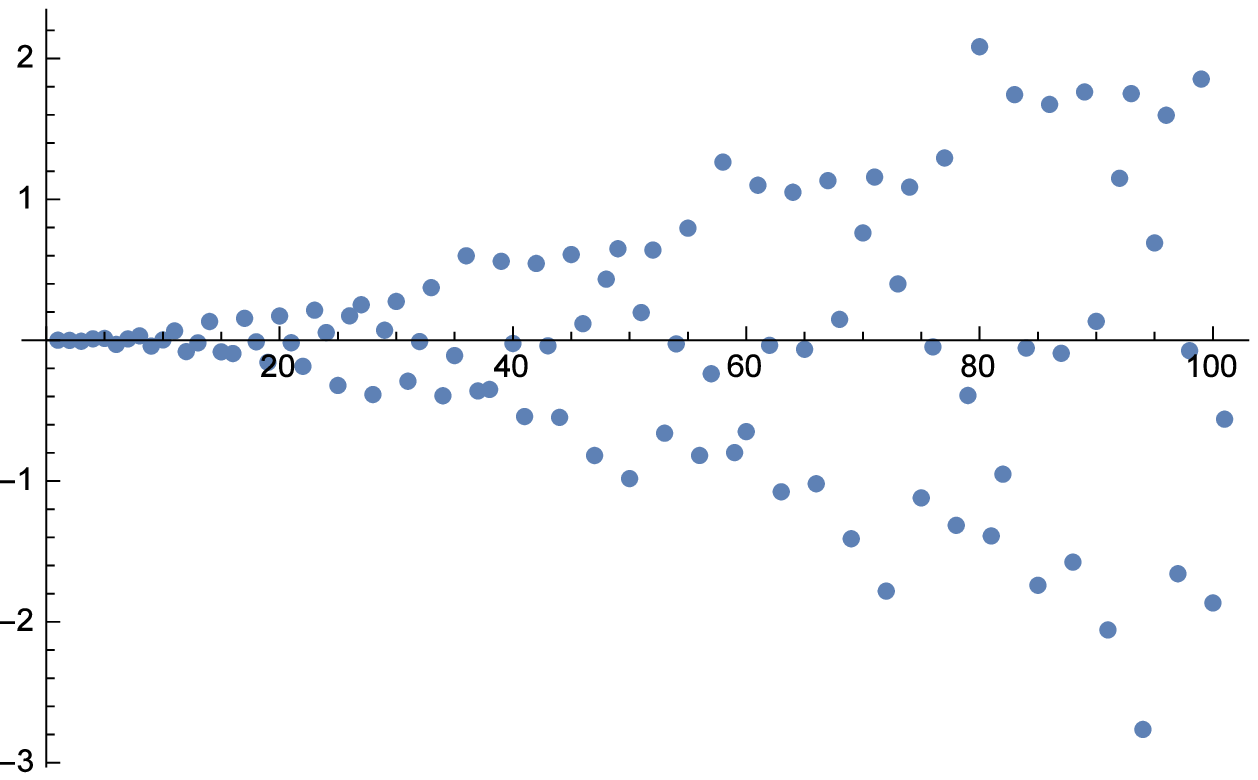}}
\end{subfigure}
\begin{subfigure}[$\kappa=4$,$\omega<M_1-M_3$]{
\includegraphics[scale=0.4]{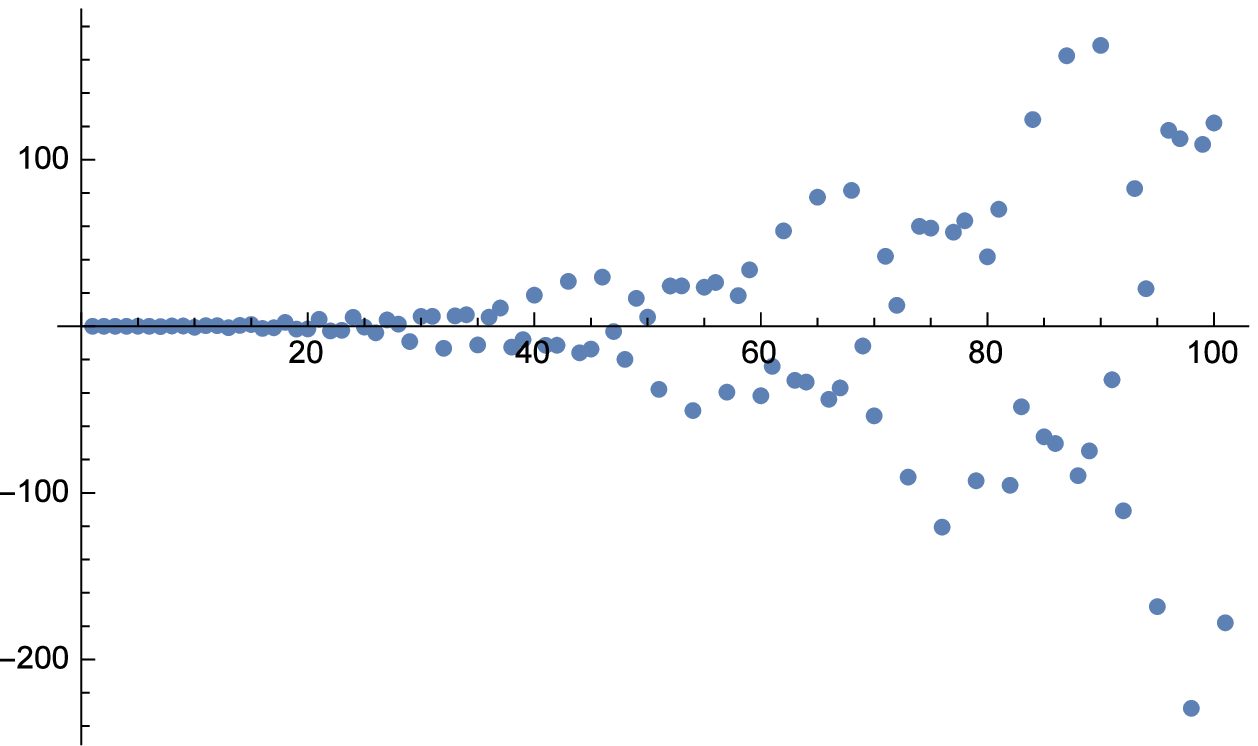}}
\end{subfigure}
\begin{subfigure}[$\kappa=4$,$\omega=M_1-M_3$]{
\includegraphics[scale=0.4]{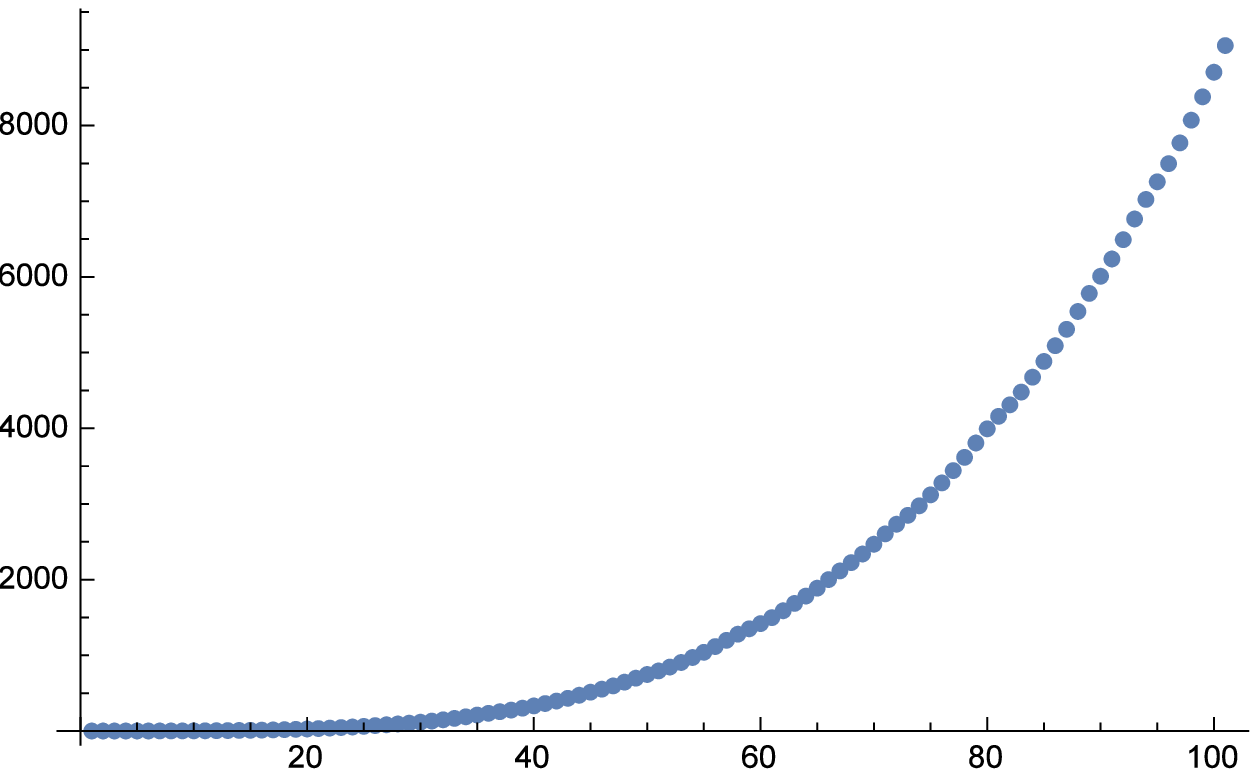}}
\end{subfigure}
\begin{subfigure}[$\kappa=4$,$\omega>M_1-M_3$]{
\includegraphics[scale=0.4]{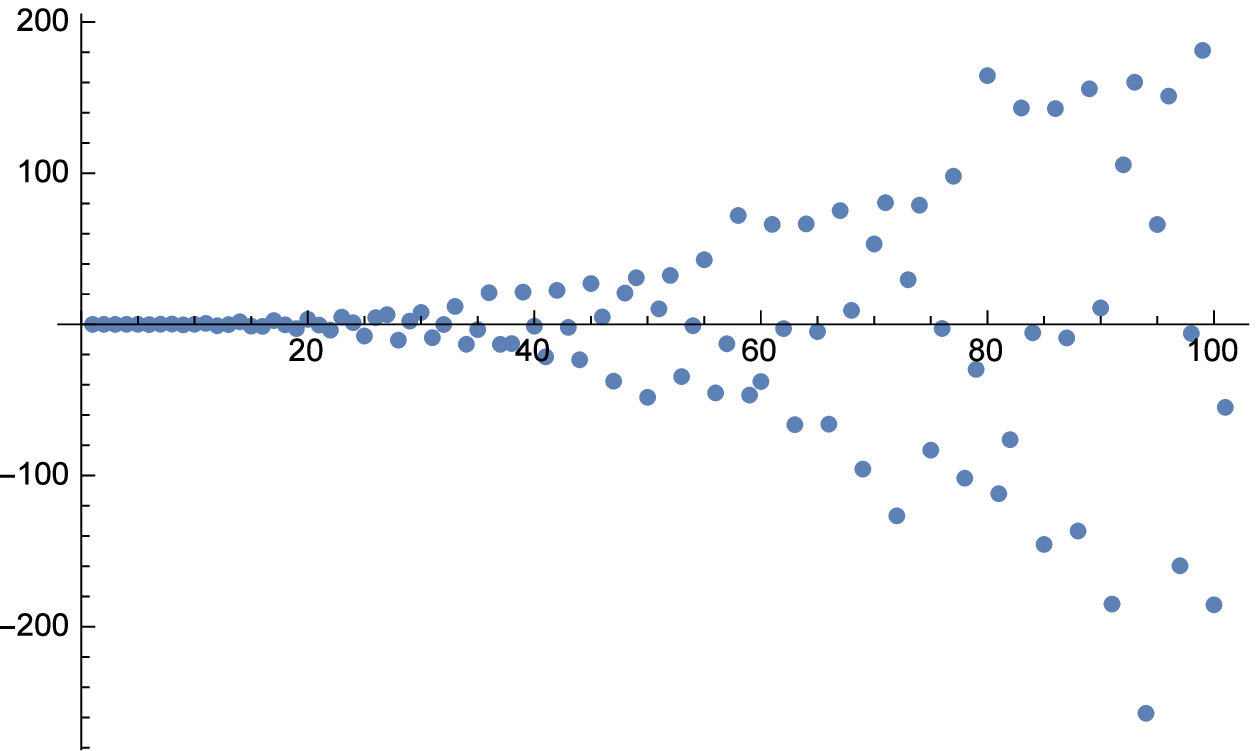}}
\end{subfigure}
\caption{\small Examples of the integrals $I^{(\kappa)}(\Lambda)$
numerically evaluated for $\kappa=1,\dots,4$ and shown as a function
of the upper integration limit $\Lambda^{-1}$. The parameters used
are $M_1=15$, $M_3=6$ and $\omega = 7,9,11$ respectively. } \label{}
\end{figure}

\begin{figure}
\centering
\begin{subfigure}[$\kappa=2$,$\omega<M_1+M_3$]{
\includegraphics[scale=0.4]{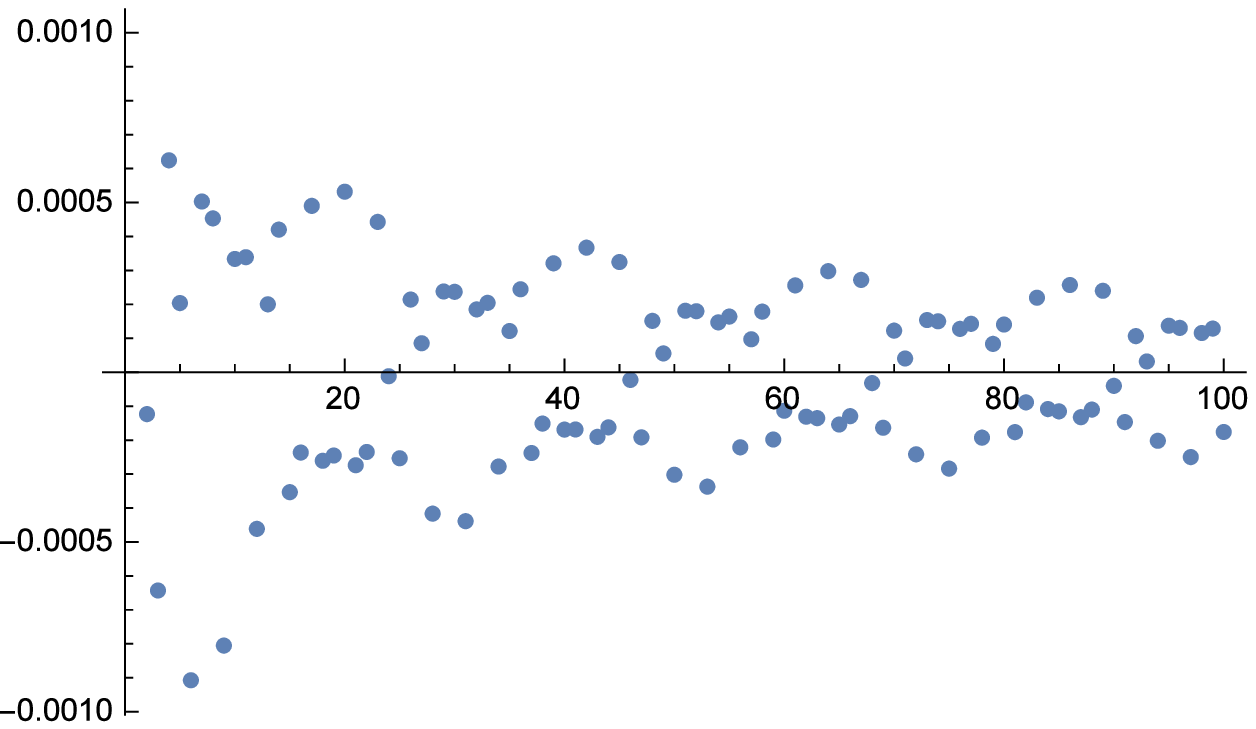}}
\end{subfigure}
\begin{subfigure}[$\kappa=2$,$\omega=M_1+M_3$]{
\includegraphics[scale=0.4]{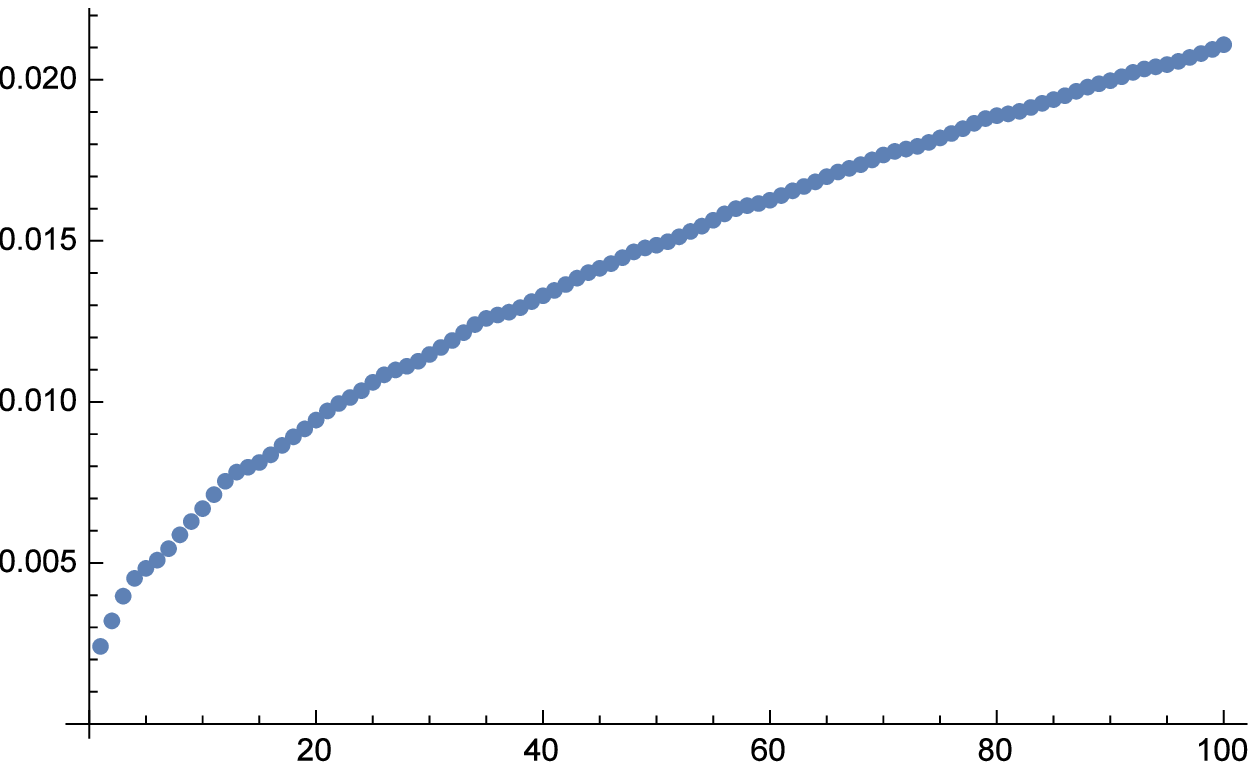}}
\end{subfigure}
\begin{subfigure}[$\kappa=2$,$\omega>M_1+M_3$]{
\includegraphics[scale=0.4]{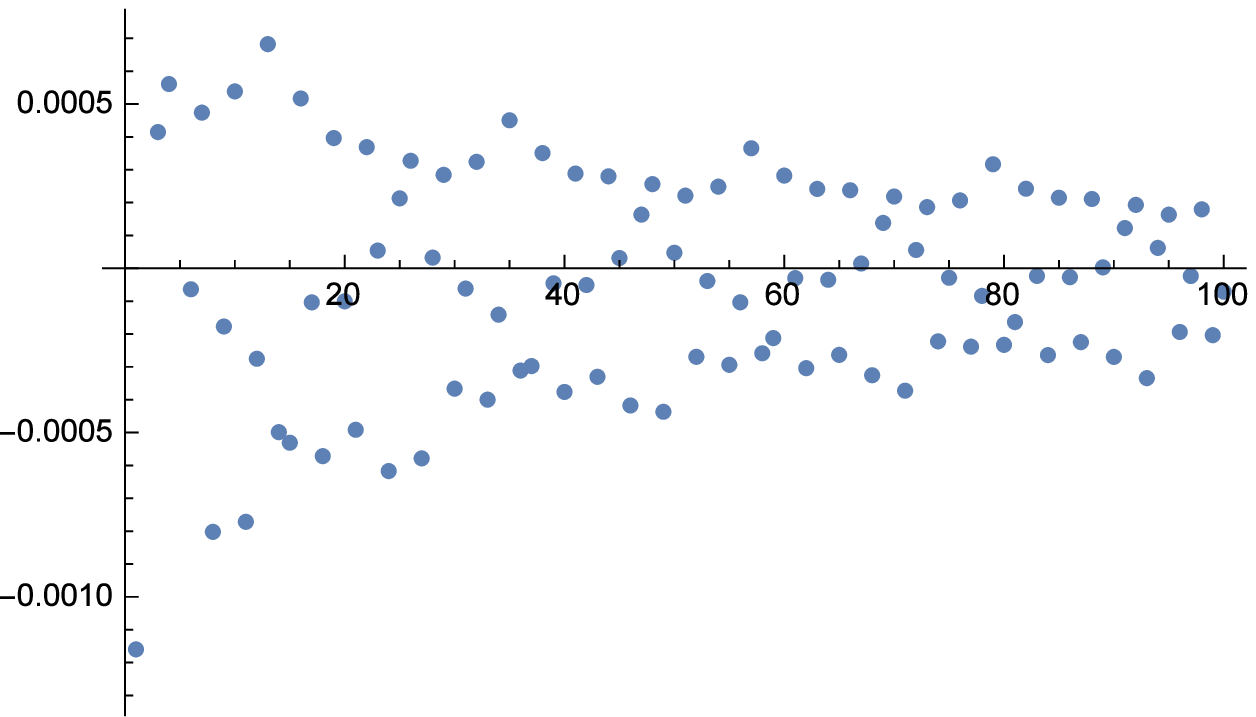}}
\end{subfigure}
\begin{subfigure}[$\kappa=3$,$\omega<M_1+M_3$]{
\includegraphics[scale=0.4]{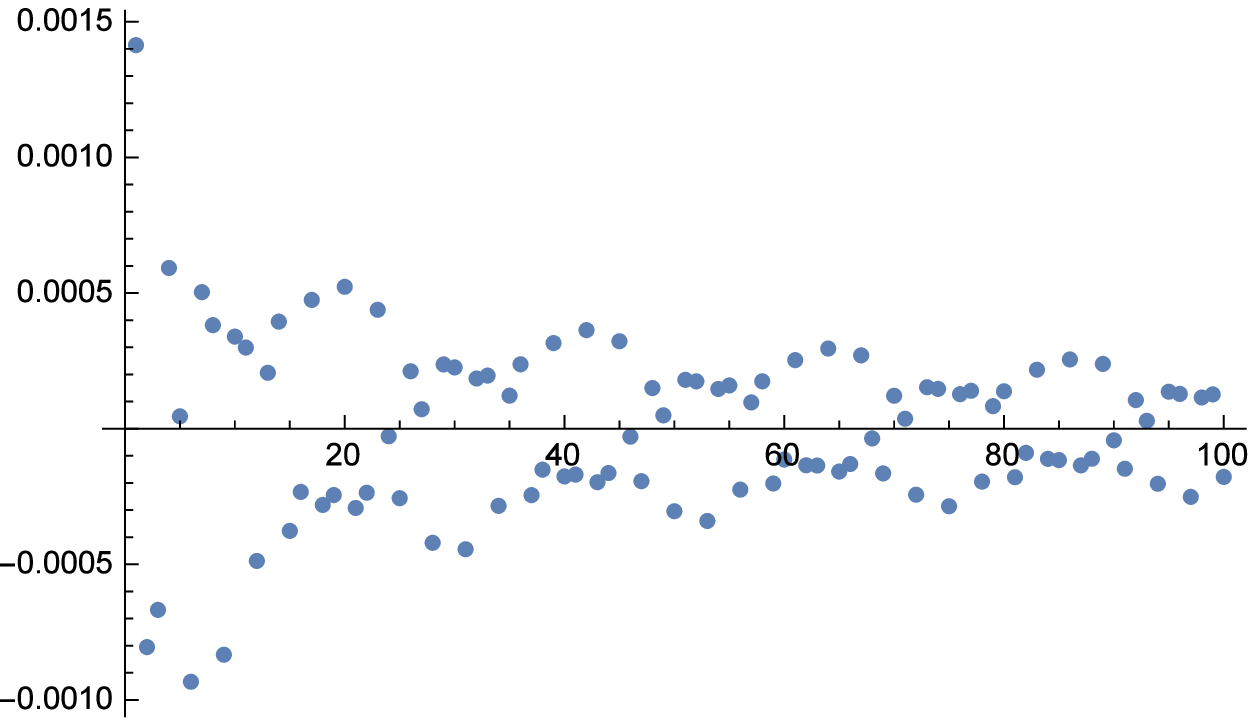}}
\end{subfigure}
\begin{subfigure}[$\kappa=3$,$\omega=M_1+M_3$]{
\includegraphics[scale=0.4]{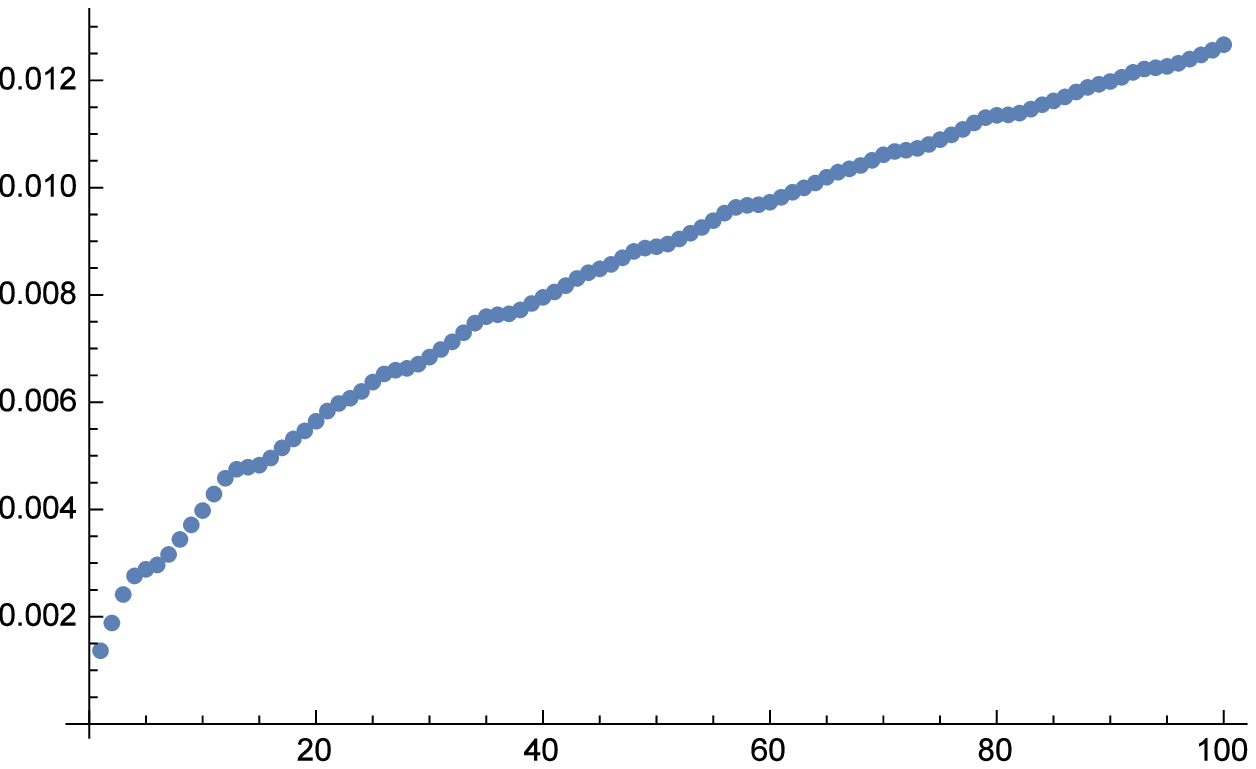}}
\end{subfigure}
\begin{subfigure}[$\kappa=3$,$\omega>M_1+M_3$]{
\includegraphics[scale=0.4]{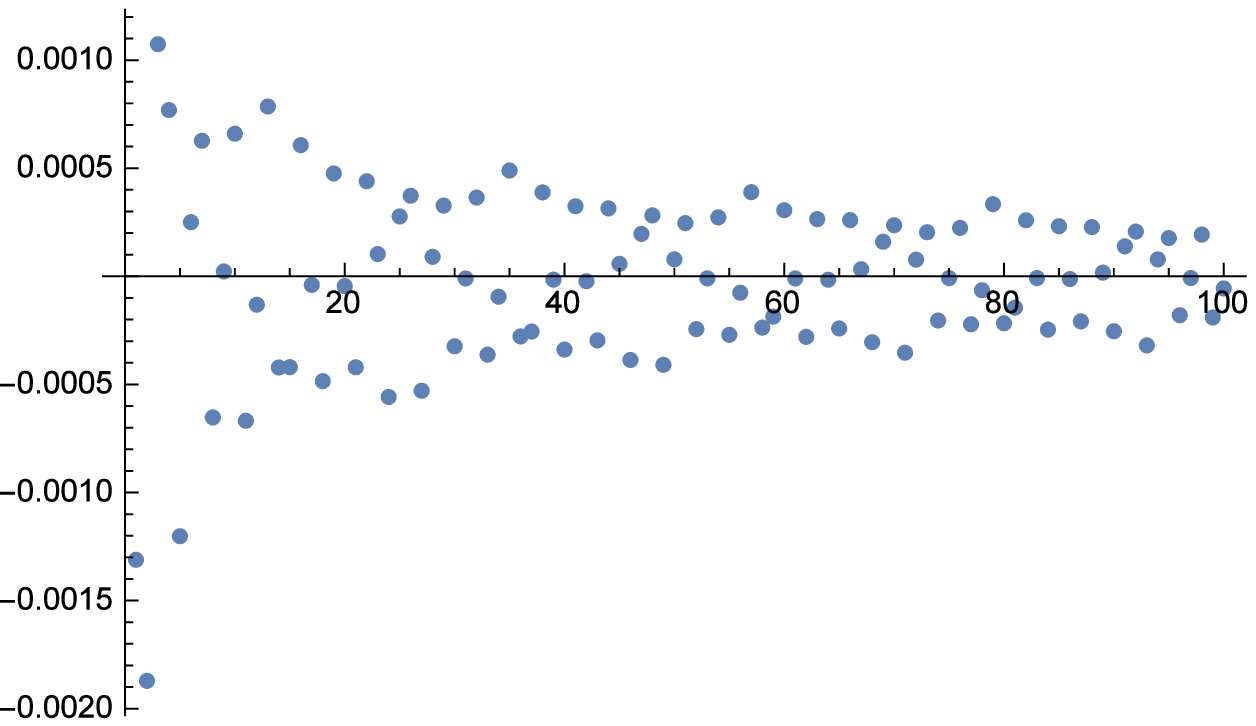}}
\end{subfigure}
\begin{subfigure}[$\kappa=4$,$\omega<M_1+M_3$]{
\includegraphics[scale=0.4]{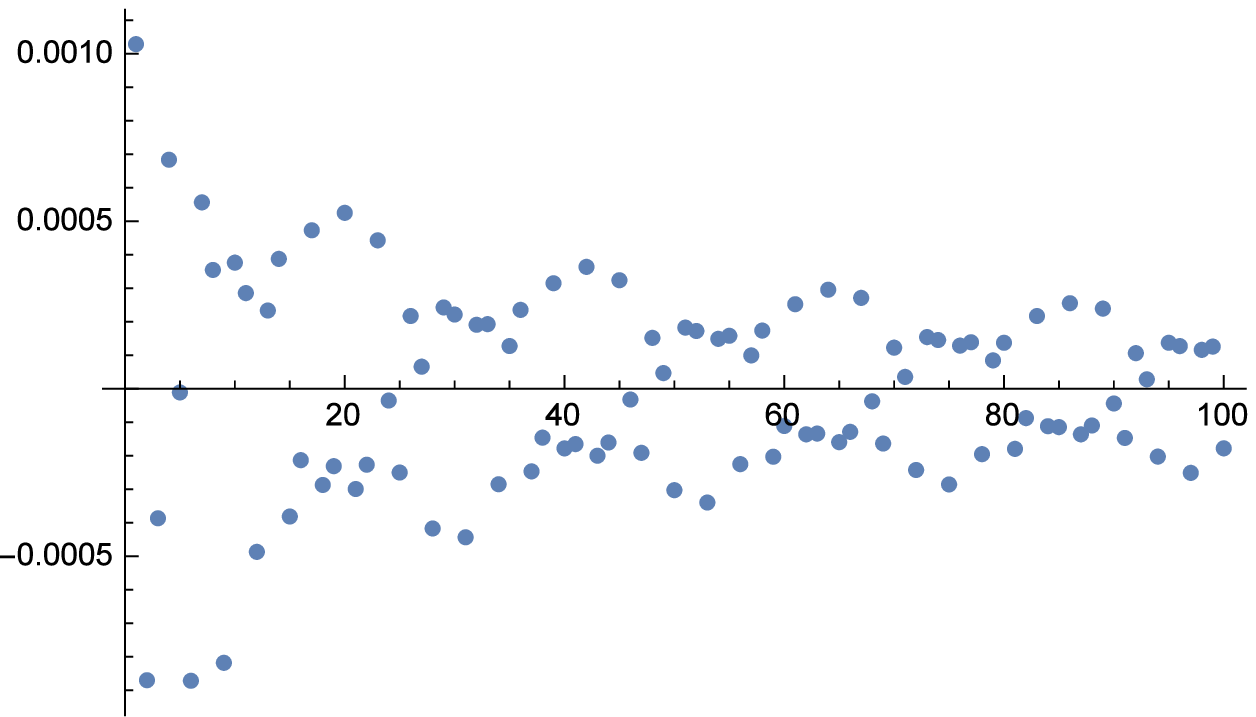}}
\end{subfigure}
\begin{subfigure}[$\kappa=4$,$\omega=M_1+M_3$]{
\includegraphics[scale=0.4]{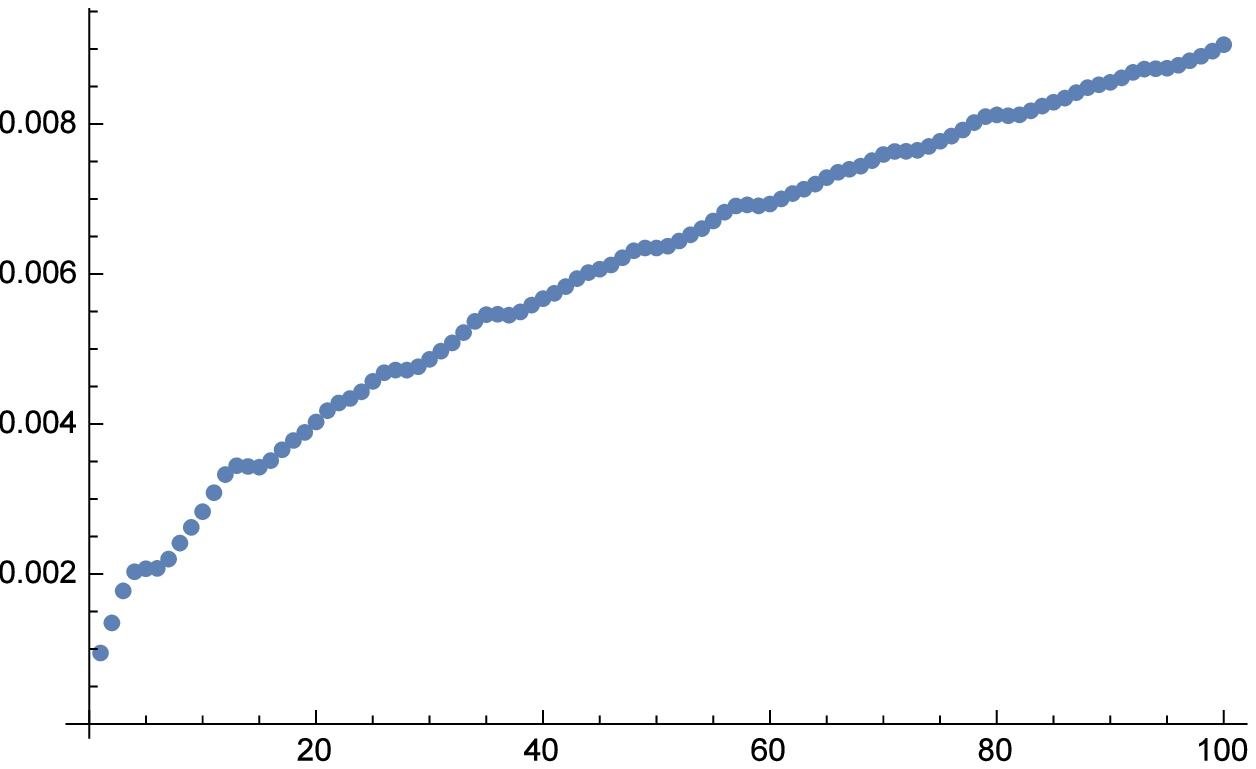}}
\end{subfigure}
\begin{subfigure}[$\kappa=4$,$\omega>M_1+M_3$]{
\includegraphics[scale=0.4]{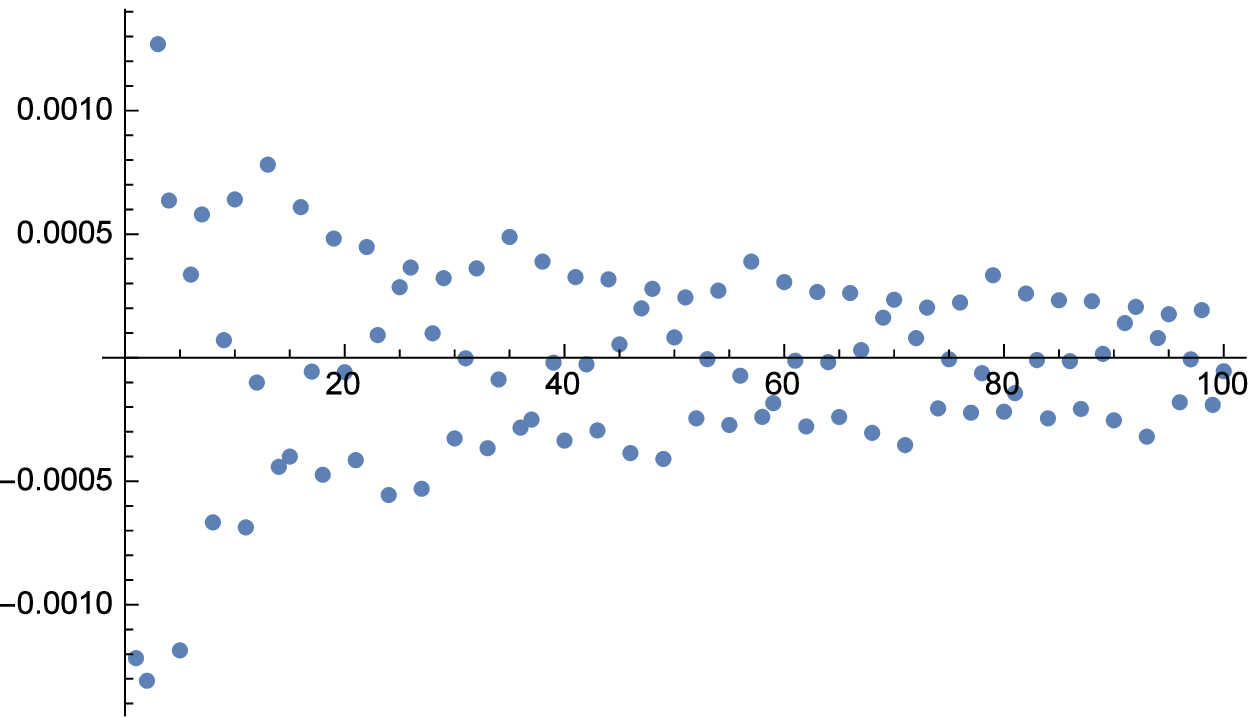}}
\end{subfigure}
\caption{\small The rescaled integrals
$\Lambda^{\kappa-1}I^{(\kappa)}(\Lambda)$ are numerically evaluated
for $\kappa=2,\dots,4$ and shown as a function of the upper
integration limit $\Lambda^{-1}$. The parameters used are $M_1=15$,
$M_3=6$ and $\omega = 7,9,11$ respectively. The different behaviors
depicted in the previous figure disappear (up to an order $1$
constant) and the results for each $\omega$ are similar for all
integrals. One can see that in all cases the integral decreases (or
gives a small constant) for all values of $\omega \neq
\omega_c=M_1-M_3$.} \label{}
\end{figure}

\begin{figure}
\centering
\begin{subfigure}[$\kappa=1,\omega=8.5$]{
\includegraphics[scale=0.23]{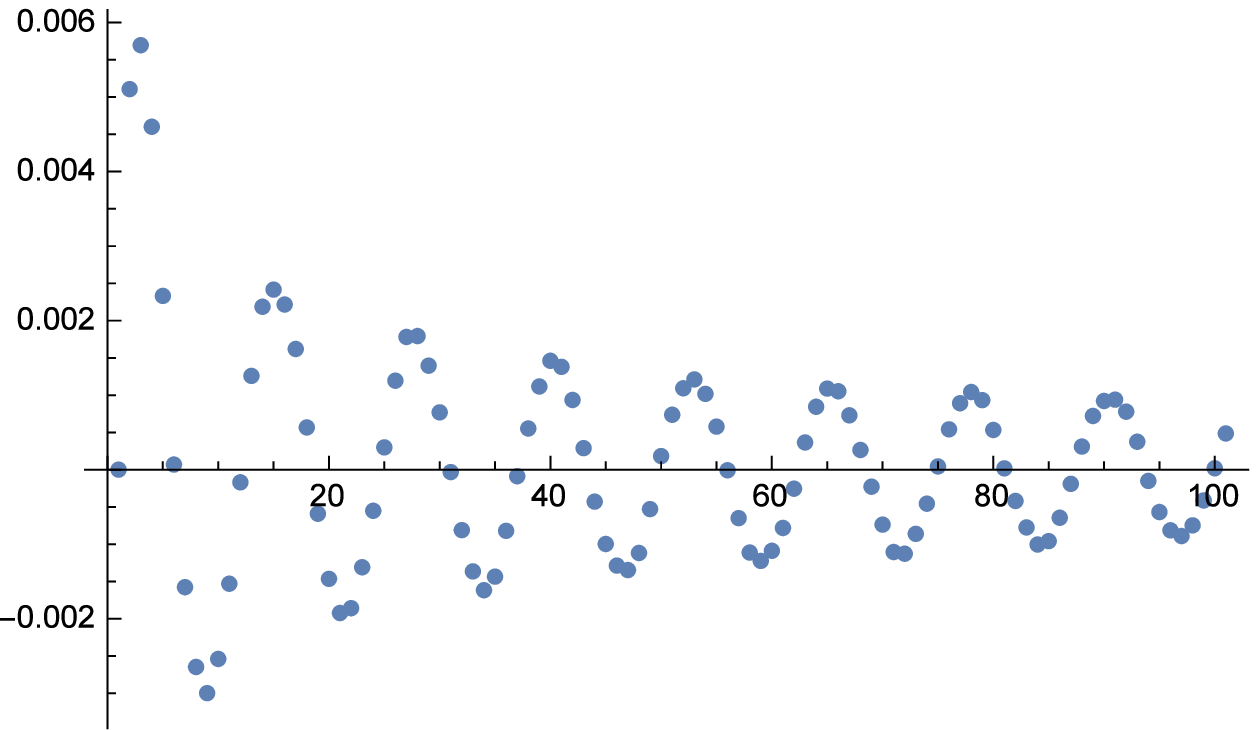}}
\end{subfigure}
\begin{subfigure}[$\kappa=1,\omega=8.75$]{
\includegraphics[scale=0.23]{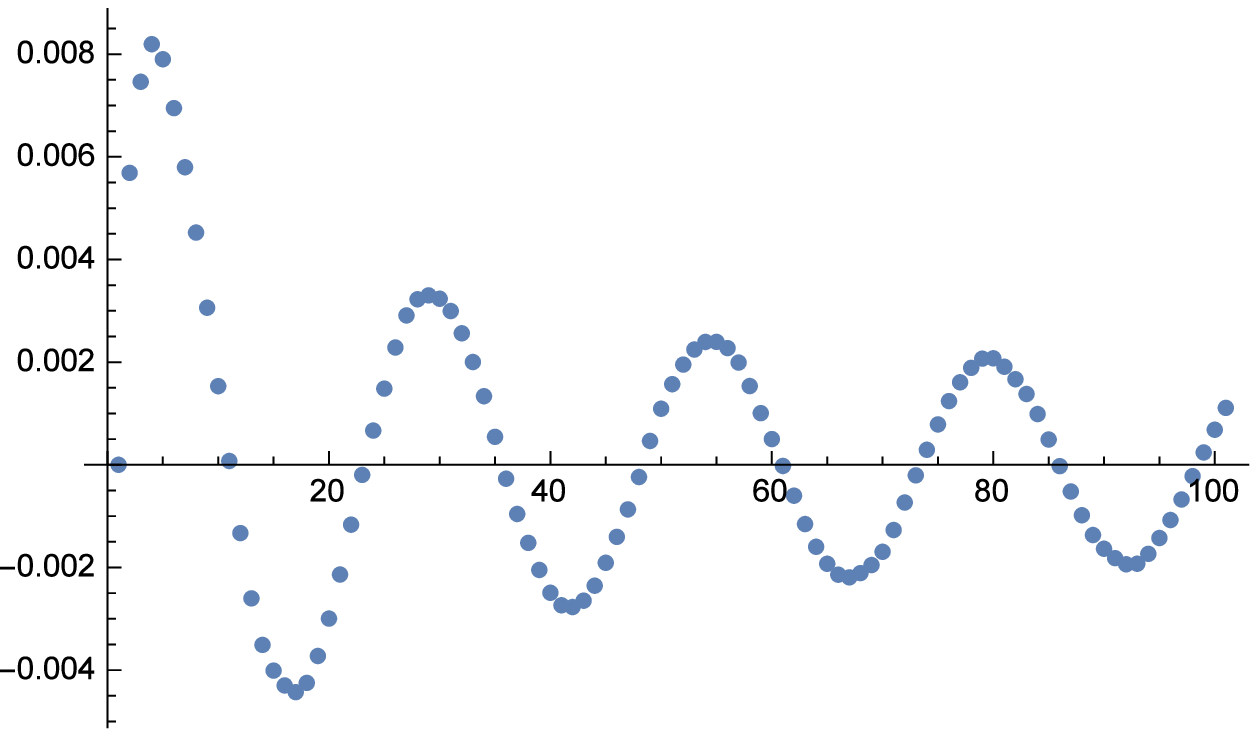}}
\end{subfigure}
\begin{subfigure}[$\kappa=1,\omega=9$]{
\includegraphics[scale=0.23]{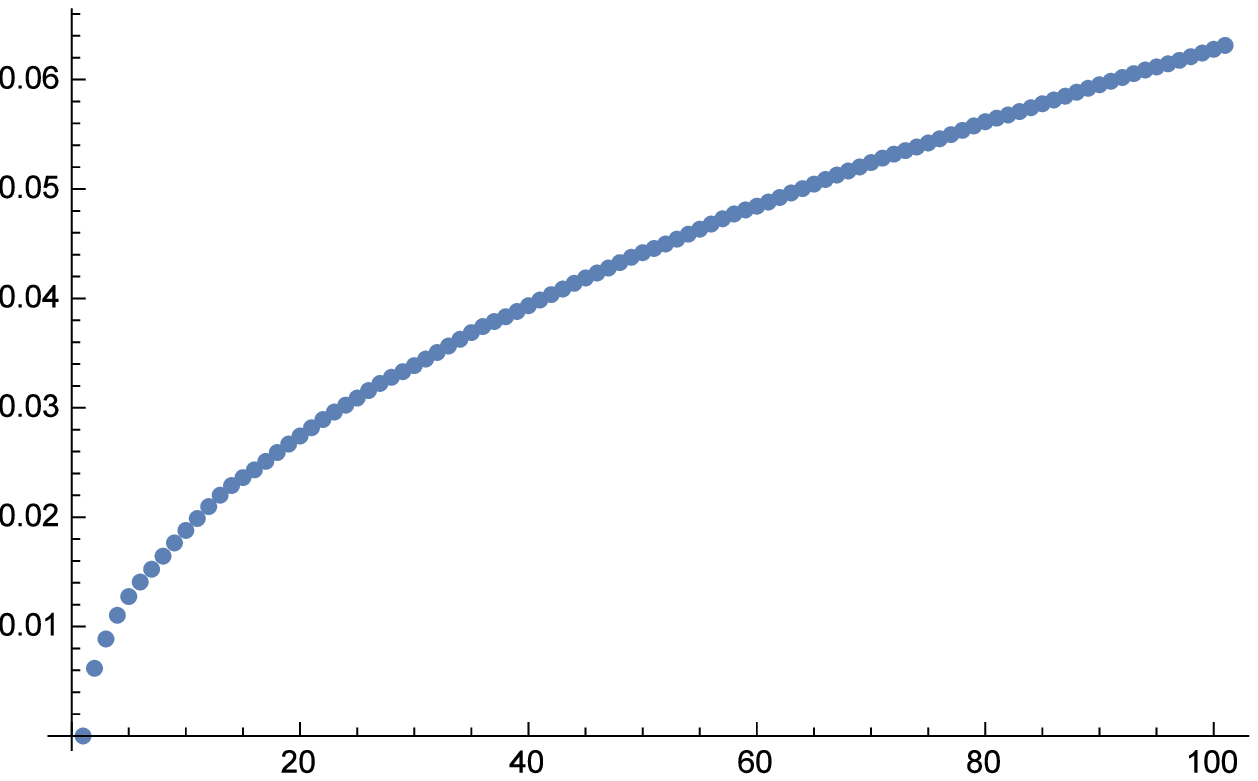}}
\end{subfigure}
\begin{subfigure}[$\kappa=1,\omega=9.25$]{
\includegraphics[scale=0.23]{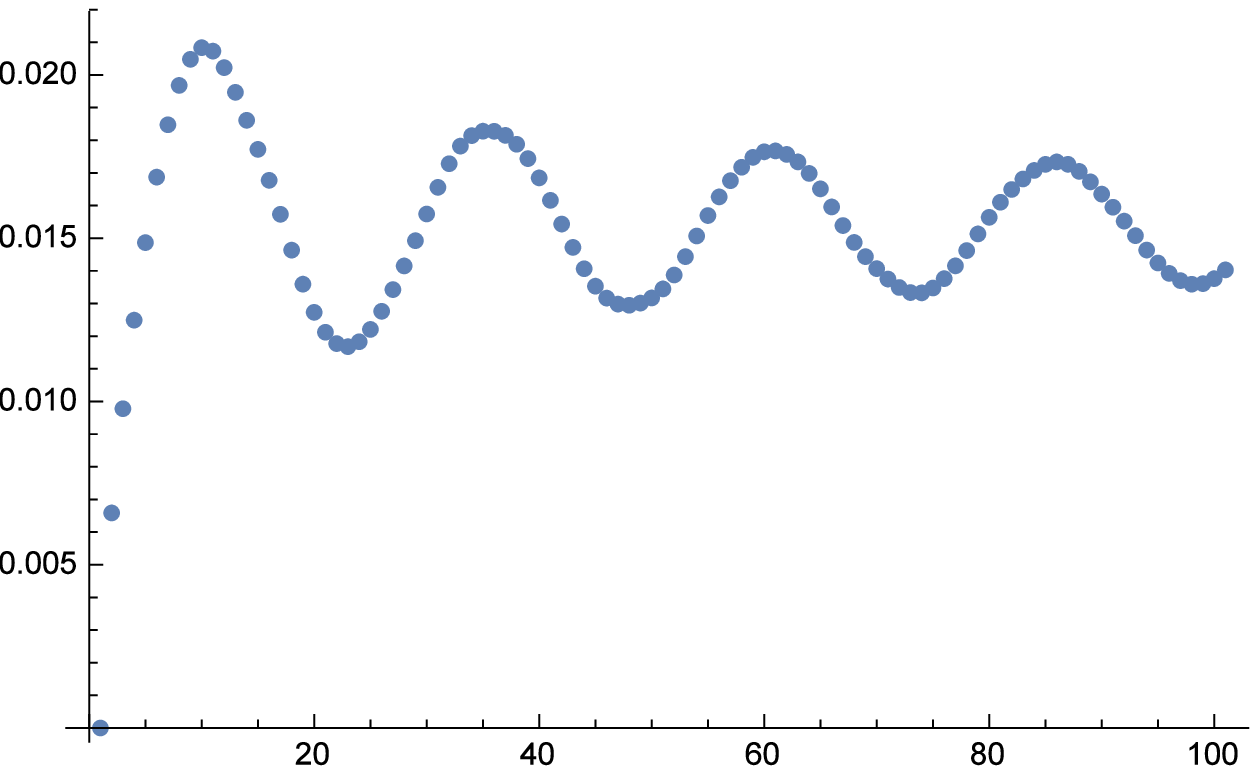}}
\end{subfigure}
\begin{subfigure}[$\kappa=1,\omega=9.5$]{
\includegraphics[scale=0.23]{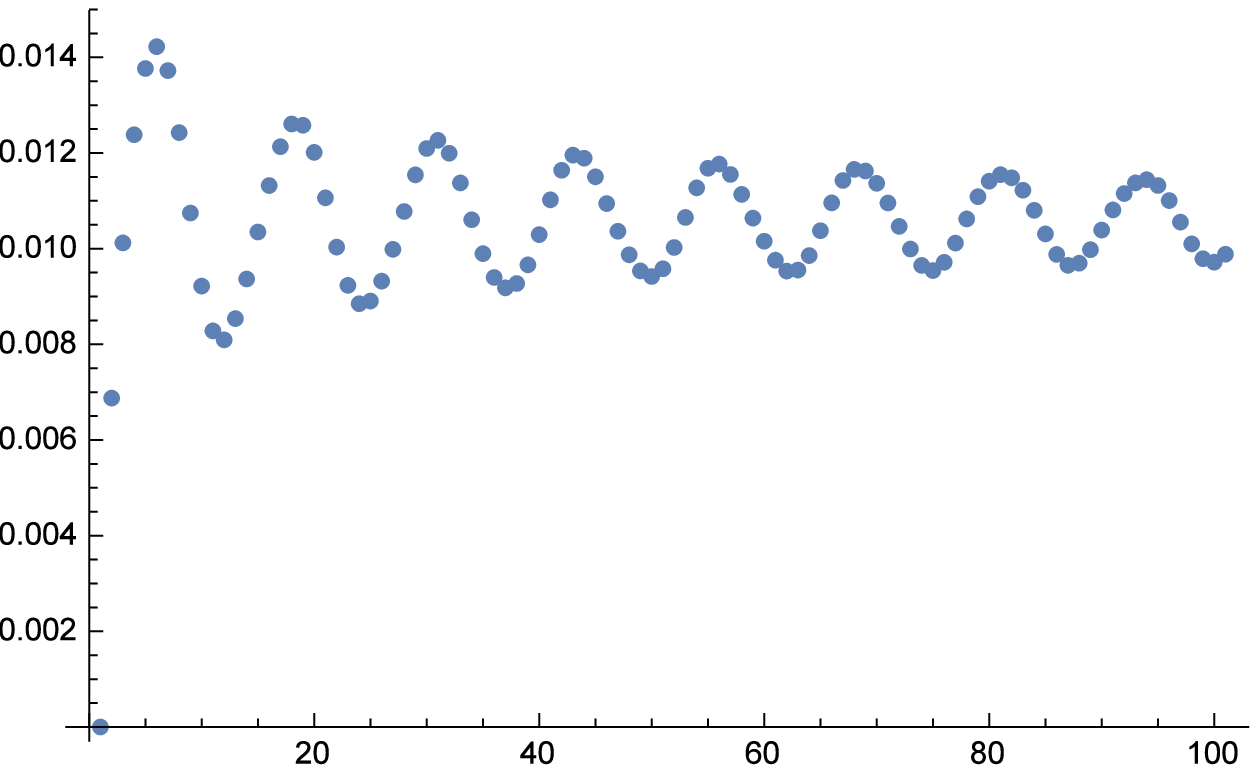}}
\end{subfigure}
\begin{subfigure}[$\kappa=4,\omega=8.5$]{
\includegraphics[scale=0.23]{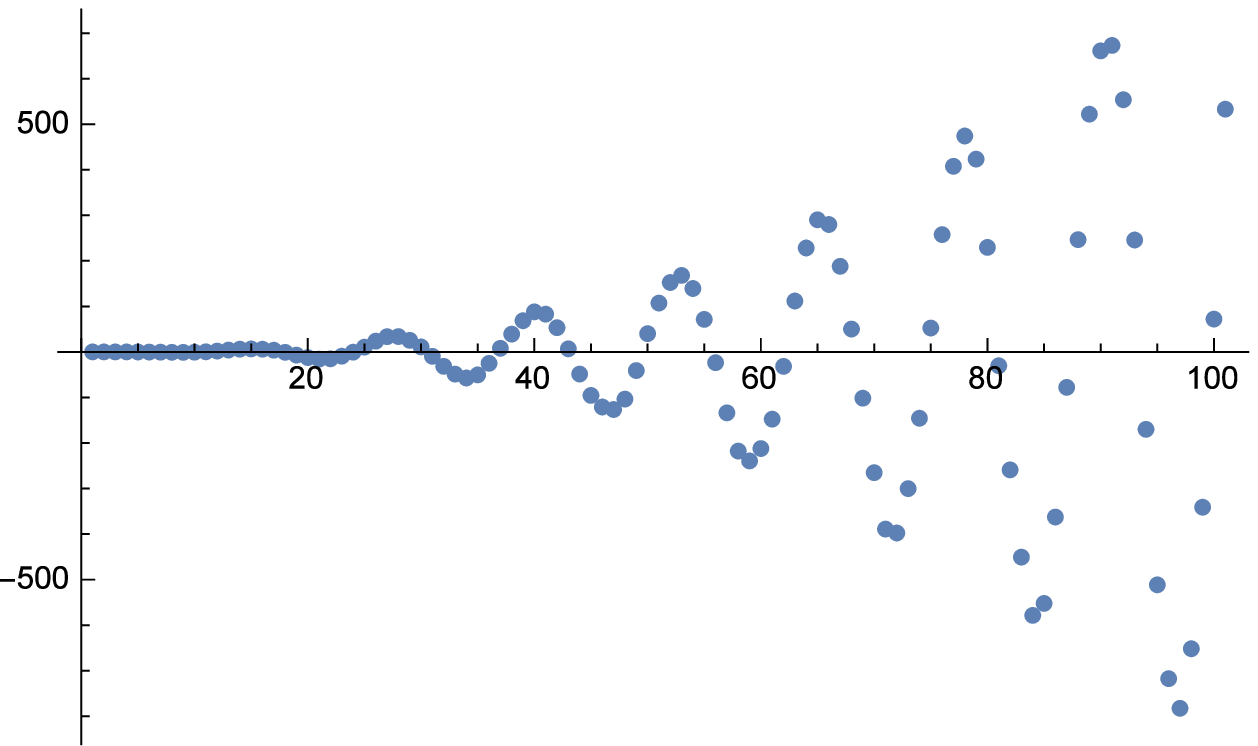}}
\end{subfigure}
\begin{subfigure}[$\kappa=4,\omega=8.25$]{
\includegraphics[scale=0.23]{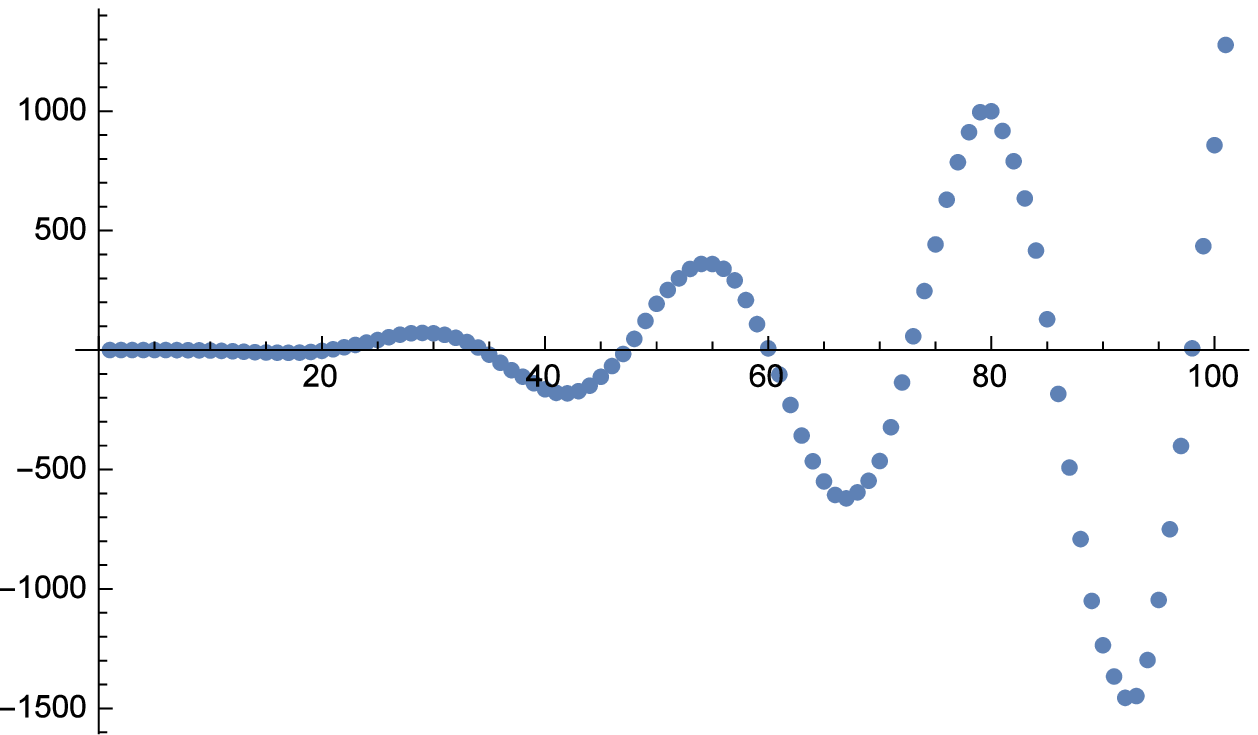}}
\end{subfigure}
\begin{subfigure}[$\kappa=4,\omega=9$]{
\includegraphics[scale=0.23]{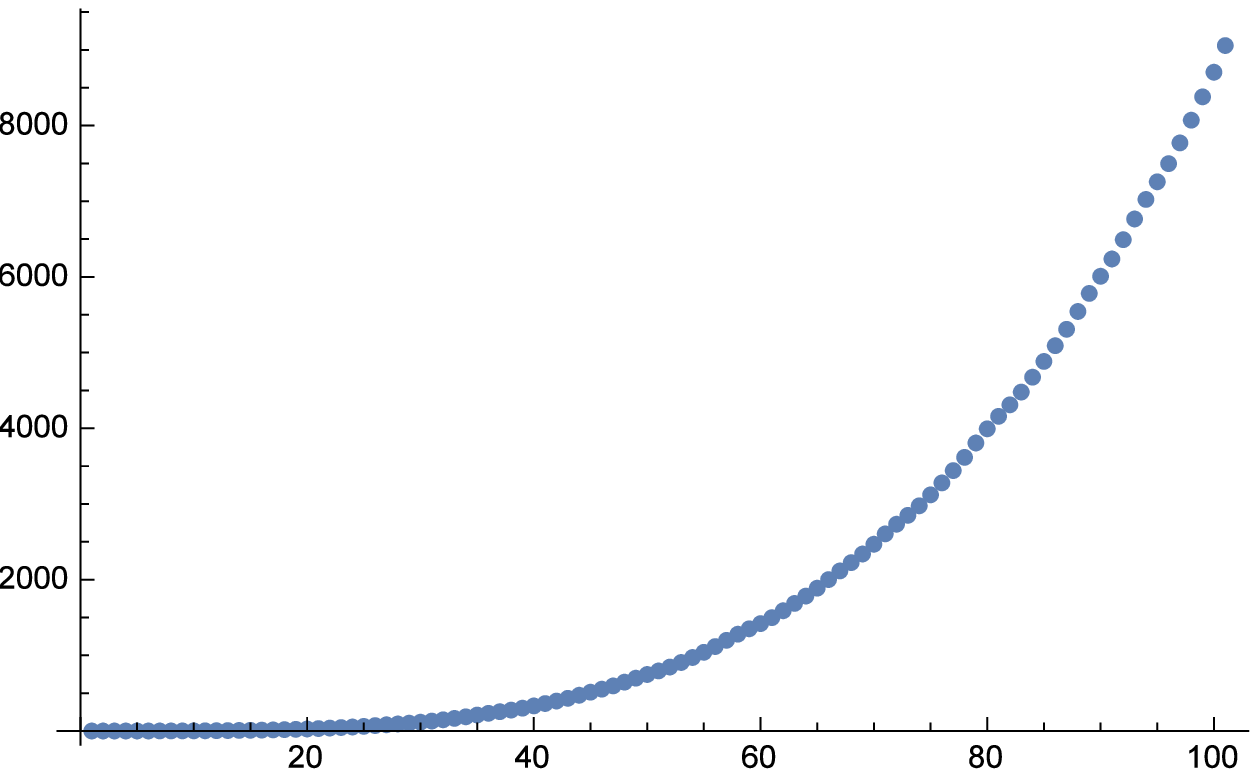}}
\end{subfigure}
\begin{subfigure}[$\kappa=4,\omega=9.25$]{
\includegraphics[scale=0.23]{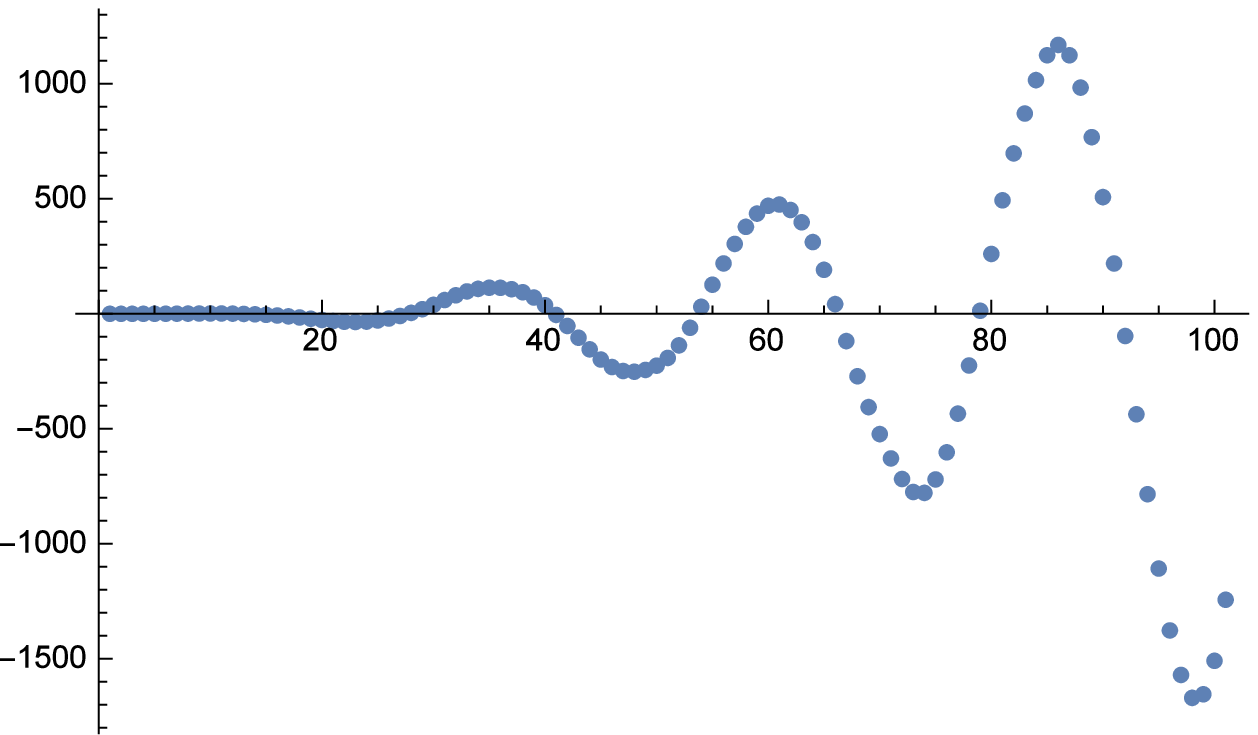}}
\end{subfigure}
\begin{subfigure}[$\kappa=4,\omega=9.5$]{
\includegraphics[scale=0.23]{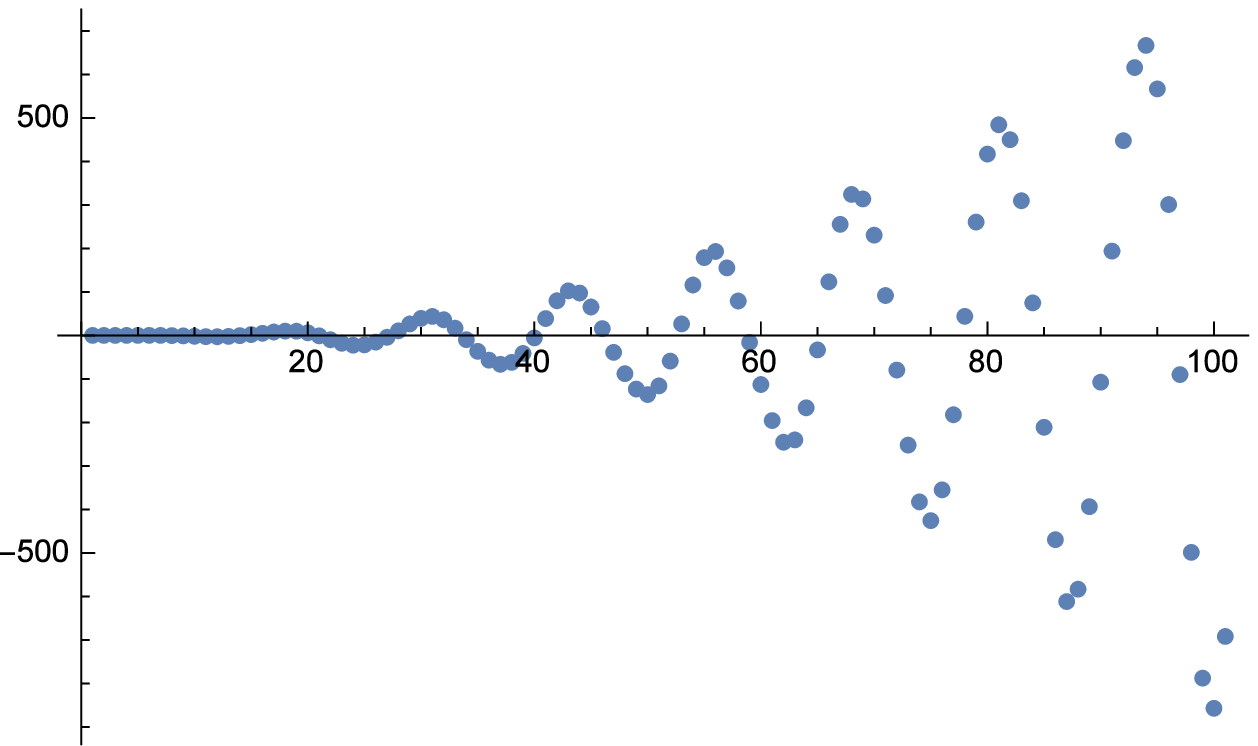}}
\end{subfigure}
\caption{\small The behavior of integrals $I^{(\kappa)}(\Lambda)$
with $\kappa=1$ and $\kappa=4$ near the peak is shown using diagrams
similar to the ones in the previous figures but with
$\omega=8.5, 8.75,9,9.25$ and
$9.5$ in both cases.} \label{}
\end{figure}

In fact, one can see directly this from (\ref{cosenocubo}) by
performing first an indefinite integral in $z$ for the four different cases,
obtaining
\begin{eqnarray}
\int dz \, z^{-1/2} \cos(P z + b) &=&
\left(\frac{2\pi}{P}\right)^{1/2}
\left[\cos(b) C(\sqrt{2\pi^{-1}P z}) - \sin(b) S(\sqrt{2\pi^{-1}P z})\right] \nn \\
\int dz \, z^{1/2} \cos(P z + b) &=& \left(\frac{2\pi}{4P^3}\right)^{1/2}
\left[-\cos(b) S(\sqrt{2\pi^{-1}P z}) - \sin(b) C(\sqrt{2\pi^{-1}P z})\right. \nn \\
&& \left. +2\sqrt{P z} \sin(P z +b) \right] \nn \\
\int dz \, z^{3/2} \cos(P z + b) &=& \frac{1}{4}\left(\frac{2\pi}{P^5}\right)^{1/2}
\left[-3\cos(b) C(\sqrt{2\pi^{-1}P z}) +3 \sin(b) S(\sqrt{2\pi^{-1}P z})\right.\nn \\
&& \left. +2\sqrt{P z} \left(3\cos(P z +b) + 2Pz\sin(Pz+b)\right) \right] \nn \\
\int dz \, z^{5/2} \cos(P z + b) &=& \frac{1}{8}\left(\frac{2\pi}{P^7}\right)^{1/2}
\left[ 15\cos(b) S(\sqrt{2\pi^{-1}P z}) + 15 \sin(b) C(\sqrt{2\pi^{-1}P z})\right. \nn \\
&& \left. +2\sqrt{P z} \left(10 P z\cos(P z +b) +
(4P^2z^2-15)\sin(Pz+b)\right) \right] \nn
\end{eqnarray}
where $P$ stands for the deviation from $\omega = M_1 \pm M_3$ and
$b$ represents some phase, while $S(x)$ and $C(x)$ are the Fresnel
sine and cosine functions, respectively. Now, evaluating these results between
$z=0$ and $z=\Lambda^{-1}$ and expanding around the peak, i.e.,
around $P=0$ one finds that, at least in this parametric region, the
different integrals are related as stated in our approximation.

The reason for this behavior is that as the integrands of the
$I^{(n)}$ integrals grow with $z$ up to $z_{max}$, the most
important contribution comes from the region $z\sim z_{max} =
\Lambda^{-1}$. This was already noticed in the study of the
normalization of the wave function of an incoming glueball in
\cite{Polchinski:2002jw}. Scaling arguments under the change
$(a,b,c) \rightarrow (ka, kb, kc)$ also agree with this analysis.
Therefore, the approximation we use is given by equations
(\ref{aproxIn}) and (\ref{aproxI4}).

~


\newpage

\begin{thebibliography}{99}


\bibitem{Kruczenski:2003be}
  M.~Kruczenski, D.~Mateos, R.~C.~Myers, D.~J.~Winters,
  ``Meson spectroscopy in AdS / CFT with flavor,''
  JHEP {\bf 0307 } (2003)  049.
  [arXiv:hep-th/0304032 [hep-th]].

\bibitem{Kruczenski:2003uq}
  M.~Kruczenski, D.~Mateos, R.~C.~Myers and D.~J.~Winters,
  ``Towards a holographic dual of large N(c) QCD,''
  JHEP {\bf 0405} (2004) 041
  [hep-th/0311270].

\bibitem{Sakai:2004cn}
  T.~Sakai, S.~Sugimoto,
  ``Low energy hadron physics in holographic QCD,''
  Prog.\ Theor.\ Phys.\  {\bf 113 } (2005)  843-882.
  [arXiv:hep-th/0412141 [hep-th]].


\bibitem{Koile:2011aa}
  E.~Koile, S.~Macaluso and M.~Schvellinger,
  ``Deep Inelastic Scattering from Holographic Spin-One Hadrons,''
  JHEP {\bf 1202}, 103 (2012)
  [arXiv:1112.1459 [hep-th]].

\bibitem{Koile:2013hba}
  E.~Koile, S.~Macaluso and M.~Schvellinger,
  ``Deep inelastic scattering structure functions of holographic spin-1 hadrons with $N_f \geq 1$,''
  JHEP {\bf 1401}, 166 (2014)
  [arXiv:1311.2601 [hep-th]].

\bibitem{Koile:2014vca}
  E.~Koile, N.~Kovensky and M.~Schvellinger,
  ``Hadron structure functions at small $x$ from string theory,''
  JHEP {\bf 1505} (2015) 001
  [arXiv:1412.6509 [hep-th]].

\bibitem{Koile:2015qsa}
  E.~Koile, N.~Kovensky and M.~Schvellinger,
  ``Deep inelastic scattering cross sections from the gauge/string duality,''
  JHEP {\bf 1512} (2015) 009
  [arXiv:1507.07942 [hep-th]].


\bibitem{Best:1997qp}
  C.~Best, M.~Gockeler, R.~Horsley, E.~M.~Ilgenfritz, H.~Perlt, P.~E.~L.~Rakow, A.~Schafer and G.~Schierholz {\it et al.},
  ``Pion and rho structure functions from lattice QCD,''
  Phys.\ Rev.\ D {\bf 56}, 2743 (1997)
  [hep-lat/9703014].

\bibitem{Brommel:2006zz}
  D.~Brommel {\it et al.} [QCDSF-UKQCD Collaboration],
  ``Quark distributions in the pion,''
  PoS LAT {\bf 2007} (2007) 140.

\bibitem{Chang:2014lva}
  L.~Chang, C.~Mezrag, H.~Moutarde, C.~D.~Roberts, J.~Rodr\'{\i}guez-Quintero and P.~C.~Tandy,
  ``Basic features of the pion valence-quark distribution function,''
  Phys.\ Lett.\ B {\bf 737} (2014) 23
  doi:10.1016/j.physletb.2014.08.009
  [arXiv:1406.5450 [nucl-th]].



\bibitem{Wijesooriya:2005ir}
  K.~Wijesooriya, P.~E.~Reimer and R.~J.~Holt,
  ``The pion parton distribution function in the valence region,''
  Phys.\ Rev.\ C {\bf 72}, 065203 (2005)
  [nucl-ex/0509012].

\bibitem{Holt:2010vj}
  R.~J.~Holt and C.~D.~Roberts,
  ``Distribution Functions of the Nucleon and Pion in the Valence Region,''
  Rev.\ Mod.\ Phys.\  {\bf 82}, 2991 (2010)
  [arXiv:1002.4666 [nucl-th]].

\bibitem{Reimer:2011}
P.~Reimer, R.~Holt and K.~Wijesooriya,
 ``The Partonic Structure of the Pion at Large-x,''
AIP Conference Proceedings 1369, 153 (2011); doi: 10.1063/1.3631531

\bibitem{Chang:2014gga}
  L.~Chang and A.~W.~Thomas,
  ``Pion Valence-quark Parton Distribution Function,''
  arXiv:1410.8250 [nucl-th].

\bibitem{Detmold:2003tm}
  W.~Detmold, W.~Melnitchouk and A.~W.~Thomas,
  ``Parton distribution functions in the pion from lattice QCD,''
  Phys.\ Rev.\ D {\bf 68}, 034025 (2003)
  [hep-lat/0303015].

\bibitem{Aicher:2011ai}
  M.~Aicher, A.~Schafer and W.~Vogelsang,
  ``Threshold-Resummed Cross Section for the Drell-Yan Process in Pion-Nucleon Collisions at COMPASS,''
  Phys.\ Rev.\ D {\bf 83}, 114023 (2011)
  [arXiv:1104.3512 [hep-ph]].

\bibitem{Aicher:2010cb}
  M.~Aicher, A.~Schafer and W.~Vogelsang,
  ``Soft-gluon resummation and the valence parton distribution function of the pion,''
  Phys.\ Rev.\ Lett.\  {\bf 105}, 252003 (2010)
  [arXiv:1009.2481 [hep-ph]].


\bibitem{Meyer:2004jc}
  H.~B.~Meyer and M.~J.~Teper,
  ``Glueball Regge trajectories and the pomeron: A Lattice study,''
  Phys.\ Lett.\ B {\bf 605} (2005) 344
  [hep-ph/0409183].

\bibitem{Meyer:2004hv}
  H.~Meyer and M.~Teper,
  ``Confinement and the effective string theory in $SU(N \rightarrow \infty)$: A Lattice study,''
  JHEP {\bf 0412} (2004) 031
  [hep-lat/0411039].

\bibitem{Meyer:2004gx}
  H.~B.~Meyer,
  ``Glueball regge trajectories,''
  hep-lat/0508002.


\bibitem{Bali:2013kia}
  G.~S.~Bali, F.~Bursa, L.~Castagnini, S.~Collins, L.~Del Debbio, B.~Lucini and M.~Panero,
  ``Mesons in large-N QCD,''
  JHEP {\bf 1306} (2013) 071
  [arXiv:1304.4437 [hep-lat]].

\bibitem{Bali:2013fya}
  G.~S.~Bali, L.~Castagnini, B.~Lucini and M.~Panero,
  ``Large-$N$ mesons,''
  PoS LATTICE {\bf 2013} (2014) 100
  [arXiv:1311.7559 [hep-lat]].


\bibitem{Polchinski:2002jw}
  J.~Polchinski and M.~J.~Strassler,
  ``Deep inelastic scattering and gauge / string duality,''
  JHEP {\bf 0305}, 012 (2003)
  [hep-th/0209211].

\bibitem{Jorrin:2016rbx}
  D.~Jorrin, N.~Kovensky and M.~Schvellinger,
  ``Towards 1/N Corrections to Deep Inelastic Scattering from the Gauge/Gravity Duality,''
  JHEP {\bf 1604} (2016) 113
  [arXiv:1601.01627 [hep-th]].

\bibitem{Gao:2014nwa}
  J.~H.~Gao and Z.~G.~Mou,
  ``Structure functions in deep inelastic scattering from gauge/string duality beyond single-hadron final states,''
  Phys.\ Rev.\ D {\bf 90}, no. 7, 075018 (2014)
  [arXiv:1406.7576 [hep-ph]].


\bibitem{Erlich:2005qh}
  J.~Erlich, E.~Katz, D.~T.~Son and M.~A.~Stephanov,
  ``QCD and a holographic model of hadrons,''
  Phys.\ Rev.\ Lett.\  {\bf 95} (2005) 261602
  [hep-ph/0501128].

\bibitem{DaRold:2005mxj}
  L.~Da Rold and A.~Pomarol,
  ``Chiral symmetry breaking from five dimensional spaces,''
  Nucl.\ Phys.\ B {\bf 721} (2005) 79
  [hep-ph/0501218].

\bibitem{Hambye:2005up}
  T.~Hambye, B.~Hassanain, J.~March-Russell and M.~Schvellinger,
  ``On the Delta I = 1/2 rule in holographic QCD,''
  Phys.\ Rev.\ D {\bf 74} (2006) 026003
  [hep-ph/0512089].

\bibitem{Hambye:2006av}
  T.~Hambye, B.~Hassanain, J.~March-Russell and M.~Schvellinger,
  ``Four-point functions and Kaon decays in a minimal AdS/QCD model,''
  Phys.\ Rev.\ D {\bf 76} (2007) 125017
  [hep-ph/0612010].

\bibitem{CaronHuot:2006te}
  S.~Caron-Huot, P.~Kovtun, G.~D.~Moore, A.~Starinets and L.~G.~Yaffe,
  ``Photon and dilepton production in supersymmetric Yang-Mills plasma,''
  JHEP {\bf 0612} (2006) 015
  [arXiv:hep-th/0607237].

\bibitem{Hassanain:2011ce}
  B.~Hassanain and M.~Schvellinger,
  ``Diagnostics of plasma photoemission at strong coupling,''
  Phys.\ Rev.\ D {\bf 85} (2012) 086007
  [arXiv:1110.0526 [hep-th]].

\bibitem{Hassanain:2011fn}
  B.~Hassanain and M.~Schvellinger,
  ``Plasma conductivity at finite coupling,''
  JHEP {\bf 1201} (2012) 114
  [arXiv:1108.6306 [hep-th]].

\bibitem{Hassanain:2010fv}
  B.~Hassanain, M.~Schvellinger,
  ``Towards 't Hooft parameter corrections to charge transport in strongly-coupled plasma,''
  JHEP {\bf 1010 } (2010)  068
  [arXiv:1006.5480 [hep-th]].

\bibitem{Hassanain:2012uj}
  B.~Hassanain and M.~Schvellinger,
  ``Plasma photoemission from string theory,''
  JHEP {\bf 1212} (2012) 095
  [arXiv:1209.0427 [hep-th]].

\bibitem{Aarts:2007wj}
  G.~Aarts, C.~Allton, J.~Foley, S.~Hands and S.~Kim,
  ``Spectral functions at small energies and the electrical conductivity in hot, quenched lattice QCD,''
  Phys.\ Rev.\ Lett.\  {\bf 99} (2007) 022002
  [hep-lat/0703008 [HEP-LAT]].

\bibitem{Hatta:2007he}
  Y.~Hatta, E.~Iancu and A.~H.~Mueller,
  ``Deep inelastic scattering at strong coupling from gauge/string duality: The Saturation line,''
  JHEP {\bf 0801} (2008) 026
  [arXiv:0710.2148 [hep-th]].

\bibitem{Hatta:2007cs}
  Y.~Hatta, E.~Iancu and A.~H.~Mueller,
  ``Deep inelastic scattering off a N=4 SYM plasma at strong coupling,''
  JHEP {\bf 0801} (2008) 063
  [arXiv:0710.5297 [hep-th]].

\bibitem{Hassanain:2009xw}
  B.~Hassanain and M.~Schvellinger,
  ``Holographic current correlators at finite coupling and scattering off a supersymmetric plasma,''
  JHEP {\bf 1004} (2010) 012
  [arXiv:0912.4704 [hep-th]].

\bibitem{Kim:1985ez}
  H.~J.~Kim, L.~J.~Romans and P.~van Nieuwenhuizen,
  ``The Mass Spectrum of Chiral N=2 D=10 Supergravity on S**5,''
  Phys.\ Rev.\ D {\bf 32} (1985) 389.
  doi:10.1103/PhysRevD.32.389

\bibitem{Myers:2006qr}
  R.~C.~Myers and R.~M.~Thomson,
  ``Holographic mesons in various dimensions,''
  JHEP {\bf 0609} (2006) 066
  [hep-th/0605017].

\bibitem{Auluck:2012}
  S.~K.~H.~Auluck,
  ``On the integral of the product of three bessel functions over an infinite domain,'' The
  Mathematica Journal, 14 (2012).

\bibitem{Nam:2012vm}
  S.~i.~Nam,
  ``Parton-distribution functions for the pion and kaon in the gauge-invariant nonlocal chiral-quark model,''
  Phys.\ Rev.\ D {\bf 86} (2012) 074005
  [arXiv:1205.4156 [hep-ph]].

\bibitem{Cutkosky:1983jd}
  R.~E.~Cutkosky,
  ``Harmonic Functions and Matrix Elements for Hyperspherical Quantum Field Models,''
  J.\ Math.\ Phys.\  {\bf 25} (1984) 939.

\bibitem{Aharony:2006rf}
  O.~Aharony, J.~Marsano and M.~Van Raamsdonk, ``Two loop partition
  function for large N pure Yang-Mills theory on a small S**3,''
  Phys.\ Rev.\ D {\bf 74} (2006) 105012
  [hep-th/0608156].

\bibitem{Libro}
  I.~S.~Gradshteyn and I.~M.~Ryzhik,
  ``Table of Integrals, Series and Products,'' Academic Press, 4th edition, 1980.

\bibitem{Hung:2010pe}
  L.~Y.~Hung and Y.~Shang,
  ``On 1-loop diagrams in AdS space,''
  Phys.\ Rev.\ D {\bf 83} (2011) 024029
  [arXiv:1007.2653 [hep-th]].




\end{thebibliography}
\end{document}